\documentclass[12pt,oneside,final]{IFUAPtesis}

\usepackage{setspace}
\usepackage{graphicx,epsfig,epsf,float,amsmath,amssymb,cite}
\usepackage[utf8]{inputenc}
\usepackage{chngcntr}
\usepackage{amsmath}
\usepackage{array}
\usepackage{float}
\usepackage{bm}
\usepackage{multirow}
\usepackage{float}
\usepackage{booktabs} 
\usepackage{multirow}
\usepackage[spanish]{babel}
\usepackage{color}

\graphicspath{ {images/} }

\counterwithin*{equation}{section}
\counterwithin*{equation}{subsection}
\counterwithin*{equation}{subsubsection}

\numberwithin{equation}{section}
\numberwithin{equation}{subsection}

\renewcommand*{\theequation}{%
  \ifnum\value{subsection}=0 %
    \thesection
  \else
    \thesubsection
  \fi
  .\arabic{equation}%
}

\begin{document}

\renewcommand{\tablename}{Tabla}
\title{Modelo de Tratamiento para Tumores en Presencia de Radiaci\'on}
\author{Diego Salda\~na Ulloa}
\department{Intituto de F\'{i}sica}
\degreemonth{Octubre} 
\degreeyear{2017}
\degree{Maestr\'ia en Ciencias}
\field{F\'isica}
\advisor{Dr. Oscar Sotolongo Costa}
\advisors {Dr. Andr\'es Fraguela Collar}
\maketitle
\tableofcontents














\newpage

\thispagestyle{plain}
\begin{center}
    
    \Large
    \vspace{0.9cm}
    \textbf{Resumen}
\end{center}
A lo largo de la historia se han documentado diversos avances sobre el entendimiento del c\'ancer. Estos avances provienen no solo de la medicina si no tambi\'en de \'areas como la fisiolog\'ia, la f\'isica, la qu\'imica y matem\'aticas. En el siglo pasado se inici\'o el estudio formal acerca de c\'omo modelar el crecimiento de c\'elulas cancer\'igenas desde la perspectiva matem\'atica. Existen varios modelos que pretenden simular este aumento en el n\'umero de c\'elulas tomando como datos iniciales ciertas caracter\'isticas del tejido a estudiar. Dentro de esos modelos, el modelo de Gompertz se apunta como un candidato id\'oneo debido a su forma sigmoidal (se sabe que un tumor tiene un tamaño l\'imite debido al consumo de los nutrientes de su entorno). De la misma forma, se han propuesto  diferentes tratamientos para combatir la enfermedad y dentro de los cuales se puede enunciar a la radioterapia. La afectaci\'on de la radiaci\'on sobre las c\'elulas puede modelarse mediante la fracci\'on de supervivencia celular que nos indica el porcentaje de c\'elulas que sobreviven o mueren tras un evento radiativo. Esta fracci\'on de supervivencia celular puede obtenerse mediante un principio de m\'axima entrop\'ia con Boltzman-Gibbs (caso extensivo) y ha sido ampliamente usada por la comunidad cient\'ifica desde el siglo pasado. En el presente trabajo combinamos el modelo de Gompertz con un nuevo t\'ermino de la fracci\'on de supervivencia celular (propuesta por O. Sotolongo et al) obtenida mediante un principio de m\'axima entrop\'ia con Tsallis (caso no extensivo) con el fin de llegar a una ecuaci\'on que modela el crecimiento/muerte celular a lo largo de los d\'ias de un tratamiento convencional de radioterapia (fraccionamiento est\'andar e hipofraccionamiento). Posteriormente se optimiza esta ecuaci\'on para encontrar las dosis \'optimas te\'oricas tales que logran reducir el tamaño de un tumor (reducir el n\'umero de c\'elulas cancer\'igenas). Para este caso se tomaron datos de  dos tipos de c\'ancer de mama: carcinoma ductal invasivo (CDI) y carcinoma ductal in situ (CDIS). Las dosis obtenidas son muy cercanas a las dosis utilizadas por los m\'edicos para ambos tipos de tumor. Adicionalmente se comprob\'o (anal\'iticamente) que el hipofraccionamiento presenta mejores resultados que el fraccionamiento est\'andar.  
















\startarabicpagination

\chapter{Introducci\'on}

De entre todas las enfermedades que afectan al ser humano el c\'ancer es una de las que, hist\'oricamente, lo han aquejado por m\'as tiempo. Su trayecto se remonta a los egipcios y se ha documentado a lo largo de muchas civilizaciones y \'epocas. El estudio formal de esta patolog\'ia data del siglo XVIII donde se realizaron las primeras observaciones de muestras de tejido canceroso bajo el microscopio \cite{49}. Desde ese tiempo y hasta la actualidad han habido grandes avances en investigaci\'on que van desde los tratamientos (que pueden ser paliativos) hasta la prevenci\'on.  

De entre esos tratamientos la radioterapia es la que, despu\'es de la cirug\'ia, ha tenido un trayecto m\'as largo y ha presentado excelentes resultados para algunos tipos de tumor. Este tipo de tratamiento fue descubierto en el siglo pasado poco despu\'es de las noticias sobre la existencia de la radiaci\'on ionizante. Al hablar de aplicaci\'on de radiaci\'on sobre el \'area afectada por un tumor tenemos que hablar de un dep\'osito de energ\'ia por unidad de masa (dosis de radiaci\'on). Esta dosis depositada, mediante un proceso de ionizaci\'on, lograr\'a provocar daños al ADN de las c\'elulas tumorales de tal modo que, eventualmente, la acumulaci\'on de multiples daños desencadenar\'a un proceso de muerte celular.

Sin embargo las dosis de radiaci\'on administradas han sido fijadas de acuerdo a la evidencia y experiencia de los m\'edicos a lo largo de los años. Emp\'iricamente se ha encontrado que dosis bajas administradas durante un largo periodo de tiempo, reportan mejores resultados tomando en cuenta la reducci\'on del tumor y el cuidado del paciente. A pesar de esto no existe herramienta alguna  que nos diga o confirme (te\'oricamente) que las dosis actualmente utilizadas son las id\'oneas o correctas.  

Es por ello que el objetivo de este trabajo consiste en presentar esa herramienta y realizar un an\'alisis sobre los resultados obtenidos (por lo menos a nivel te\'orico) tomando en cuenta las dosis utilizadas actualmente y tambi\'en proponer unas dosis que ser\'an consideradas \'optimas ya que lograr\'an reducir el n\'umero de c\'elulas cancer\'igenas. Dichas dosis \'optimas ser\'an resultado de una optimizaci\'on matem\'atica.

\section{El c\'ancer como enfermedad}

Como parte fundamental del desarrollo de la presente tesis, conviene saber algunas caracter\'isticas fundamentales del problema que estamos enfrentando. Por ello en este cap\'itulo describiremos los principios del proceso de inducci\'on al c\'ancer, el cual inicia con la aparici\'on de tumores malignos.

\subsection{C\'elulas cancer\'igenas y c\'ancer}

A lo largo de la existencia de un ser vivo, las c\'elulas de su organismo siguen un ciclo ordenado de divisi\'on y muerte celular. Este proceso de crecimiento es relativamente m\'as r\'apido durante su juventud. Al llegar a la etapa adulta el organismo contin\'ua con el proceso de divisi\'on a fin de reparar c\'elulas ``viejas'' o dañadas. 
El c\'ancer es una enfermedad donde las c\'elulas del organismo comienzan un proceso de crecimiento descontrolado. Dicho crecimiento se debe a ciertos defectos de c\'elulas que en lugar de sufrir el proceso de apoptosis, tienden a replicarse y nuevas c\'elulas comienzan a desarrollarse. 

Las c\'elulas cancer\'igenas presentan caracter\'isticas muy diferentes a las c\'elulas sanas ya que contin\'uan creciendo en n\'umero (sin morir), pueden llegar a formar masas de tejido anormal y persisten en su replicaci\'on aunque el agente externo que lo provoc\'o haya desaparecido. De igual modo, las c\'elulas cancerosas pueden transportarse por el organismo teniendo la capacidad de ``contagiar'' otros tejidos y diseminar la enfermedad. Una c\'elula sana se convierte en c\'elula cancer\'igena cuando ocurre una alteraci\'on en su ADN, interviniendo de manera directa en el proceso de replicaci\'on celular (crecimiento anormal y descontrolado). Si una c\'elula con el ADN alterado se replica, dar\'a lugar a copias id\'enticas con el ADN alterado. Este crecimiento anormal puede llevar a la formaci\'on de tumores si el n\'umero de c\'elulas cancer\'igenas es completamente grande. Cabe destacar que no todos los tumores son malignos, existen algunos considerados benignos. Los tumores benignos pueden ocasionar problemas dentro del \'area en que se desarrollan sin embargo no podr\'an diseminar la enfermedad (met\'astasis).  El c\'ancer como enfermedad engloba a todo un conjunto de enfermedades clasificadas de acuerdo al tejido u \'organo que afectan \cite{25}.

Podemos englobar al c\'ancer dentro de los principales tres tipos que existen: los sarcomas, los carcinomas y las leucemias y linformas. Los sarcomas son tumores malignos provenientes de tejido conjuntivo (como huesos), muscular y vasos sangu\'ineos. Los carcinomas son tumores formados a partir de tejido epitelial. Y por \'ultimo, la leucemia es un tipo de c\'ancer que afecta el proceso de producci\'on de leucocitos en la sangre mientras que los linfomas son tumores malignos de los ganglios linf\'aticos. 

\subsection{Factores causantes de c\'ancer}

Al d\'ia de hoy, las investigaciones sobre los factores que causan c\'ancer continuan en pie, se han podido indentificar varios de estos factores, aunque se pueden resumir en dos tipos: factores hereditarios y factores externos. Los factores hereditarios se refieren al origen gen\'etico del c\'ancer por herencia. Existen genes llamados protooncogenes (que estimulan el crecimiento celular) y genes supresores tumorales que en condiciones normales previenen y regulan la replicaci\'on celular \cite{25}. Alguna falla dentro de este tipo de genes o incluso la ausencia de ellos, propician a que el individuo pueda desarrollar la enfermedad. Esta falla generalmente esta relacionada con las mutaciones en el ADN, la rotura de cromosomas y el traslado cromos\'omico \cite{25}.  

Dentro de los factores externos se incluyen: los factores qu\'imicos, agentes radiativos y las infecciones o virus. Los factores qu\'imicos provocan que la continua exposici\'on a esas sustancias provoquen daños a nivel celular. En la literatura existe todo un tratamiento de porqu\'e y c\'omo estos factores afectan el organismo, y a su vez la incidencia de los mismos \cite{25}. Los agentes radiativos son las radiaciones ionizantes, los fotones atraviesan la c\'elula y pueden provocar daño al ADN como rotura de los cromosomas. Entre menor sea la longitud de onda de dicha radiaci\'on existe un mayor riesgo de daño celular, es por ello que se utlizan rayos X en el tratamiento de tumores.
Las infecciones o virus pueden desencadenar la creaci\'on y multiplicaci\'on de c\'elulas cancer\'igenas. Existen muchas evidencias que de distintos tipos de virus y bacterias estan relacionados con oncogenes y factores de crecimiento que estimulan el desarrollo de c\'elulas tumorales (para m\'as informaci\'on ver \cite{25}).

\subsection{Carcinog\'enesis}

A continuaci\'on describiremos el proceso de desarrollo del c\'ancer. 
Una c\'elula sana tiene la capacidad de convertirse en c\'elula cancer\'igena (cambiar su fenotipo) si es expuesta constantemente a un agente carcin\'ogeno, de tal forma que se produzca una mutaci\'on del ADN. La c\'elula alterada puede replicarse si los mecanismos de supresi\'on tumoral fallan. De este modo la c\'elula maligna contin\'ua replic\'andose acumulando errores en cada etapa de duplicaci\'on, por lo que el riesgo para del desarrollo de la enfermedad es inminente. 

El proceso de carcinog\'enesis puede dividirse en tres etapas: \textbf{iniciaci\'on}, un cambio permanente y heredable que ocurre a nivel del ADN de la c\'elula; \textbf{promoci\'on}, etapa de crecimiento tisular que da lugar a la formaci\'on de tumores, por consecuencia de alteraciones gen\'eticas las c\'elulas alteradas proliferan y contin\'uan acumulando daños; \textbf{progresi\'on}, cuando las c\'elulas cancer\'igenas siguen proliferando invadiendo tejidos adyacentes y metastizando otras partes del organismo \cite{26}.

Las c\'elulas tumorales presentan un consumo mayor de ox\'igeno, debido a ello contienen factores gen\'eticos que estimulan la angiog\'enesis tumoral \cite{25}. La angiog\'enesis tumoral est\'a relacionada a la continua proliferaci\'on de c\'elulas tumorales en el organismo (met\'astasis). Cabe destacar que una de cada diez mil c\'elulas que logre llegar al torrente sanguineo o el sistema linf\'atico, ser\'a capaz de desencadenar un cuadro de met\'astasis \cite{25,26}.  La met\'astasis generalmente es el proceso final de la enfermedad, cerca del 90\% de las muertes debidas a c\'ancer son debido a crecimientos de tumores en zonas alejadas del tumor primario (solo el 10\% se debe a tumores primarios) \cite{26}.

\subsection{Tratamientos contra la enfermedad}

En general los m\'etodos de tratamiento que existen para combatir el c\'ancer son: cirug\'ia, radioterapia y quimioterapia. La cirug\'ia es el m\'etodo con m\'as historial y es usado para eliminar tumores tempranos localizados as\'i como en combinaci\'on de tratamientos. La quimioterapia consiste en la aplicaci\'on de sustancias qu\'imicas antineopl\'asicas. Las sustancias atacan el ADN tanto de c\'elulas sanas como c\'elulas tumorales. De este modo se produce muerte celular de ambas partes. La radioterapia consiste en la aplicaci\'on de radiaci\'on ionizante en el \'area del tumor. La radiaci\'on afecta directamente al ADN de la c\'elula cancer\'igena provoc\'andole la muerte. Cabe destacar que dependiendo del tipo de tumor se escoge el tratamiento adecuado. Algunos tumores requieren la utilizaci\'on de los tres tipos de tratamientos: radioterapia y/o quimioterapia para reducir el tamaño del tumor y finalmente cirug\'ia para extirparlo completamente del organismo. A partir de ahora nos enfocaremos en la utilizacion de la radiaci\'on como tratamiento. 

\subsection{Radioterapia Oncol\'ogica}

Como se mencion\'o anteriormente, la radioterapia consiste en la aplicaci\'on de radiaci\'on ionizante en el \'area del tumor. La radiaci\'on est\'a compuesta por flujos continuos de part\'iculas fermi\'onicas (electrones, protones o neutrones) o fotones.   Esta radiaci\'on se denomina ionizante debido a que tiene la capacidad de ionizar \'atomos con los que interactuan (arrancar electrones de ellos). Los electrones arrancados dentro de la c\'elula salen disparados y pueden llegar a interaccionar con otros electrones para arrancarlos de su orbital. Este proceso puede continuar a fin de ionizar m\'as \'atomos y llegar a formar clusters de ionizaci\'on \cite{21}. La radiaci\'on ionizante produce lesiones en las mol\'eculas de ADN y pueden provocar: ruptura de cadenas simples o dobles, alteraci\'on de las bases, destrucci\'on de az\'ucares y formaci\'on de d\'imeros. Una vez que una c\'elula es alcanzada por una cantidad suficiente de radiaci\'on, se produce cualquiera de los daños descritos anteriormente. La c\'elula comienza el proceso de cit\'olisis: su membrana celular se descompone y pierde su capacidad de divisi\'on celular. Llegados a este punto podemos hablar de muerte celular como la p\'erdida irreversible en la capacidad de reproducci\'on celular \cite{19}. La muerte celular, despu\'es de un evento de radiaci\'on, se puede producir por diferentes tipos: apoptosis, autofagia, necrosis, senescencia (envejecimiento) celular y cat\'astrofe mit\'otica [21]. La mayor\'ia de la muerte celular por radiaci\'on ionizante ocurre en etapas tard\'ias de la cat\'astrofe mit\'otica y no inicialmente en respuesta a la radiaci\'on \cite{21}.   

\section{El Modelo de Gompertz}

El modelado matem\'atico se ha utilizado para describir el sistema inmune \cite{4} y como es su interacción con las células tumorales, con y sin tratamiento. Algunos modelos han sido usados para observar el papel del retraso de la respuesta del sistema inmune \cite{6}, tratamientos de radioviroterapia \cite{7}, quimioterapia \cite{8} e incluso cirugía \cite{9}. Estos modelos trabajan principalmente en la interacci\'on del tumor con el sistema inmune.

\subsection{Modelo de crecimiento tumoral}\label{sct:2.2}

Es importante conocer la din\'amica del crecimiento de c\'elulas canc\'erigenas a fin de proponer nuevas metodolog\'ias de tratamiento o mejorar las t\'ecnicas actuales. Es por ello que el crecimiento tumoral ha sido objeto de estudio desde el siglo pasado \cite{26}. A lo largo de los años, el peligro que conlleva la adquisici\'on de esta enfermedad as\'i como el aumento en la tasa de pacientes con c\'ancer, encamin\'o las investigaciones tanto m\'edicas como biol\'ogicas al desarrollo de una teor\'ia que explicara el proceso de crecimiento de neoplasias, es decir, la carcinog\'enesis, que fue descrita brevemente en el cap\'itulo anterior. De la misma forma, surgieron muchos modelos matem\'aticos que predec\'ian la tasa de crecimiento de c\'elulas cancer\'igenas a lo largo de un tiempo determinado.

El m\'as simple de todos los modelos consiste en suponer que el crecimiento tumoral sigue un comportamiento exponencial, es decir que las c\'elulas se dividen constantemente independientemene del tamaño del tumor \cite{28} por lo que se satisface

\begin{equation}
    \frac{dx}{dt}=rx,
\end{equation}
donde $x=x(t)$ representa el n\'umero de c\'elulas cancer\'igenas en un instante de tiempo $t$ y $r$ es la tasa de crecimiento celular. Recordemos que la soluci\'on a esta ecuaci\'on es

\begin{equation}
    x(t)=x_0 e^{rt},
\end{equation}
con $x_0$ representando el n\'umero de c\'elulas cancer\'igenas iniciales al tiempo $t=0$. 

Este modelo fue el primero en considerarse para aplicaciones m\'edicas \cite{28}. Algo interesante que podemos observar de este modelo es que nos permite predecir a manera de esbozo, el tiempo de duplicaci\'on para el tamaño del tumor. Si la soluci\'on al modelo anterior es

\begin{equation}
    2 x_0 = x_0 e^{rt},
\end{equation}
es decir, el doble de c\'elulas cancer\'igenas iniciales. Consideramos que ya que el modelo nos predice el n\'umero de c\'elulas a lo largo de una escala amplia de tiempo, entonces es factible pensar en la ecuaci\'on anterior. Despejando para $t$ obtenemos

\begin{equation}
    t=\frac{ln(2)}{r}.
\end{equation}

La ecuaci\'on anterior representa el tiempo que tarda el n\'umero de c\'elulas cancer\'igenas iniciales en duplicarse. Adem\'as este resultado nos sirve para saber cual es la tasa de crecimiento celular $r$. Por supuesto que este escenario es ideal y no representa realmente lo que sucede en varios tipos de tumores, sin embargo fue un primer paso para el desarrollo de modelos m\'as complejos. 

Para describir el crecimiento tumoral hay que tener ciertas premisas, como el crecimiento exponencial comentado anteriormente, de ese modo, un modelo m\'as general que el modelo exponencial consiste en tener una ecuaci\'on diferencial con potencias 
\begin{equation}
    \frac{dx}{dt}=r x(t)^{\mu}.
\end{equation}

Observamos que la soluci\'on corresponder\'a a la del modelo exponencial para $\mu=1$. La soluci\'on para este caso, considerando la potencia, es

\begin{equation}
    x(t)=\left( x_{0}^{1-\mu} + (1-\mu) rt \right)^{1/(1-\mu)},
\end{equation}
donde $x_0$ es el n\'umero de c\'elulas cancer\'igenas iniciales y $r$ la tasa de crecimiento. La complejidad de este modelo respecto del anterior nos puede llevar al descubrimiento por comparaci\'on, de procesos de crecimiento para diferentes tipos de tumores. Para el caso descrito por este modelo, existe un an\'alisis en \cite{29}. 

Otros tipos de modelos se usan para describir una variedad diferente de tumores. Existen ejemplos bastante analizados en \cite{24}, la decisi\'on por trabajar con uno u otro depende de las caracter\'isticas (biol\'ogicas) que quieran analizarse (tambi\'en depende de las caracter\'isticas matem\'aticas del modelo). De igual manera se pueden crear modelos completamente nuevos para el crecimiento tumoral, solo hay que ser precisos con las premisas biol\'ogicas que estar\'an dadas por las caracter\'isticas que observemos del comportamiento celular y posteriormente compararlo con datos experimentales. 

\section{ Modelo de Gompertz para el crecimiento tumoral}

De todos los modelos existentes analizaremos uno que es de relevancia debido a sus buenas predicciones para varios tipos de tumor \cite{30}, el llamado modelo de Gompertz.  Este modelo fue propuesto en 1925 por Benjamin Gompertz para explicar el crecimiento poblacional de individuos. Un siglo despu\'es se utiliz\'o para describir crecimientos poblacionales de c\'elulas, espec\'ificamente c\'elulas cancer\'igenas \cite{30,31} (que conjuntamente forman un tumor). El modelo de Gompertz no fue creado con la intenci\'on de explicar otros tipos de crecimientos que no fueran el de los individuos de una comunidad, sin embargo debido a sus caracter\'isticas se redescubri\'o y al d\'ia de hoy se sigue utilizando como modelo de investigaciones. 

Existen muchas formas del modelo de Gompertz, cada una difiere de los par\'ametros a considerarse. Lo que define que un modelo sea considerado como Gompertz es la forma de la soluci\'on de ese modelo, debe tener un comportamiento sigmoidal (en forma de "S"). De ah\'i que podamos decir que la curva de Gompertz pertenece al tipo de las llamadas funciones sigmoides. 

\subsection{Modelo Generalizado de Dos Par\'ametros}

En este caso se considera que la tasa de cambio en el n\'umero de c\'elulas cancer\'igenas es una diferencia entre la tasa de crecimiento y la tasa de degradaci\'on (muerte celular). Adem\'as la tasa de cambio es proporcional al tama\~no del tumor \cite{23} 

\begin{equation}
    \frac{dx}{dt}=ax^{\mu}-bx^{\nu}, 
\end{equation}
donde $x=x(t)$ es el n\'umero de c\'elulas cancer\'igenas en un instante $t$ y $a,b,\mu, \nu$ son par\'ametros relacionados con la tasa de crecimiento. En \cite{23} se describe la metodolog\'ia para obtener el llamado modelo de Gompertz generalizado

\begin{equation}
    \frac{dx}{dt}=ax^{\mu}-bx^{\mu}\text{ln}x, 
\end{equation}

si $\mu=1$ obtenemos el modelo de Gompertz

\begin{equation}
    \frac{dx}{dt}=ax-bx\text{ln}x .
\end{equation}

\subsection{Modelo Gen\'erico}

En este tipo de modelo se supone que la tasa de cambio del n\'umero de c\'elulas es proporcional a una funci\'on que incrementa con el tama\~no y otra que disminuye con el tama\~no

\begin{equation}
    \frac{dx}{dt}=\frac{r}{N^n}x^{1-np} (N^n -x^n)^{1+p} 
\end{equation}

donde $N$ es la capacidad m\'axima de carga (el tamaño m\'aximo que puede alcanzar el tumor) y $r,n,p$ constantes relacionadas con la tasa de crecimiento. De acuerdo con \cite{33} mediante un sencillo cambio de variable en la soluci\'on de la ecuaci\'on anterior, podemos obtener el llamado modelo Hyper-Gompertz que est\'a dado por

\begin{equation}
    \frac{dx}{dt}= r x \left( \text{ln} \frac{N}{x} \right)^{1+p}. 
\end{equation}

Si en la ecuaci\'on anterior $p=0$ obtenemos de nuevo el modelo de Gompertz 

\begin{equation}\label{eq12}
    \frac{dx}{dt}=r x \text{ln}\frac{N}{x}. 
\end{equation}

Hay que destacar que todos estos modelos son equivalentes respecto de su comportamiento en el crecimiento, la elecci\'on de uno u otro depender\'a del an\'alisis que se lleve a cabo.

En el presente trabajo utilizaremos el modelo de Gompertz simple, el cual est\'a dado por

\begin{equation} \label{31}
    \frac{dx(t)}{dt}= r x(t) ln \left( \frac{N}{x(t)} \right),
\end{equation}

donde $x(t)$ es el n\'umero de c\'elulas cancer\'igenas en el instante de tiempo $t$; $N$ es la capacidad de carga del sistema, es decir, el tamaño o n\'umero m\'aximo de c\'elulas que se pueden alcanzar con los nutrientes disponibles y $r$ es la constante relacionada a la habilidad proliferativa de las c\'elulas, una tasa de crecimiento.

\section{Resoluci\'on del modelo de Gompertz}

Procederemos a resolver la ecuaci\'on diferencial (\ref{31}). Definimos el cambio de variable $u=x/N$ y sustituimos en la ecuaci\'on,

\begin{equation}
    \frac{d}{dt} u = -r u ln \left( u \right),
\end{equation}

separando variables obtenemos,

\begin{equation}
    \int \frac{du}{u ln(u)} = - \int r dt,
\end{equation}

haciendo de nuevo un cambio de variable $w=ln(u)$ y $dw=du/u$,

\begin{equation}
    \int \frac{dw}{w}= - \int rdt.
\end{equation}

De modo que

\begin{equation}
    ln | ln(u) |=-rt+c,
\end{equation}

resolviendo para $u$ llegamos a

\large
\begin{equation}
    u=e^{ce^{-rt}}.
\end{equation}

\normalsize
Por lo que la soluci\'on general es 

\begin{equation} \label{36}
    x=N\left( e^{c} \right)^{e^{-rt}}.
\end{equation}

Al detectar un tumor cancer\'igeno lo primero que se evalua es el tamaño de este, por lo que $x(0)=X_0$ c\'elulas cancer\'igenas es una condici\'on inicial que resulta de la detecci\'on de un tumor. Aplicando esta condici\'on inicial en (\ref{36}) obtenemos la soluci\'on anal\'itica al modelo de Gompertz

\large
\begin{equation} \label{37}
    x(t)=N e^{ ln \left( \frac{X_0}{N} \right) e^{-rt} }.
\end{equation}

\normalsize
Este resultado nos da un estimado del n\'umero de c\'elulas cancer\'igenas a lo largo de una escala de tiempo $t$ por lo que se pueden hacer predicciones para determinar el comportamiento de un tumor basados en datos experimentales. Al aplicar un ajuste de los datos experimentales (de varios tumores de un mismo tipo) con la soluci\'on (\ref{37}) se puede llegar a caracterizar el tumor.

\begin{figure}[h!]
    \centering
    \includegraphics[scale=0.5]{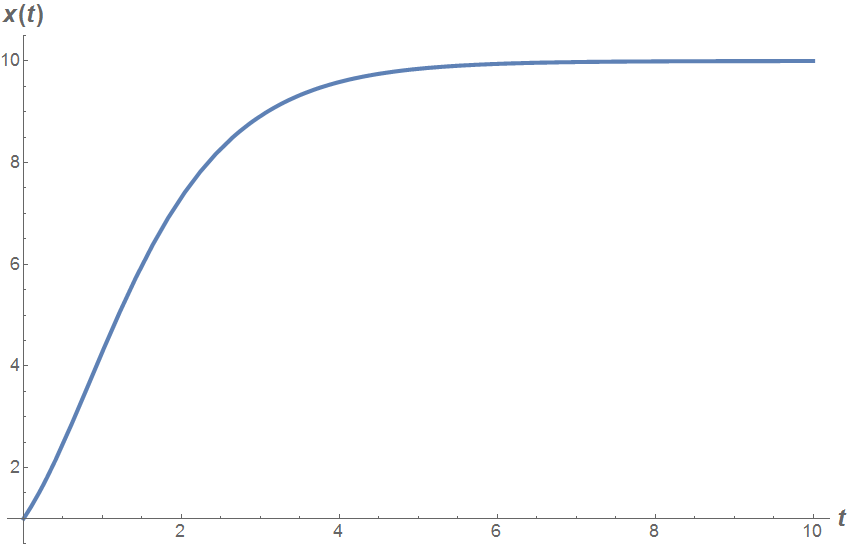}
    \caption{Ejemplo de crecimiento de c\'elulas cancer\'igenas para los valores especificados en la pagina anterior. Las unidades de $t$ no han sido especificadas pues depende de las unidades de $r$}
    \label{g1}
\end{figure}

La siguiente gr\'afica nos muestra la soluci\'on del modelo para $X_0=1$, $r=1$ y $N=10$. Los valores anteriores solo son ilustrativos para apreciar el potencial del modelo. Debido a esta forma sigmoidal de la ecuaci\'on de Gompertz, se ha considerado una candidata id\'onea para explicar procesos de crecimiento biol\'ogicos. Desde hace bastante tiempo se conoce que el crecimiento de cualquier tejido tiene un l\'imite m\'aximo del cual no puede pasar. Las investigaciones han encontrado que el c\'ancer se comporta de la misma manera, existe un tamaño m\'aximo (n\'umero de c\'elulas m\'aximo) a partir del cual el tumor deja de crecer \cite{32}. Obviamente esto entra como la capacidad de carga $N$ en el modelo.

Observamos que de acuerdo al modelo de Gompertz, el comportamiento de un tumor depender\'a del n\'umero inicial de c\'elulas malignas, la tasa de crecimiento $r$ y la capacidad de carga $N$. En instancia, solo mediante un ajuste se podr\'ia conocer estos valores pero lo que nos interesa es utilizar este modelo como m\'etodo de predicci\'on m\'as que como m\'etodo de ajuste.

\section{Modelo de Gompertz y Radiaci\'on}

El modelo de Gompertz representa una potente herramienta mediante la cual podemos caracterizar diferentes tipos de tumores. Una vez caracterizados, es \'util usar los datos obtenidos para mejorar los m\'etodos de diagn\'ostico o los tratamientos actuales. Sin embargo un enfoque diferente pero con potencial de arrojar buenos resultados en t\'erminos de diagn\'ostico-tratamiento, consistir\'ia en unir un modelo de crecimiento tumoral con un modelo de daño celular basado en uno de los tratamientos m\'as comunes al diagnosticarse un tumor, la radioterapia (radiaci\'on).

En el apartado anterior discutimos brevemente el proceso de muerte celular debido a la incidencia de radiaci\'on en un conjunto de c\'elulas (tejido). Ahora estamos interesados en el n\'umero o fracci\'on de c\'elulas que mueren debido a esa radiaci\'on.

\subsection{Curvas de Supervivencia}

Para analizar la fracci\'on de c\'elulas supervivientes bajo una dosis de radiaci\'on debemos recurrir a las curvas de supervivencia. Estas curvas son una representaci\'on gr\'afica accerca de c\'omo la fracci\'on de c\'elulas supervivientes ($F_S$) disminuyen conforme la dosis ($D$) incrementa, algunas curvas de supervivencia para distintos tipos de c\'elula pueden encontrarse en \cite{19}.
La fracci\'on de supervivencia celular puede modelarse matem\'aticamente tomando datos experimentales del laboratorio. Una part\'icula ionizante que atraviese una c\'elula puede desencadenar una serie de eventos letales que incrementar\'an conforme la dosis aumente, de este modo, la fracci\'on de supervivencia celular se comporta como una funci\'on exponencial y est\'a dada por \cite{19}

\begin{equation}
    F_S = e^{-D/D_0}
\end{equation}

La fracci\'on de supervivencia celular se comporta como una distribuci\'on de Poisson teniendo en cuenta que una dosis de radiaci\'on puede producir $n$ eventos letales. $F_S$ es un n\'umero entre $0$ y $1$. Encontrar la fracci\'on de supervivencia celular para una cierta dosis de radiaci\'on ha sido uno de los objetivos fundamentales de la radiobiolog\'ia  \cite{19,20}. 
Existen muchos modelos matem\'aticos sobre la forma de $F_S$ teniendo en cuenta el n\'umero de colisiones y objetivos, es decir, la interacci\'on de las part\'iculas de radiaci\'on (fotones o electrones) con los \'atomos de las mol\'eculas que forman el ADN de las c\'elulas. Tenemos, por ejemplo, el modelo lineal (colisi\'on \'unica y objetivo \'unico donde la muerte celular proviene del daño a un solo objetivo debido a un evento de ionizaci\'on)

\begin{equation}
    F_S = e^{-\alpha D},
\end{equation}

 y $\alpha$ es un coeficiente lineal que refleja la radiosensibilidad celular. El modelo lineal cumple la propiedad aditiva
 
 \begin{equation}
 F_S (D_1+D_2)=F_S (D_1) F_S (D_2),
 \end{equation}
 
 que nos indica que los efectos de la radiaci\'on son acumulativos \cite{52}. Tambi\'en existe el modelo lineal cuadr\'atico (colisi\'on doble y objetivo \'unico, donde la muerte celular se debe a dos subeventos letales debido la interacci\'on de dos part\'iculas independientes con un solo objetivo)

\begin{equation}\label{LQME}
    F_S = e^{-\alpha D - \beta D^2}
\end{equation}

 y $\beta$ es un coeficiente cuadr\'atico relacionado con el mecanismo de reparaci\'on celular \cite{11}. Para este modelo se cumple que
 
 \begin{equation}
 F_S(D_1+D_2) < F_S (D_1) F_S (D_2),
 \end{equation}
 
 por lo que en la transici\'on del modelo lineal al modelo cuadr\'atico, la propiedad aditiva se pierde y el principio de superposici\'on no se cumple \cite{20}. Esto representa un problema ya que cualquier tipo de radiaci\'on continua sobre un tejido vivo, debe permitir que la dosis pueda separarse en intervalos finitos y de ese modo el efecto resultante sobre el tejido debe ser el mismo \cite{52}. Lo anterior sugiere que otra forma de la fracci\'on de supervivencia celular ser\'ia necesaria tal que lograra cumplir la propiedad aditiva. Adicionalmente, se sabe que el modelo lineal y el lineal cuadr\'atico ajustan con los datos experimentales para dosis bajas y dosis moderadas, respectivamente.  Es por ello que las curvas de supervivencia, tomando uno u otro modelo, diferir\'an de c\'omo ajustan mejor a los datos experimentales \cite{20,21}.

\subsection{Principio de M\'axima Entrop\'ia y Supervivencia Celular}

En \cite{20} se analiza que si $F_S$ puede verse como una funci\'on de probabilidad, entonces el encontrar una nueva forma para la fracci\'on de supervivencia celular puede tratarse mediante la aplicaci\'on de un principio de m\'axima entrop\'ia. El principio de m\'axima entrop\'ia consiste en calcular su m\'aximo mediante el m\'etodo de los coeficientes de Lagrange \cite{18, 20}.  Si $p(E)$ es la densidad de probabilidad de muerte celular y $E(D)$ una forma adimensional de la dosis de radiaci\'on (efecto tisular introducido en radiobiolog\'ia \cite{21}) tambi\'en expresado como $E=-ln(F_S)$. Entonces, la entrop\'ia de Boltzmann-Gibbs puede expresarse como

\begin{equation}
    S=\int_{\Omega} p(E) ln \left[ p(E) \right]dE,
\end{equation}
donde $\Omega$ representa todos los estados de E. Para aplicar el principio de m\'axima entrop\'ia se imponen algunas condiciones como la normalizaci\'on

\begin{equation}
    \int_{\Omega} p(E)dE=1
\end{equation}

y la existencia del valor medio (dado que $p(E)$ es una densidad de probabilidad)

\begin{equation}
    \int_{\Omega} p(E)EdE= <E>.
\end{equation}

Tomando las tres expresiones anteriores y aplicando un principio de m\'axima entrop\'ia, obtenemos el modelo lineal de radiobiolog\'ia (ver ap\'endice B).
Desde hace tiempo se sabe que la Mec\'anica Estad\'istica de Boltzmann-Gibbs (BG) logra describir los problemas s\'i es que las interacciones entre las partes del sistema son de corto alcance o d\'ebiles. Paralelamente, la entrop\'ia en este caso se dice que es extensiva ya que $ S(A+B)=S(A)+S(B) $. Sin embargo, si en el sistema, todas las partes  est\'an fuertemente correlacionadas, BG ya no puede aplicarse y es necesario otro enfoque. Se debe buscar otra definici\'on de entrop\'ia que tome en cuenta las interacciones entre los componentes y que recobre, en alg\'un punto, la entrop\'ia de BG (nos referimos a un enfoque no extensivo que recupere la extensividad en algun l\'imite). Para este caso, el principio de m\'axima entrop\'ia (MaxEnt) se realizar\'a con la entrop\'ia de Tsallis. Esta forma de la entrop\'ia fue propuesta en 1988 por Constantino Tsallis como una generalizaci\'on de la entrop\'ia de Boltzmann-Gibbs. Desde entonces se ha usado en el an\'alisis de diferentes problemas de \'indole natural \cite{20}. 

La entrop\'ia de Tsallis se define

\begin{equation}
    S_q = \frac{1}{q-1} \left[ 1-\int_{\Omega} p^{q} (E) dE \right],
\end{equation}

donde $q$ es el \'indice de no extensividad. En \cite{20} se dice que la entrop\'ia de Tsallis es una buena candidata como herramienta para la obtenci\'on de una nueva forma de $F_S$ debido a su no extensividad. Esto es que $S_q (A+B)= S_q(A) + S_q (B) +(1-q) S_q (A) S_q (B)$ adem\'as, en la transici\'on del modelo lineal al cuadr\'atico, $F_S$ pierde la propiedad aditiva, como ya se habia comentado.

Para aplicar el principio de m\'axima entrop\'ia debemos postular la existencia de una cantidad de dosis absorbida $D_0$ despu\'es de la cual ninguna c\'elula sobrevive (equivalentemente $E_0 = \beta_0 D_0$ que es una forma adimensional de la dosis de radiaci\'on). Ya que $\Omega$ son los estados accesibles de E, entonces la entrop\'ia de Tsallis quedar\'ia 

\begin{equation}
    S_q = \frac{1}{q-1} \left[ 1-\int_{0}^{E_0} p^q (E) dE \right],
\end{equation}

la condici\'on de normalizaci\'on

\begin{equation}
    \int_{0}^{E_0} p(E)dE=1
\end{equation}

y el valor q-medio (que es finito)

\begin{equation}
    \int_{0}^{E_0} p^q (E)dE=<E>_q \ \ < \infty.
\end{equation}

Para calcular el m\'aximo de $S_q$ bajo las condiciones anteriores debemos utilizar el m\'etodo de los multiplicadores de Lagrange. El desarrollo de este paso puede verse en el ap\'endice A, obtenemos un valor para $p(E)$ y teniendo en cuenta que la fracci\'on de supervivencia celular es

\begin{equation}
    F_S (E)= \int_{E}^{E_0} p(x)dx,
\end{equation}

desarrollando la integral con el resultado para $p(E)$ del ap\'endice A, obtenemos

\begin{equation}\label{2.5.2.8}
    F_S (E) = \left( 1-\frac{E}{E_0} \right)^{2-q/1-q}= \left( 1-\frac{D}{D_0} \right), ^{\gamma}
\end{equation}

con $q<1$ para $E<E_0$,  $F_s (D)=0$ $\forall D \geq D_0$, $\gamma=\frac{2-q}{1-q}$ y $D_0 = E_0/\beta$. La ecuaci\'on anterior representa la fracci\'on de supervivencia celular $F_S$ en t\'erminos de cantidades que podemos medir experimentalmente ($\gamma$, $D$ y $D_0$). Esta ecuaci\'on fue propuesta por primera vez en \cite{20} y ah\'i tambi\'en se hace una comparaci\'on de las curvas de supervivencia para diferentes tipos de c\'elula as\'i como de este modelo con el modelo lineal cuadr\'atico \ref{LQME}. En \cite{20} se concluye que este nuevo modelo para la fracci\'on de supervivencia celular ajusta para todos los rangos de radiaci\'on inclusive para altas dosis donde el modelo lineal cuadr\'atico presenta fallas, debido a esto lo consideramos un excelente candidato para el presente trabajo.  

Algo importante a destacar es que definiendo expresiones del tipo \cite{52, 53}

\begin{equation}
exp_{\gamma} (x) = \left[ 1+ \frac{x}{\gamma} \right]^{\gamma},
\end{equation}

$\gamma$-exponente; su funci\'on inversa, 
\begin{equation}
ln_{\gamma} (exp_{\gamma} (x)) = x
\end{equation}

$\gamma$-logaritmo y el producto de dos n\'umeros $x$ e $y$ como

\begin{equation}
x \otimes_{\gamma} y = exp_{\gamma} \left[ln_{\gamma} (x) + ln_{\gamma}(y)\right]=\left[ x^{1/ \gamma} + y^{1/ \gamma} -1 \right]^{\gamma}
\end{equation}

$\gamma$-producto, podemos encontrar que la nueva forma de la fracci\'on de supervivencia celular cumple el principio de superposici\'on anteriormente descrito: si las dosis son aditivas entonces la fracci\'on de supervivencia celular es multiplicativa. Esto es

\begin{equation}
F_S (ND) = \left[\bigotimes_{i=1}^{N} \right]_{\gamma} F_S (D_i),
\end{equation}

donde $\left[\bigotimes_{i=1}^{N}\right]$ denota la m\'ultiple aplicaci\'on del $\gamma$-producto \cite{52}. Por lo que es una raz\'on m\'as para considerar esta nueva forma de la fracci\'on de supervivencia celular como una mejor candidata para ser utilizada en radiobiolog\'ia y modelos de interaccion c\'elula-radiaci\'on.

\section{Presentaci\'on del Contenido}

Una vez delimitadas las herramientas a utilizar durante el desarrollo de este trabajo, podemos continuar y exponer lo relevante de cada cap\'itulo con el fin de presentar una idea general.

En el cap\'itulo 2 el modelo de crecimiento tumoral y el de la fracci\'on de supervivencia celular son combinados con el objetivo de obtener un modelo de predicci\'on acerca de como crecen y mueren las c\'elulas cancer\'igenas a lo largo de los d\'ias de un tratamiento convencional de radioterapia.

El cap\'itulo 3 se enfoca en minimizar el n\'umero de c\'elulas cancer\'igenas sujeto a una serie de restricciones referente a las dosis. Con la expresi\'on obtenida en el cap\'itulo anterior para el modelo de predicci\'on se logra minimizar este n\'umero de c\'elulas cancer\'igenas mediante m\'etodos de busqueda directa. Dichos m\'etodos simplifican las operaciones computacionales que se deben llevar a cabo para obtener los resultados.

En el cap\'itulo 4 se presentan los resultados de las optimizaciones computacionales para las dosis. Como ejemplo se tom\'o el caso de c\'ancer de mama, particularmente carcinoma ductal invasivo (CDI) y carcinoma ductal in situ (CDIS). En primer lugar se muestra como se reduce el n\'umero de c\'elulas cancer\'igenas de estos tumores tomando en cuenta los m\'etodos actuales de tratamiento (dosificaci\'on est\'andar). Posteriormente, para ambos tipos de tumor, se obtienen las dosis \'optimas tales que disminuyen el n\'umero de c\'elulas cancer\'igenas. Estos resultados se muestran para cada tipo de tumor y para cada m\'etodo de optimizaci\'on. Al final de cap\'itulo se toma un m\'etodo de optimizaci\'on basado en gradientes para comparar con los resultados obtenidos.

Finalmente en el c\'apitulo 5 se exponen las conclusiones pertinentes. Adicionalmente se incluyen tres ap\'endices donde se muestra la obtenci\'on de las fracciones de supervivencia celular tomando en cuenta la entrop\'ia de Tsallis y la de Boltzman-Gibbs as\'i como varios m\'etodos de optimizaci\'on matem\'atica.

\chapter{Modelo de Predicci\'on Sobre la Proliferaci\'on y Muerte Celular Bajo una Dosis de Radiaci\'on}

Hemos obtenido una forma de cuantificar el n\'umero de c\'elulas que mueren bajo radiaci\'on, ahora lo aplicaremos al modelo de Gompertz de forma que obtengamos una especie de modelo de predicci\'on para el n\'umero de c\'elulas cancer\'igenas que son incididas por cierta dosis de radiaci\'on $D$.  

La forma de unir estos dos modelos es la siguiente. Supongamos el modelo de Gompertz ~\eqref{31}, sin un proceso de control el n\'umero de c\'elulas cancer\'igenas crecer\'a hasta un valor l\'imite, como ya lo hemos comentado, y la soluci\'on obedecer\'a la ecuaci\'on ~\eqref{37}, digamos un $x_0 (t)$. La ecuaci\'on de Gompertz tendr\'a la forma

\begin{equation}
   \dot{x_0} (t)=r x_0 (t) ln \left( \frac{N}{x_0 (t)} \right) 
\end{equation}

Sin embargo, si aplicamos una cierta dosis de radiaci\'on $D_1$ despu\'es de ser detectado el tumor y queremos modelar nuevamente su comportamiento a lo largo del tiempo mediante Gompertz (obtener, por ejemplo, $x_1(t)$), el n\'umero de c\'elulas iniciales a tomarse en cuenta (condici\'on inicial para resolver la ecuaci\'on diferencial) ya no corresponder\'a a $x_0 (t_0=0)=X_0$, m\'as bien ser\'a la fracci\'on de supervivencia celular $F_S (D_1)$ multiplicada por el n\'umero de c\'elulas cancer\'igenas en el instante en que es aplicada la dosis de radiaci\'on, es decir 

\begin{equation}
    x_1 (t_1) = \left(1-\frac{D_1}{D_{max}} \right)^{\gamma} x_{0} (t_1),
\end{equation}

la ecuaci\'on de Gompertz para este caso es

\begin{equation}
    \dot{x_1} (t)=r x_1 (t) ln \left( \frac{N}{x_1 (t)} \right). 
\end{equation}

$x_1 (t)$ modelar\'a un nuevo comportamiento de las c\'elulas cancer\'igenas. Aplicando de nuevo una dosis de radiaci\'on, ahora $D_2$, la condici\'on inicial se ve nuevamente modificada y ser\'a $F_S(D_2)$ multiplicada por el n\'umero de c\'elulas cancer\'igenas que existan en ese instante

\begin{equation}
    x_2 (t_2) = \left(1-\frac{D_2}{D_{max}} \right)^{\gamma} x_{1} (t_2),
\end{equation}

y la ecuaci\'on general de Gompertz es

\begin{equation}
    \dot{x_2} (t)=r x_2 (t) ln \left( \frac{N}{x_2 (t)} \right). 
\end{equation}

Podemos seguir haciendo la iteraci\'on y obtener una forma para la ecuaci\'on de Gompertz de la $n$-\'esima sesi\'on de un tratamiento 

\begin{equation} \label{26}
    \dot{x_n} (t)=r x_n (t) ln \left( \frac{N}{x_n (t)} \right), 
\end{equation}

con condici\'on inicial

\begin{equation} \label{2.7}
    x_n (t_n) = \left(1-\frac{D_n}{D_{max}} \right)^{\gamma} x_{n-1} (t_n).
\end{equation}

En este caso $t \in [t_n, t_{n+1}]$ con $n=1,2,3,...$.

Para resolver, hacemos un cambio de variable (al igual que en el cap\'itulo anterior), $u_n (t) = x_n (t)/N$

\begin{equation}
    \frac{d}{dt} u_n (t) = -r u_n (t) ln \left( u_n (t) \right),
\end{equation}

separando variables obtenemos,

\begin{equation}
    \int \frac{d u_n}{u_n (t) ln(u_n (t))} = - \int r dt,
\end{equation}

aplicando de nuevo un cambio de variable $w_n=ln(u_n)$ y $dw_n=du_n / u_n$,

\begin{equation}
    \int \frac{dw_n}{w_n}= - \int rdt.
\end{equation}

Llegamos a

\begin{equation}
    ln | ln(u_n) |=-rt+c,
\end{equation}

resolviendo para $u_n$ 

\large
\begin{equation}
    u_n (t)=e^{ce^{-rt}}.
\end{equation}

\normalsize
la soluci\'on general es 

\begin{equation} \label{213}
    x_n (t)=N\left( e^{c} \right)^{e^{-rt}}.
\end{equation}

Tomando la condici\'on inicial (\ref{2.7}) tenemos

\large
\begin{equation}
    x_n (t_n)= \alpha_n x_{n-1} (t_n) = N (e^c)^{e^{-rt_n}}
\end{equation}

\normalsize
donde $\alpha_n \equiv (1-D_n/D_{max})^{\gamma}$. Despejando para $e^c$  

\large
\begin{equation}
    e^{c}= \left( \frac{\alpha_n x_{n-1} (t_n)}{N} \right)^{e^{rt_n}}.
\end{equation}

\normalsize
Finalmente, sustituyendo en (\ref{213}) obtenemos

\Large
\begin{equation} \label{2.6.16}
    x_n (t) = N e^{ln\left( \frac{\alpha_n x_{n-1} (t_n)}{N} \right) e^{- r(t-t_n)}}
\end{equation}

\normalsize
Esta ecuaci\'on nos da el comportamiento del n\'umero de c\'elulas cancer\'igenas en un tiempo $t$ para una $n$-\'esima sesi\'on de tratamiento con radiaci\'on. A continuaci\'on se muestra un esquema representativa de la propuesta del modelo de predicci\'on.

\begin{figure}[h!]
    \centering
    \includegraphics[scale=0.45]{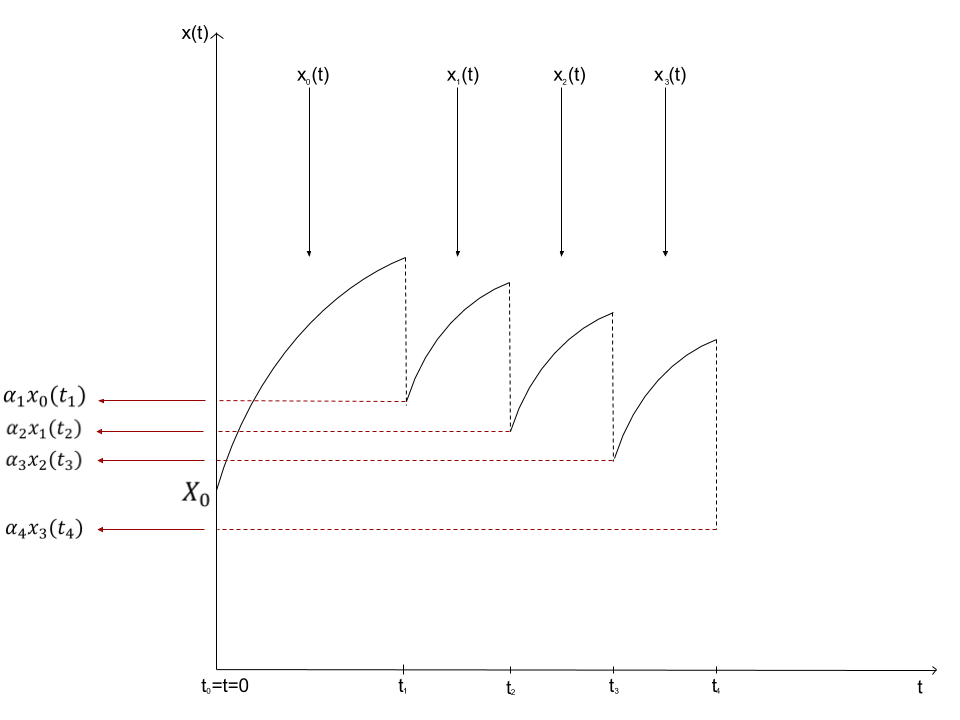}
    \caption{Gr\'afica de la ejemplificaci\'on del modelo de predicci\'on}
    \label{g2}
\end{figure}

De acuerdo a la gr\'afica anterior, inicialmente el tumor se detecta en $t_0=t=0$ con $X_0$ c\'elulas cancer\'igenas (tamaño del tumor), un modelo de Gompertz puede plantearse para este caso dando una curva $x_0(t)$. En un tiempo $t_1$ se aplica una dosis $D_1$ de radiaci\'on lo que hace que el n\'umero de c\'elulas cancer\'igenas disminuya (debido a los mecanismos de muerte celular provocados por la radiaci\'on ) hasta un valor $\alpha_1 x_0 (t_1)$. Despu\'es de la dosis de radiaci\'on las c\'elulas supervivientes continuar\'an replic\'andose, el modelado del crecimiento se har\'a nuevamente con una curva de Gompertz para $x_1(t)$ y la condici\'on inicial ahora ser\'a $x_1(t_1)=\alpha_1 x_0 (t_1)$. Posteriormente se aplica otra dosis de radiaci\'on $D_2$ que provoca la disminuci\'on de c\'elulas cancer\'igenas hasta un valor $\alpha_2 x_1(t_2)$, las c\'elulas se replicar\'an de nuevo y obedecer\'an una curva de Gompertz dada por $x_2(t)$, la condici\'on inicial de c\'elulas cancer\'igenas para obtener esta gr\'afica ser\'a $x_2(t_2)=\alpha_2 x_1 (t_2)$. La iteraci\'on se continuar\'a aplicando $n$ veces.

El objetivo fundamental de esta propuesta es reducir el tamaño del tumor mediante este proceso controlado de aplicaciones sucesivas de radiaci\'on. Lo novedoso de esta propuesta es que se puede predecir el n\'umero de aplicaciones de dosis de radiaci\'on para reducir el tamaño del tumor hasta un punto en el que sea conveniente (un cirujano pudiera extraerlo, por ejemplo). 

Uno de los inconvenientes que se presenta es la determinaci\'on de los par\'ametros $N$, $r$, $D_{max}$, $t_{min}$ (tiempo m\'inimo que deber\'ia de durar el tratamiento a fin de reducir el tamaño del tumor) y $x_{umbral}$ (el tamaño/n\'umero del tumor/c\'elulas cancer\'igenas ideal u objetivo  despu\'es de ser reducido por las aplicaciones sucesivas de radiaci\'on). La identificaci\'on  de estos par\'ametros es crucial para considerar este modelo como algo \'optimo en t\'erminos de predicci\'on. En el cap\'itulo siguiente nos enfocaremos en una idea de c\'omo determinar estos par\'ametros.


\chapter{Minimizaci\'on del N\'umero de C\'elulas Cancer\'igenas}

En el cap\'itulo anterior se introdujo la propuesta del modelo de tratamiento para predecir en que n\'umero de sesi\'on del tratamiento cl\'inico un tumor cancer\'igeno podr\'ia reducir su tamaño (n\'umero de c\'elulas cancer\'igenas) hasta un valor prefijado por el M\'edico. Como hemos analizado la propuesta del modelo se basa en la fracci\'on de c\'elulas cancer\'igenas que sobreviven bajo una dosis de radiaci\'on y que obedecen una ley de crecimiento tipo Gompertz. A continuaci\'on nos enfocaremos en c\'omo reducir el tamaño del tumor (n\'umero de c\'elulas cancer\'igenas) mediante un m\'etodo de optimizaci\'on matem\'atica.

\section{Optimizaci\'on Matem\'atica}

Si un sistema puede ser modelado mediante una serie de ecuaciones diferenciales, algo que nos preguntamos es ¿C\'omo podemos controlar dicho sistema a fin de obtener el mejor resultado posible sujeto a ciertas restricciones o metas? Esta idea puede ser abordada mediante la optimizaci\'on matem\'atica, que se enfoca en encontrar un valor \'optimo para estos sistemas. Los sistemas analizados pueden corresponder a diversas areas: ciencias biol\'ogicas, econom\'ia, f\'isica, etc. No hay limitante m\'as que el sistema pueda ser modelado mediante una serie de ecuaciones, de ah\'i lo robusto de esta area de estudio. 

Por ejemplo, podemos utilizar lo anterior para tomar decisiones sobre situaciones biol\'ogicas complejas. Supongamos que somos parte del comite la OMS o de la CDC de Estados Unidos, encargados de hacer frente a una enfermedad y nos preguntamos ¿Cual es el porcentaje de la poblaci\'on que deberia ser vacunada, conforme el tiempo avanza, en un modelo epid\'emico a fin de minimizar el n\'umero  de infectados y el costo de implementaci\'on de la vacuna? \'o ¿C\'omo minimizar la poblaci\'on de un virus manteniendo bajos los niveles de un f\'armaco? \cite{34}. Este tipo de interrogantes pueden ser resueltas mediante varios tipos de modelos de optimizaci\'on. Debido a que el objetivo de este trabajo no consiste en exponer una introducci\'on al tema y dado que existe una vasta literatura de esta \'area, el lector interesado en detalles m\'as t\'ecnicos sobre este tipo de problemas puede consultar \cite{34} y su bibliograf\'ia. 

Nosotros nos enfocaremos en la resoluci\'on del problema de c\'elulas cancer\'igenas y citaremos las herramientas necesarias para ello. Este enfoque resulta m\'as conciso y adecuado a nuestros intereses.

\section{Optimizaci\'on del Problema}

Recordemos el modelo con el que estamos tratando, la ecuaci\'on diferencial de Gompertz para la n-\'esima sesi\'on de tratamiento 

\begin{equation} 
    \dot{x_n} (t)=r x_n (t) ln \left( \frac{N}{x_n (t)} \right), 
\end{equation}
con condici\'on inicial

\begin{equation} \label{27}
    x_n (t_n) = \left(1-\frac{D_n}{D_{max}} \right)^{\gamma} x_{n-1} (t_n),
\end{equation}
y que, como ya hemos visto, tiene la siguiente soluci\'on

\Large
\begin{equation} \label{eqn:3.2.3}
    x_n (t) = N e^{ln\left( \frac{\alpha_n x_{n-1} (t_n)}{N} \right) e^{- r(t-t_n)}}.
\end{equation}

\normalsize
Como un primer paso, nos enfocaremos en determinar el m\'inimo de c\'elulas cancer\'igenas para un n\'umero $n$ de sesi\'on determinado. Trabajaremos con la forma de la ecuaci\'on ~\eqref{eqn:3.2.3}. Para nuestro fin, debemos restringir el problema, es decir, nuestro objetivo ser\'a el determinar el valor m\'inimo de c\'elulas cancer\'igenas para una determinada sesi\'on de un tratamiento y con los tiempos entre sesiones prefijados. De acuerdo a la literatura, la radioterapia oncol\'ogica lleva a cabo los tratamientos en intervalos de un d\'ia a lo largo de un n\'umero determinado de d\'ias, siendo esto de Lunes a Viernes con tiempos de recuperaci\'on en S\'abado y Domingo \cite{35}. La dosis total general es de 70 Gy y la dosis para cada d\'ia del tratamiento es alrededor de 2 Gy.\\
De la misma forma en que se ha delimitado los tiempos entre sesiones del tratamiento, se impone una restricci\'on sobre las dosis de radiaci\'on: la suma de todas las dosis administradas ser\'a menor o igual a una dosis m\'axima. Tomaremos esta dosis m\'axima, la cual depender\'a del tipo de tumor, como la dosis total habitualmente administrada para cierto tipo de c\'ancer. El motivo de esta elecci\'on es debido a que existen estudios cl\'inicos basados en el descubrimiento emp\'irico sobre c\'omo responden los pacientes bajo cierto regimen de dosis de radiaci\'on \cite{35}. Si bien es cierto que una dosis m\'axima mayor a la dosis total habitual para cierto tumor nos ayudar\'ia a reducir considerablemente el n\'umero de c\'elulas cancer\'igenas (tamaño del tumor) no podemos ignorar los efectos biol\'ogicos que producir\'ia.\\
Si nos enfoc\'aramos en minimizar una expresi\'on de la forma de ~\eqref{eqn:3.2.3} y ya que a partir de ello obtendremos valores para las diferentes dosis de radiaci\'on, nos encontrar\'iamos con el problema de que nuestros resultados depender\'ian de la variable no especificada $t$ (todos los dem\'as par\'ametros deben estar especificados a fin de optimizar y obtener un n\'umero: $N$, $t_n$, $\alpha_n$, $x_{n-1}(t_n)$). Esto puede ser solucionado f\'acilmente al recordar la figura (\ref{g2}). De modo que queremos obtener los valores para las dosis en los tiempos $t_n$, donde inicia la evoluci\'on de cada $x_n$. Ya que el tiempo es un valor prefijado y corresponde a d\'ias de tratamiento (1 sesi\'on por d\'ia y el tiempo corre desde el instante 0 hasta el 1 d\'ia) adem\'as de que en $x_n (t_n)$ comienza el comportamiento de cada modelo, podemos decir que $t_n=0$. \\
Adicionalmente en la figura (\ref{g2}) observamos que el $x_{n-1}$ y el $x_{n}$ comparten un mismo $t_n$, raz\'on por la cual se aprecian las l\'ineas punteadas en negro. El $t$ correr\'a de 0 a 1 y nos interesa el comportamiento en $t=t_n$, es decir, el \'ultimo punto o instante del tiempo para cada modelo de Gompertz construido, pero como hemos dicho que el $t_n$ corresponde al punto inicial en que comienza la evoluci\'on del modelo, entonces toma el valor 0. Otra forma m\'as sencilla de entenderlo es que queremos obtener las dosis \'optimas en los puntos en que $t=t_n$(ver figura (\ref{g2})). \\ 
Una vez delimitadas las restricciones y definiendo

\Large
\begin{equation}
    f(D_1, D_2, ..., D_n) \equiv N e^{ln\left( \frac{\alpha_n (D_n) x_{n-1} (t_n )}{N} \right) e^{- r(t_n-t_n)}},
\end{equation}
\normalsize
podemos enunciar el problema

\begin{equation} \label{eqn:3.2.5}
   \text{min} \ \ \  f(D_1, D_2, ..., D_n),
\end{equation}
sujeto a 
\begin{equation}\label{eqn:3.2.6}
    \sum_{i=1}^{n} D_{i} \leq D_{max},
\end{equation}
donde $n$ toma los valores de $n=1,2,3,...$ y el $t_n $ corresponde a los tiempos previos de las sesiones (el modelo depende de los resultados de las sesiones previas), estos tiempos no ser\'an 0 ya que solo nos interesa minimizar la funci\'on de la $n$-\'esima sesi\'on en su punto inicial de evoluci\'on, $t=t_n=0$, como ya mencionamos. A continuaci\'on describiremos que t\'ecnicas podemos usar para atacar el problema.

\subsection{Condiciones de Kuhn-Tucker}

Observando r\'apidamente el problema enunciado en las ecuaciones ~\eqref{eqn:3.2.5} y ~\eqref{eqn:3.2.6} podemos establecer una analog\'ia con el m\'etodo de Lagrange para resolver problemas de optimizaci\'on cl\'asicos. Sin embargo el m\'etodo de Lagrange es utilizado para problemas con restricciones de igualdad, en nuestro caso trabajamos con restricciones de desigualdad. Las condiciones de Kuhn-Tucker tambi\'en son llamadas condiciones de Karush-Kuhn-Tucker y son una generalizaci\'on del m\'etodo de los multiplicadores de Lagrange.
Estamos interesados en la soluci\'on de un problema como el siguiente

\begin{equation}\label{eqn:3.2.1.1}
\begin{array}{ccccc}
    \text{min} &  & f_0 (\boldsymbol{x}) & &\\
    
    \text{s.a.}&  & f_i (\boldsymbol{x}) & \leq & 0 \  \ (i=1,2,...,k),
\end{array}
\end{equation}

donde podemos obtener una funci\'on de Lagrange

\begin{equation}\label{eqn:3.2.1.2}
    \psi (\boldsymbol{x},\boldsymbol{\lambda})=f_0 (\boldsymbol{x}) + \sum \limits_{i=1}^{k} \lambda_i f_i (\boldsymbol{x}).
\end{equation}

Se puede extender el m\'etodo cl\'asico de Lagrange para un problema del tipo de ~\eqref{eqn:3.2.5} y ~\eqref{eqn:3.2.1.1} mediante el teorema enunciado en el ap\'endice C. \vspace{15pt}

Las ecuaciones ~\eqref{3.2.1.6} y ~\eqref{3.2.1.7} corresponden a las condiciones de KT para un problema con variables no negativas.\\
Ahora veamos como se aplicar\'ian las condiciones de KT para nuestro problema, en este caso solo tenemos una restricci\'on ``general'' correspondiente a la suma de las dosis y varias restricciones ``particulare'' correspondientes a su no negatividad. El primer bloque de las condiciones ~\eqref{3.2.1.6} quedar\'ia

\begin{equation}\label{3.2.1.8}
    \dfrac{\partial f(D_1,...,D_n)}{\partial  D_n} + \lambda_1 \dfrac{\partial (D_1+...+D_n-D_{max})}{\partial D_n} - \mu_n = 0 
\end{equation}

y recordemos que el $n$ toma el valor del n\'umero de sesi\'on del tratamiento. Las dem\'as condiciones ser\'ian

\begin{equation} \label{3.2.1.9}
    \begin{array}{rccc}
         D_1+...+D_n-D_{max}& \leq &0 & \\
         \lambda_1 ( D_1+...+D_n-D_{max})&= &0& \\
         -D_n & \leq & 0 & \ \ (n=1,2,...) \\ 
         \mu _n (-D_n) & = & 0 & \ \ (n=1,2,...)  \\
         \lambda_1 & \geq&0 &\\
         \mu_n & \geq & 0&
    \end{array}
\end{equation}

En las ecuaciones ~\eqref{3.2.1.9}, $n$ toma el valor del numero de sesi\'on, sin embargo si observamos la forma de ~\eqref{3.2.1.7} nos daremos cuenta que debemos poner cada una de las variables no negativas que entran dentro del problema, por ejemplo, si $n=2$ tendremos 6 desigualdades $-D_1 \leq 0$, $-D_2 \leq 0$, $\mu_1(-D_1) \leq 0$, $\mu_2(-D_2) \leq 0$, $\mu_1\geq 0$, $\mu_2\geq 0$ adicional a ~\eqref{3.2.1.8} y las igualdades y desigualdades relacionadas con el multiplicador $\lambda_1$. Conforme el valor de $n$ aumente tendremos m\'as igualdades y desigualdades en nuestro sistema. La soluci\'on del sistema de ecuaciones e inecuaciones ~\eqref{3.2.1.8} y ~\eqref{3.2.1.9} nos dar\'a los valores de las dosis que buscamos $D_1, D_2, ..., D_n$.

\subsection{M\'etodo de Nelder-Mead}

Podemos dividir los m\'etodos de optimizaci\'on dentro de dos tipos: m\'etodos basados en gradientes, que usan las primeras o segundas derivadas de la funci\'on objetivo como ingrediente para obtener una soluci\'on y m\'etodos de busqueda directa los cuales no involucran el uso de derivadas si no m\'as bien se basan en la comparaci\'on de sus resultados y como su nombre lo indica, buscan los puntos \'optimos a lo largo de todo su espacio. El m\'etodo de Nelder-Mead pertenece a los de busqueda directa y utiliza una aproximaci\'on "geom\'etrica" para encontrar los puntos \'optimos de una funci\'on. La construcci\'on del algoritmo utiliza el concepto de simplex. Un simplex es una especie de politopo, una generalizaci\'on de un pol\'igono en $n$ dimensiones, en este caso el pol\'itopo corresponder\'a a la generalizaci\'on de un tri\'angulo. En dos dimensiones el simplex ser\'a un tri\'angulo, en tres dimensiones ser\'a un tetraedro y asi sucesivamente. En el ap\'endice C se desarrolla la teor\'ia del m\'etodo de Nelder-Mead.

Para nuestro problema de entrada vemos que el simplex tendr\'a $n+1$ v\'ertices ya que nos interesa conocer las $D_n$ dosis que minimizan el n\'umero de c\'elulas cancer\'igenas. No construiremos explicitamente el procedimiento secuencial para nuestro problema debido a que el n\'umero de iteraciones puede ser muy grande, en cambio esta tarea la dejaremos a la parte computacional en el siguiente cap\'itulo. Algo a tomar en cuenta es que el algoritmo de Nelder-Mead es utilizado para problemas sin restricciones, sin embargo existe un enfoque para trabajar con problemas de optimizaci\'on (minimizaci\'on para nuestros intereses) con restricciones dentro de Nelder-Mead y se realiza mediante el \textbf{m\'etodo de penalizaci\'on} el cual describiremos a continuaci\'on. \\
El problema que queremos atacar corresponde a ~\eqref{3.2.1.5}, las restricciones estan en forma de desigualdad. Si queremos usar el m\'etodo de Nelder-Mead para resolver este problema, debemos hacer uso del m\'etodo de penalizaci\'on que se basa en la penalizaci\'on de los puntos que se encuentran fuera de la regi\'on factible o donde las constricciones no se cumplen. Mediante el m\'etodo de penalizaci\'on transformamos un problema de optimizaci\'on con restricciones a una serie de problemas sin restricciones pero cuya soluci\'on converge a la soluci\'on del problema con restricciones \cite{37}. Para pasar de un problema con restricciones a uno sin restricciones agregamos la funci\'on de penalizaci\'on a la funci\'on objetivo, la funci\'on de penalizaci\'on va multiplicada por un coeficiente de penalizaci\'on. La medida de esta penalizaci\'on es cero cuando no se violan las constricciones y diferente de cero cuando las constricciones no se satisfacen.
\\
Regresando al problema ~\eqref{3.2.1.5} podemos expresar las restricciones como $g_i (\boldsymbol{x})\leq 0$ (en este caso $i \in I$, donde $I$ es el conjunto de todas las restricciones) de modo que transformamos el problema a

\begin{equation}\label{3.2.2.8}
\begin{array}{ccc}
    \text{min} &  & \Phi_k (\boldsymbol{x}) = f(\boldsymbol{x}) + \delta_k \sum \limits_{i \in I} c(g_i(\boldsymbol{x}))
\end{array}
\end{equation}

donde $k=1,2,3,...$ indica el numero de iteraci\'on, $\delta$ es el coeficiente de penalizaci\'on y 

\begin{equation}
c(g_i(\boldsymbol{x}))= \text{max}(0,g_i (\boldsymbol{x}))^2
\end{equation}

la funci\'on de penalizaci\'on que elegiremos para nuestro caso, esta funci\'on de valor absoluto es comunmente usada como funci\'on de penalizaci\'on \cite{38}. Durante cada iteraci\'on $k$ se elegir\'a un par\'ametro de penalizaci\'on, se resolver\'a el problema de minimizaci\'on sin restricciones ~\eqref{3.2.2.8} y la soluci\'on se usar\'a como estimaci\'on inicial para la siguiente iteraci\'on. El proceso se repetir\'a hasta encontrar una soluci\'on. La elecci\'on del par\'ametro de penalizaci\'on se puede hacer como se describe en \cite{38}: dado $\delta_k, \ k=1,2,...$ con $\delta_k > 0$ y $\delta_{k+1} > \delta_k$, se resuelve el problema ~\eqref{3.2.2.8} para cada k a fin de obtener un $\boldsymbol{x}^k$ que ser\'a la soluci\'on \'optima. \textbf{Primero} seleccionamos par\'ametros de crecimiento $\eta >1$ y de parada $\epsilon>0$ y escogemos un valor inicial para $\delta_0$. Escogemos un punto inicial $x_0 \in \boldsymbol{x}$ que viole al menos una restricci\'on y formulamos 

\begin{equation}
\Phi_k (\boldsymbol{x}) = f(\boldsymbol{x}) + \delta_0 \sum \limits_{i \in I}c(g_i(\boldsymbol{x}))
\end{equation}

y el contador inicia con k=1. Como \textbf{segundo paso}, iniciamos con $\boldsymbol{x}_{k-1}$ y resolvemos el problema de minimizaci\'on con Nelder-Mead para $\Phi_{k-1} (\boldsymbol{x})$. A la soluci\'on la llamamos $\boldsymbol{x}_k$ y determinamos que restricciones son violadas. \textbf{Finalmente}, si la distancia entre $\boldsymbol{x}_{k-1}$ y $\boldsymbol{x}_{k}$ es m\'as pequeña que $\epsilon$, ($|| \boldsymbol{x}_{k-1}- \boldsymbol{x}_{k} ||$ $< \epsilon$) \'o la diferencia entre dos funciones objetivos sucesivas es menor que $\epsilon$, ($|f( \boldsymbol{x}_{k-1})- f(\boldsymbol{x}_{k}) |$ $< \epsilon$) el proceso debe deternerse y obtener un $\boldsymbol{x}_k$ que ser\'a un estimado de la soluci\'on \'optima. Si lo anterior no se cumple entonces añadimos el par\'ametro de crecimiento $\eta$ a $\delta_k$ ($\eta \delta_k$), formulamos un nuevo $\Phi_{k} (\boldsymbol{x})$ basandonos en que restricciones se violan en $\boldsymbol{x}_k$, $k=k+1$ y repetimos nuevamente desde el \textbf{segundo paso}.\\
Con esta nueva implementaci\'on podemos resolver nuetro problema de minimizaci\'on, la descripcci\'on expl\'icita aplicada al problema no se plasmar\'a en el trabajo ya que las iteraciones pueden ser demasiadas (como ya habiamos mencionado), en cambio dejamos la soluci\'on a la parte computacional y cuyos resultados se analizar\'an en el siguiente cap\'itulo.

\subsection{Evoluci\'on Diferencial}

Concluiremos este cap\'itulo con el m\'etodo de evoluci\'on diferencial. Pertenece tambi\'en al tipo de optimizadores de busqueda directa, se caracteriza por trabajar con variables reales, es r\'apido, robusto y se conoce por ser un muy buen optimizador global \cite{39}. En el ap\'endice C se describe la metodolog\'ia de este m\'etodo.

De igual manera, este tipo de optimizador, por si solo, no hace alusi\'on al uso de restricciones, sin embargo, el m\'etodo de penalizaci\'on es un m\'etodo universal para optimizadores de busqueda, como los ya mencionados. Debido a ello podemos hacer uso de Evoluci\'on Diferencial para resolver nuestro problema de minimizaci\'on ~\eqref{3.2.1.5}, solo hay que pasar el problema con restricciones a un problema sin restricciones como se explic\'o en el m\'etodo de penalizaci\'on y en el paso donde hay que resolver el problema de minimizaci\'on sin restricciones en lugar de hacer uso de Nelder-Mead, hacemos uso de Evolucion Diferencial. La construcci\'on expl\'icita de la resoluci\'on de este problema se omite debido a que se deja a la parte computacional y ser\'a abordada en el siguiente cap\'itulo.

Cabe recalcar que al solucionar nuestro problema de minimizaci\'on con Evoluci\'on Diferencial, el m\'etodo solo hace uso de las variables, la "forma" de la funci\'on objetivo solo se utiliza en el paso de \textbf{selecci\'on}, es decir, solo utilizamos la informaci\'on total de la funci\'on para evaluar si el vector de p\'arametros/variables cumple el criterio. De ah\'i que los m\'etodos de busqueda directa sean m\'etodos robustos en optimizaci\'on matem\'atica.
\chapter{Aplicaci\'on del Modelo de Tratamiento}

En el cap\'itulo anterior expusimos la metodolog\'ia para resolver nuestro problema de minimizaci\'on, basados en el desarrollo de diferentes m\'etodos optimizadores. En este cap\'itulo aplicaremos la metodolog\'ia expuesta a un problema tangible de c\'ancer y compararemos con los modelos de tratamiento usados actualmente.

\section{Aplicaci\'on a c\'ancer de mama}

De entre todos los tumores que existen nos enfocaremos en c\'ancer de mama por ser uno de los que est\'a aumentando a nivel mundial debido a factores involucrados con el estilo de vida moderno y uno de los tipos de cancer m\'as recurrentes en mujeres (y causantes de muerte) \cite{41} de igual manera existe una basta literatura al respecto. Sin embargo, hay que aclarar que la metodolog\'ia desarrollada en este trabajo es aplicable para cualquier tipo de tumor. 

Como primer paso nos enfocaremos en determinar los par\'ametros fijos de la ecuaci\'on ~\eqref{eqn:3.2.3}. En primer lugar necesitamos conocer la capacidad de carga $N$ del sistema , es decir, el tamaño maximo del tumor o n\'umero m\'aximo de c\'elulas cancer\'igenas que puede alcanzar este. En \cite{42} se hace, entre otras cosas, un an\'alisis para la identificaci\'on de la capacidad de carga basados en los datos obtenidos de 250 mujeres con c\'ancer de mama sin tratamiento y encontraron que la capacidad de carga correspond\'ia a $N=1\times 10^{12}$ c\'elulas cancer\'igenas. Esta claro que hablar de tamaño m\'aximo de un tumor resulta un poco atrevido sin embargo se toma $N$  m\'as bien como el tamaño letal del tumor, que para nuestros intereses no resulta contraproducente ya que no nos interesa acercarnos a ese l\'imite letal y por ello tomar este valor de $N$ resulta factible. Adem\'as de que si el paciente muere a causa de llegar a este l\'imite letal, el tumor no podr\'ia seguir creciendo.  

A partir de este n\'umero nosotros podemos encontrar el tamaño del tumor ya que se sabe que hay alrededor de $10^9$ c\'elulas cancer\'igenas en 1 $\text{cm}^3$  \cite{43}, este dato solo lo usaremos como constante de transformaci\'on ya que resulta m\'as \'util y sencillo trabajar con unidades de n\'umero de c\'elulas.   

Otro par\'ametro que debe ser determinado es la tasa de crecimiento $r$. Una ecuaci\'on sencilla puede ser obtenida mediante los tiempos de duplicaci\'on del tumor, esta metodolog\'ia fue brevemente descrita en el cap\'itulo 1. Tomemos la soluci\'on de la ecuaci\'on de Gompertz para la $n$-\'esima sesi\'on del tratamiento, sea $t_0$ el tiempo que tarda en duplicarse el tumor y $X_0$ el tamaño inicial del tumor entonces se cumple que

\Large
\begin{equation}
   N e^{ ln \left( \frac{X_0}{N} \right) e^{-rt_0} }=2X_0
\end{equation}

\normalsize
Despejando $r$ de la ecuaci\'on anterior obtenemos

\begin{equation}
   r=-\dfrac{1}{t_0} ln \left[ \dfrac{ln\left(2\right)}{ln\left( X_0 / N \right)} +1 \right]
\end{equation}

De acuerdo con \cite{44} el tiempo de duplicaci\'on de un carcinoma mamario, en promedio, es de 212 d\'ias. Tenemos que sustituir $t_0=212$ d\'ias en la ecuaci\'on anterior y tomar el valor establecido para $N$, $r$ corresponder\'a a los casos de tamaño inicial del tumor que consideraremos (1 $\text{cm}^3$, 3 $\text{cm}^3$ y 5 $\text{cm}^3$) por lo que tendremos tres valores para esta constante. $r$ tiene unidades de 1 / d\'ias

Una \'ultima constante que debe ser determinada es $\gamma$ de la ecuaci\'on ~\eqref{2.7} y que est\'a relacionada con la fracci\'on de supervivencia celular. En \cite{20} se detalla la metodolog\'ia para obtener esta constante tomando como datos las fracciones o probabilidades de supervivencia de diferentes tejidos bajo ciertas dosis de radiaci\'on, es decir, las curvas de supervivencia. Aplicando el logaritmo a ambos lados de la ecuaci\'on ~\eqref{2.5.2.8} obtenemos la ecuaci\'on de una l\'inea recta cuya pendiente sera $\gamma$, obteniendo la gr\'afica Ln-Ln de las curvas de supervivencia podemos obtener la l\'inea que ajusta a la ecuaci\'on antes mencionada, de ah\'i obtenemos el valor de $\gamma$. Para este trabajo consideramos las curvas de supervivencia de \cite{45} referentes a dos tipos de c\'ancer de mama (de pacientes con reincidencia por lo que se consideran muy agresivos) : carcinoma ductal invasivo (CDI) y carcinoma ductal in situ (CDIS) en condiciones de Normoxia (oxigenaci\'on normal)

\begin{figure}[h!]
    \centering
    \includegraphics[scale=0.75]{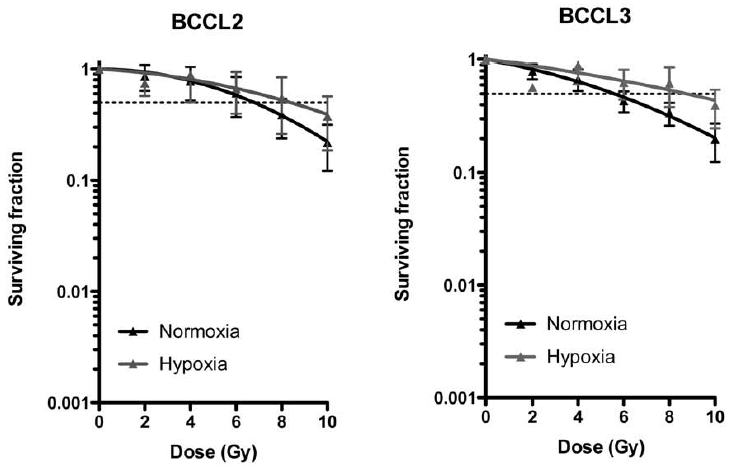}
    \caption{Curva de supervivencia para CDI (izquierda) y CDIS (derecha) obtenidas de \cite{45}}
    \label{CDIbccl2}
\end{figure}

Mediante el an\'alisis ya mencionado obtenemos el valor de la constante $\gamma_{CDI} = 8.46$ y $\gamma_{CDIS}=9.16$.

\begin{figure}[h!]
    \centering
    \includegraphics[scale=0.5]{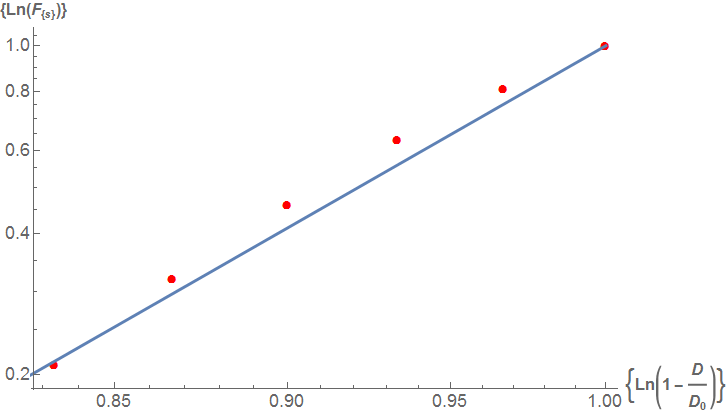}
    \caption{Gr\'afica Ln-Ln para la fracci\'on de supervivencia de CDI, $\gamma_{CDI} = 8.46$. Representa un buen ajuste dado que las curvas de supervivencia de \cite{45} presentan una gran desviaci\'on reflejada por las barras de error que se pueden apreciar en ese \'articulo. Hay que destacar que las muestras provienen directamente de pacientes con este tumor.}
    \label{CDIbccl2}
\end{figure}

\newpage
\begin{figure}[h!]
    \centering
    \includegraphics[scale=0.5]{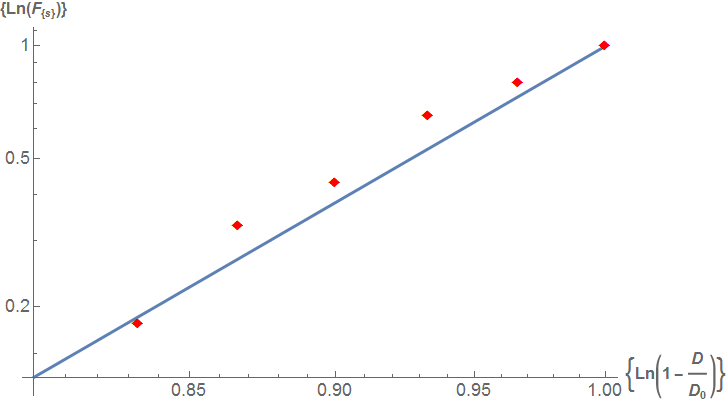}
    \caption{Gr\'afica Ln-Ln para la fracci\'on de supervivencia de CDIS, $\gamma_{CDIS} = 9.16$. Al igual que en el caso anterior se considera un buen ajuste debido a la presencia de las barras de error en \cite{45}. Las muestras tambien provienen de pacientes con este tumor.}
    \label{CDISbccl3}
\end{figure}

\subsection{Tratamiento con Radioterapia Actual}

El tratamiento con radioterapia que generalmente es usado para cualquier tipo de tumor se basa en el fraccionamiento de la dosis total, la cual debe ser administrada al paciente a lo largo de los d\'ias del tratamiento. Este tipo de programa se puede clasificar en: fraccionamiento est\'andar e hipofraccionamiento \cite{46}. Ambos son los tratamientos m\'as comunes para diferentes tipos de c\'ancer incluido el de mama. 

Para c\'ancer de mama, el fraccionamiento est\'andar divide una dosis total de 50 Gy a lo largo de 25 d\'ias (de Lunes a Viernes con descanso en fin de semana) por lo que el tratamiento se completa en 5 semanas. Mientras que el hipofraccionamiento (tambi\'en para cancer de mama) divide una dosis total de 40 Gy en 15 d\'ias (de Lunes a Viernes con descanso en fin de semana). Existe literatura al respecto sobre c\'omo se deber\'ia optar por un tipo de programa u otro (\cite{46} y su bibliograf\'ia) pero resumiendo, debe tomarse en cuenta el riesgo a desarrollar fibrosis por el tratamiento (si existe riesgo se opta por un tratamiento est\'andar y si no existe se opta por hipofraccionamiento), a pesar que la sugerencia de esta dosis est\'a documentada \cite{47} en la pr\'actica los radio onc\'ologos evaluan el caso del paciente y pueden optar por dosis totales mayores o menores.

Con esto en mente podemos hacer uso del modelo desarrollado en este trabajo (ecuaci\'on ~\eqref{2.6.16}) y observar como el n\'umero de c\'elulas cancer\'igenas disminuye tomando ambos tipos de tratamiento. Los datos que necesitamos para el modelo fueron delimitados en la secci\'on anterior. El \'unico dato faltante era respecto de las dosis que entran en la ecuaci\'on de la fracci\'on de supervivencia celular, a continuaci\'on compararemos ambos modelos suponiendo una dosis m\'axima $D_{max}=60$ Gy (la elecci\'on de esta dosis se hizo ya que los tratamientos habituales para c\'ancer de mama se hacen en dosis de 40-60 Gy \cite{47}, a pesar de no ser una dosis global m\'axima en la pr\'actica si representa nuestra dosis l\'imite y podemos tomarla como la dosis despu\'es de la cual ninguna c\'elula sobrevive o para los fines de nuestro modelo la dosis que no queremos rebasar debido a los efectos que ocasionar\'ia sobre el paciente) tanto para CDI como CDIS, supondremos un tamaño inicial del tumor de 1 $\text{cm}^3$ ($1\times 10^9$ celulas cancer\'igenas) este tamaño de tumor se consideraria de etapa I \cite{26}.

Con los datos una vez delimitados obtenemos los siguientes resultados (todas las simulaciones sobre el tratamiento y la optimizaci\'on de las dosis fueron realizadas en Mathematica\textsuperscript{\textregistered}).


\begin{table}[h!]
\centering
\begin{tabular}{c|c|l|l|l|l|l|l|c|l|l|l|l|l|l|}
\cline{2-15}
\multirow{3}{*}{\textbf{}} & \multicolumn{7}{c|}{\multirow{3}{*}{\textbf{\begin{tabular}[c]{@{}c@{}}Fraccionamiento\\ Est\'andar\end{tabular}}}} & \multicolumn{7}{c|}{\multirow{3}{*}{\textbf{Hipofraccionamiento}}} \\
 & \multicolumn{7}{c|}{} & \multicolumn{7}{c|}{} \\
 & \multicolumn{7}{c|}{} & \multicolumn{7}{c|}{} \\ \hline
\multicolumn{1}{|c|}{\multirow{4}{*}{\textbf{\begin{tabular}[c]{@{}c@{}}CDI\\ (N\'umero de c\'elulas cancer\'igenas\\ al final del tratamiento)\end{tabular}}}} & \multicolumn{7}{c|}{\multirow{4}{*}{781,853}} & \multicolumn{7}{c|}{\multirow{4}{*}{$3.21\times 10^6$}} \\
\multicolumn{1}{|c|}{} & \multicolumn{7}{c|}{} & \multicolumn{7}{c|}{} \\
\multicolumn{1}{|c|}{} & \multicolumn{7}{c|}{} & \multicolumn{7}{c|}{} \\
\multicolumn{1}{|c|}{} & \multicolumn{7}{c|}{} & \multicolumn{7}{c|}{} \\ \hline
\multicolumn{1}{|c|}{\multirow{4}{*}{\textbf{\begin{tabular}[c]{@{}c@{}}CDIS\\ (N\'umero de c\'elulas cancer\'igenas\\ al final del tratamiento)\end{tabular}}}} & \multicolumn{7}{c|}{\multirow{4}{*}{432,199}} & \multicolumn{7}{c|}{\multirow{4}{*}{$1.99\times 10^6$}} \\
\multicolumn{1}{|c|}{} & \multicolumn{7}{c|}{} & \multicolumn{7}{c|}{} \\
\multicolumn{1}{|c|}{} & \multicolumn{7}{c|}{} & \multicolumn{7}{c|}{} \\
\multicolumn{1}{|c|}{} & \multicolumn{7}{c|}{} & \multicolumn{7}{c|}{} \\ \hline
\end{tabular}
\caption{Resultados para el tratamiento con radioterapia actual, usando el modelo ~\eqref{2.6.16} para un tumor inicial de 1 $\text{cm}^3$ \'o $1\times 10^9$ c\'elulas}
\label{resultsCDICDIS1}
\end{table}

Lo que podemos observar es que el n\'umero de c\'elulas cancer\'igenas se redujo bastante m\'as con el fraccionamiento est\'andar (algo de esperarse debido a que la dosis total es mayor) que con el hipofraccionamiento. Con el fraccionamiento est\'andar se lle\'go a ordenes de magnitud de 0.781 $\text{mm}^3$ para CDI y de 0.432 $\text{mm}^3$ para CDIS. Usando hipofraccionamiento se logro reducir el tumor inicialmente de 1 $\text{cm}^3$ a $3.21 \ \text{mm}^3$ para CDI y a $1.99 \ \text{mm}^3 $ para CDIS lo cual tambien supone excelentes resultados.

Mediante la ecuaci\'on ~\eqref{2.6.16} tambi\'en podemos obtener una gr\'afica como la de la figura 2.3 con el fin de observar como van aumentando y disminuyendo las c\'elulas cancer\'igenas durante cada d\'ia del tratamiento. Para este caso, $r=4.9878 \times 10^{-4} \frac{1}{\text{dias}}$

\begin{figure}[h!]
    \centering
    \includegraphics[scale=0.6]{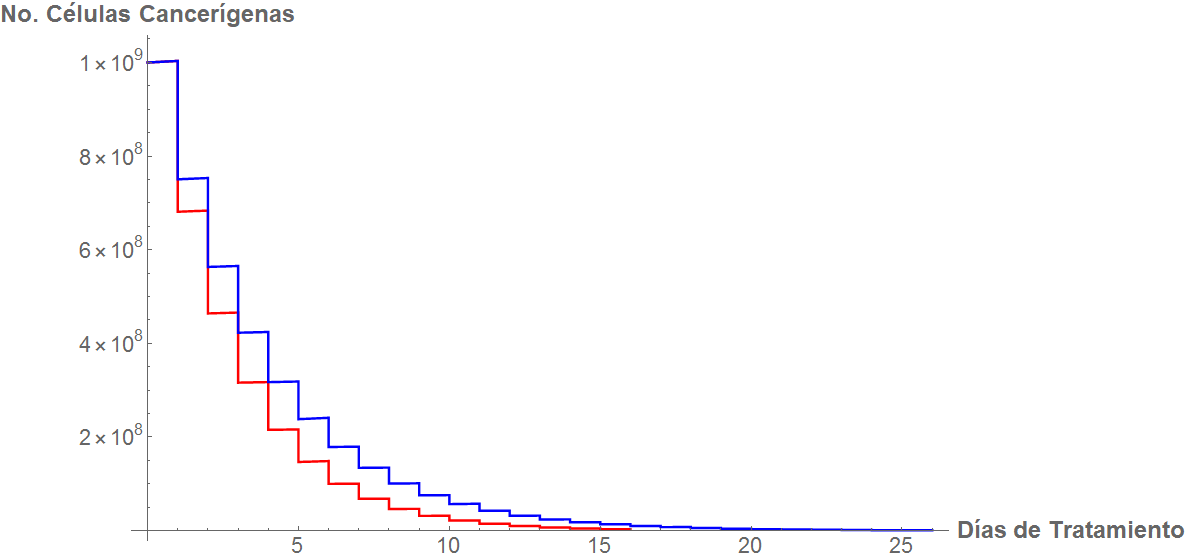}
    \caption{No. de c\'elulas cancer\'igenas durante el fraccionamiento est\'andar (2 Gy por d\'ia) para CDI (l\'inea azul) y durante el hipofraccionamiento (2.66 Gy por d\'ia, l\'inea roja), tamaño inicial del tumor de 1 $\text{cm}^3$}
    \label{BCCL2FraccStand50gy}
\end{figure}

\newpage
\begin{figure}[h!]
    \centering
    \includegraphics[scale=0.6]{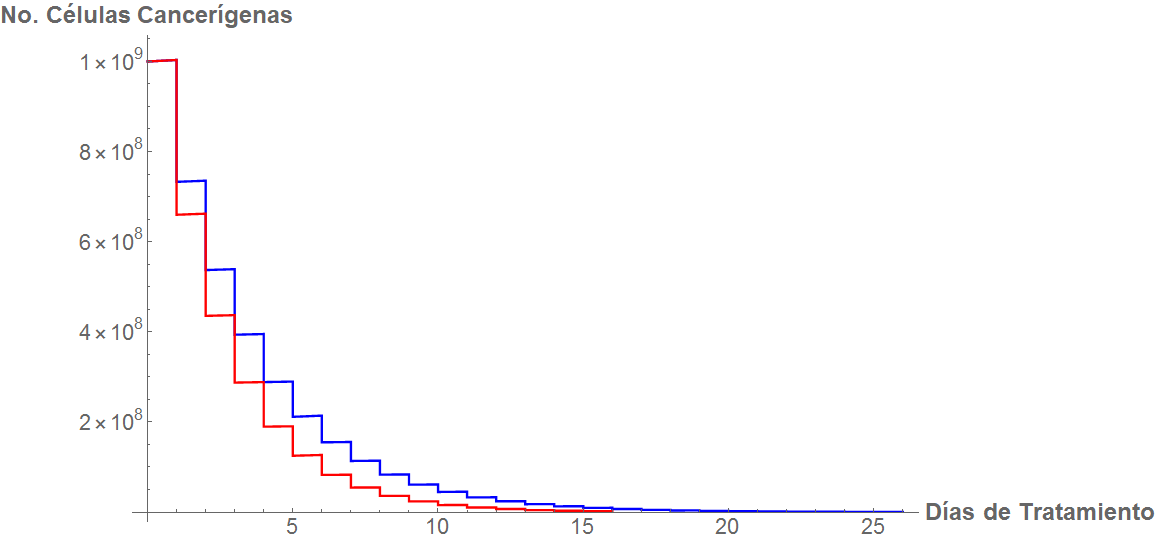}
    \caption{No. de c\'elulas cancer\'igenas durante el fraccionamiento est\'andar (2 Gy por d\'ia) para CDIS (l\'inea azul) y durante el hipofraccionamiento (2.66 Gy por d\'ia, l\'inea roja), tamaño inicial del tumor de 1 $\text{cm}^3$}
    \label{BCCL3FraccStand50gy}
\end{figure}

De entrada, podemos observar que el hipofraccionamiento presenta mejores resultados ya que logra reducir el n\'umero de c\'elulas cancerosas en un menor tiempo y solo con el aumento de 0.66 Gy de dosis cada d\'ia (respecto del fraccionamiento estandar). Lo anterior se encuentra en perfecta concordancia con lo encontrado clinicamente para el Hipofraccionamiento \cite{51}.

Ahora supongamos un tumor etapa II de tamaño intermedio, es decir unos 3 $\text{cm}$ de largo (en su mayor extensi\'on) \cite{26} (la medida del largo, ancho y profundidad independientes unas de otras) y tambi\'en supongamos que tiene una forma tal que su tamaño total es de 3 $\text{cm}^3$. 
 
Con el an\'alisis anterior obtenemos   

\begin{table}[h!]
\centering
\begin{tabular}{c|c|l|l|l|l|l|l|c|l|l|l|l|l|l|}
\cline{2-15}
\multirow{3}{*}{\textbf{}} & \multicolumn{7}{c|}{\multirow{3}{*}{\textbf{\begin{tabular}[c]{@{}c@{}}Fraccionamiento\\ Est\'andar\end{tabular}}}} & \multicolumn{7}{c|}{\multirow{3}{*}{\textbf{Hipofraccionamiento}}} \\
 & \multicolumn{7}{c|}{} & \multicolumn{7}{c|}{} \\
 & \multicolumn{7}{c|}{} & \multicolumn{7}{c|}{} \\ \hline
\multicolumn{1}{|c|}{\multirow{4}{*}{\textbf{\begin{tabular}[c]{@{}c@{}}CDI\\ (N\'umero de c\'elulas cancer\'igenas\\ al final del tratamiento)\end{tabular}}}} & \multicolumn{7}{c|}{\multirow{4}{*}{$2.34\times 10^6$}} & \multicolumn{7}{c|}{\multirow{4}{*}{$9.66\times 10^6$}} \\
\multicolumn{1}{|c|}{} & \multicolumn{7}{c|}{} & \multicolumn{7}{c|}{} \\
\multicolumn{1}{|c|}{} & \multicolumn{7}{c|}{} & \multicolumn{7}{c|}{} \\
\multicolumn{1}{|c|}{} & \multicolumn{7}{c|}{} & \multicolumn{7}{c|}{} \\ \hline
\multicolumn{1}{|c|}{\multirow{4}{*}{\textbf{\begin{tabular}[c]{@{}c@{}}CDIS\\ (N\'umero de c\'elulas cancer\'igenas\\ al final del tratamiento)\end{tabular}}}} & \multicolumn{7}{c|}{\multirow{4}{*}{$1.29\times 10^6$}} & \multicolumn{7}{c|}{\multirow{4}{*}{$6\times 10^6$}} \\
\multicolumn{1}{|c|}{} & \multicolumn{7}{c|}{} & \multicolumn{7}{c|}{} \\
\multicolumn{1}{|c|}{} & \multicolumn{7}{c|}{} & \multicolumn{7}{c|}{} \\
\multicolumn{1}{|c|}{} & \multicolumn{7}{c|}{} & \multicolumn{7}{c|}{} \\ \hline
\end{tabular}
\caption{Resultados para el tratamiento con radioterapia actual, usando el modelo ~\eqref{2.6.16} para un tumor inicial de 3 $\text{cm}^3$ \'o $3\times 10^9$ c\'elulas}
\label{resultsCDICDIS2}
\end{table}

\newpage
El aumento y disminuci\'on del tamaño del tumor se puede observar a continuaci\'on, $r=5.9934 \times 10^{-4} \frac{1}{\text{dias}}$

\begin{figure}[H]
    \centering
    \includegraphics[scale=0.6]{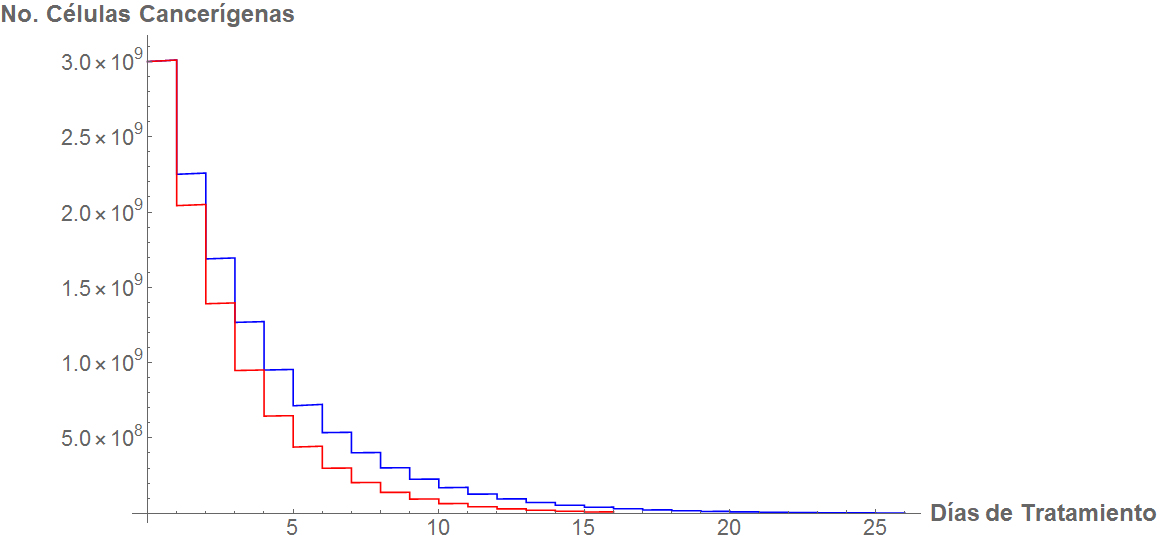}
    \caption{No. de c\'elulas cancer\'igenas durante el fraccionamiento est\'andar (2 Gy por d\'ia) para CDI (l\'inea azul) y durante el hipofraccionamiento (2.66 Gy por d\'ia, l\'inea roja), tamaño inicial del tumor de 3 $\text{cm}^3$}
    \label{BCCL2FraccStand50gy3cm}
\end{figure}

\begin{figure}[H]
    \centering
    \includegraphics[scale=0.6]{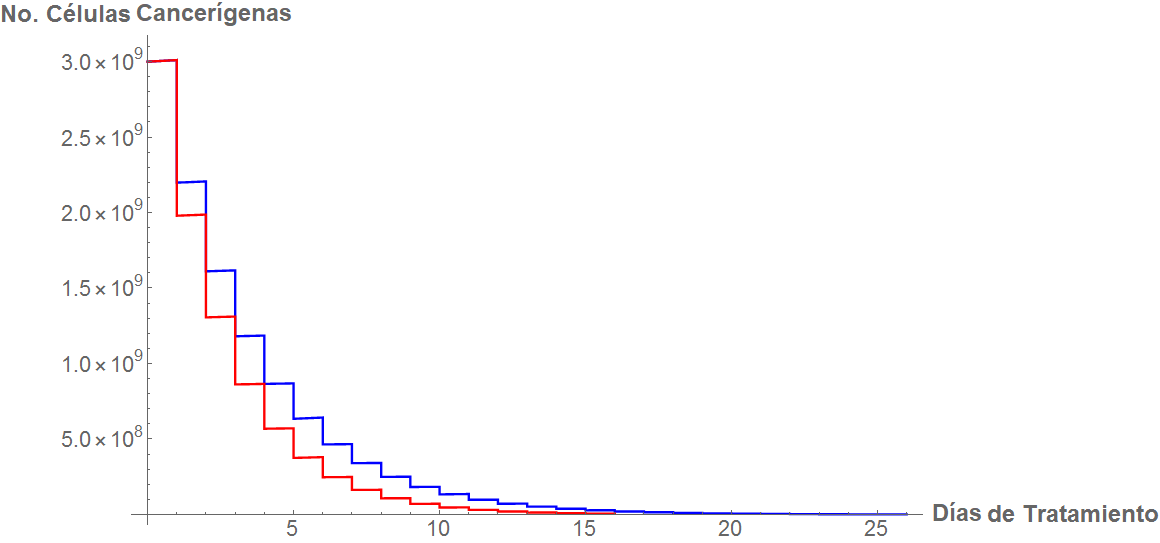}
    \caption{No. de c\'elulas cancer\'igenas durante el fraccionamiento est\'andar (2 Gy por d\'ia) para CDIS (l\'inea azul) y durante el hipofraccionamiento (2.66 Gy por d\'ia, l\'inea roja), tamaño inicial del tumor de 3 $\text{cm}^3$}
    \label{BCCL3FraccStand50gy3cm}
\end{figure}

De igual manera que en el caso anterior, el hipofraccionamiento supone un mejor tipo de tratamiento respecto de la disminuci\'on de c\'elulas cancerosas.

Para tener m\'as datos a comparar apliquemos el modelo ahora a un tumor etapa III de gran tamaño, 5 $\text{cm}^3$ de volumen total inicial \cite{26}, $r=6.6136 \times 10^{-4} \frac{1}{\text{dias}}$

\begin{table}[h!]
\centering
\begin{tabular}{c|c|l|l|l|l|l|l|c|l|l|l|l|l|l|}
\cline{2-15}
\multirow{3}{*}{\textbf{}} & \multicolumn{7}{c|}{\multirow{3}{*}{\textbf{\begin{tabular}[c]{@{}c@{}}Fraccionamiento\\ Est\'andar\end{tabular}}}} & \multicolumn{7}{c|}{\multirow{3}{*}{\textbf{Hipofraccionamiento}}} \\
 & \multicolumn{7}{c|}{} & \multicolumn{7}{c|}{} \\
 & \multicolumn{7}{c|}{} & \multicolumn{7}{c|}{} \\ \hline
\multicolumn{1}{|c|}{\multirow{4}{*}{\textbf{\begin{tabular}[c]{@{}c@{}}CDI\\ (N\'umero de c\'elulas canc\'erigenas\\ al final del tratamiento)\end{tabular}}}} & \multicolumn{7}{c|}{\multirow{4}{*}{$3.91\times 10^6$}} & \multicolumn{7}{c|}{\multirow{4}{*}{$1.61\times 10^7$}} \\
\multicolumn{1}{|c|}{} & \multicolumn{7}{c|}{} & \multicolumn{7}{c|}{} \\
\multicolumn{1}{|c|}{} & \multicolumn{7}{c|}{} & \multicolumn{7}{c|}{} \\
\multicolumn{1}{|c|}{} & \multicolumn{7}{c|}{} & \multicolumn{7}{c|}{} \\ \hline
\multicolumn{1}{|c|}{\multirow{4}{*}{\textbf{\begin{tabular}[c]{@{}c@{}}CDIS\\ (N\'umero de c\'elulas cancer\'igenas\\ al final del tratamiento)\end{tabular}}}} & \multicolumn{7}{c|}{\multirow{4}{*}{$2.16\times 10^6$}} & \multicolumn{7}{c|}{\multirow{4}{*}{$1\times 10^7$}} \\
\multicolumn{1}{|c|}{} & \multicolumn{7}{c|}{} & \multicolumn{7}{c|}{} \\
\multicolumn{1}{|c|}{} & \multicolumn{7}{c|}{} & \multicolumn{7}{c|}{} \\
\multicolumn{1}{|c|}{} & \multicolumn{7}{c|}{} & \multicolumn{7}{c|}{} \\ \hline
\end{tabular}
\caption{Resultados para el tratamiento con radioterapia actual, usando el modelo ~\eqref{2.6.16} para un tumor inicial de 5 $\text{cm}^3$ \'o $5\times 10^9$ c\'elulas}
\label{resultsCDICDIS3}
\end{table}

\begin{figure}[H]
    \centering
    \includegraphics[scale=0.6]{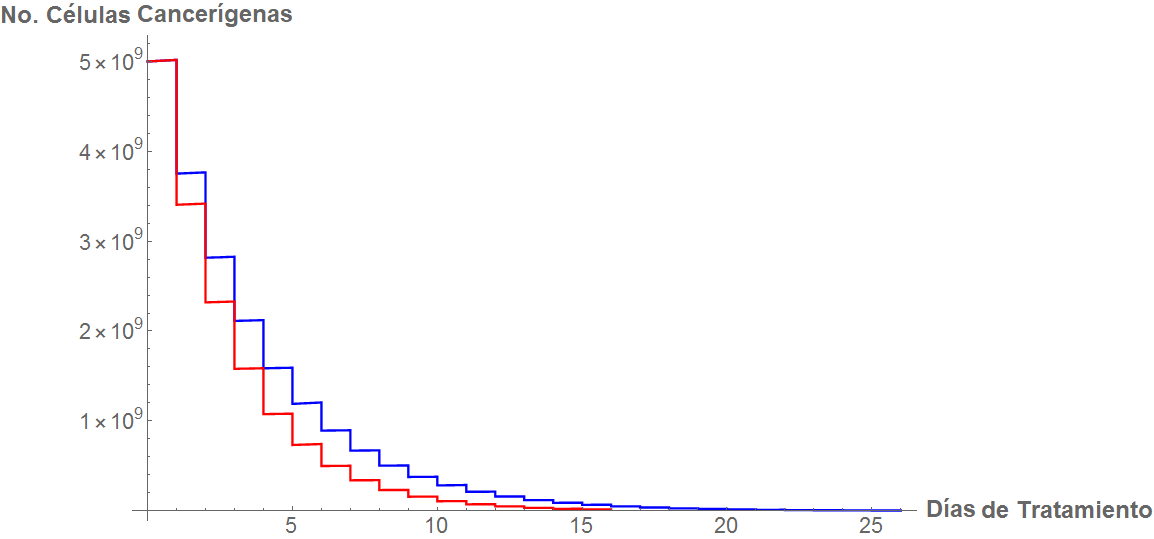}
    \caption{No. de c\'elulas cancer\'igenas durante el fraccionamiento est\'andar (2 Gy por d\'ia) para CDI (l\'inea azul) y durante el hipofraccionamiento (2.66 Gy por d\'ia, l\'inea roja), tamaño inicial del tumor de 5 $\text{cm}^3$}
    \label{BCCL2FraccStand50gy5cm}
\end{figure}

\begin{figure}[H]
    \centering
    \includegraphics[scale=0.6]{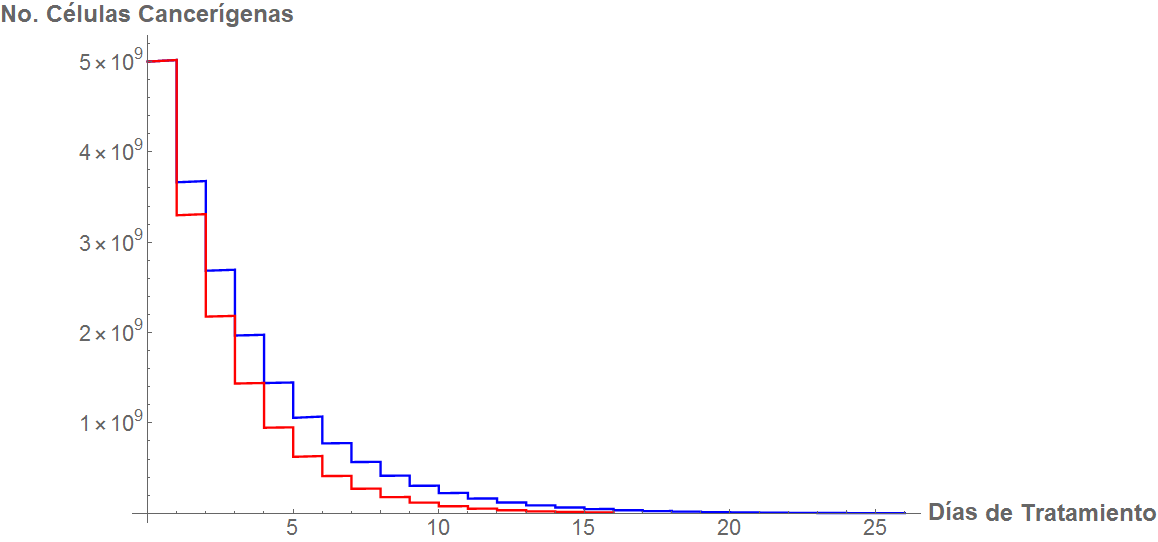}
    \caption{No. de c\'elulas cancer\'igenas durante el fraccionamiento est\'andar (2 Gy por d\'ia) para CDIS (l\'inea azul) y durante el hipofraccionamiento (2.66 Gy por d\'ia, l\'inea roja), tamaño inicial del tumor de 5 $\text{cm}^3$}
    \label{BCCL3FraccStand50gy5cm}
\end{figure}

\subsection{Tratamiento con Optimizaci\'on de las Dosis Fijadas por los M\'edicos}

En la secci\'on anterior describimos la metodolog\'ia que nos llev\'o a c\'omo reducir el n\'umero de c\'elulas cancer\'igenas tomando como restricciones que las dosis fueran positivas y la suma de estas fuera menor o igual a una dosis m\'axima ~\eqref{eqn:3.2.5}, ~\eqref{eqn:3.2.6} y ~\eqref{3.2.1.5}. La aplicaci\'on de los distintos procedimientos nos llev\'o a elegir los m\'etodos de busqueda directa en lugar de los basados en gradientes debido a que su implementaci\'on computacional (de los de busqueda directa) requiere menos recursos por lo que computacionalmente resulta \'optimo. La implementaci\'on de KKT requiere un an\'alisis riguroso respecto de su c\'odigo, no es el objetivo de este trabajo presentar un an\'alisis de c\'omo KKT se podr\'ia aplicar de manera eficiente a la resoluci\'on de este problema, sino m\'as bien una comparaci\'on de los distintos m\'etodos que pueden utilizarse adem\'as de la respectiva optimizaci\'on del problema.  

Dicho lo anterior y adicionalmente a las restricciones ya impuestas, agregaremos una restricci\'on m\'as relacionada al n\'umero de c\'elulas cancer\'igenas que queramos alcanzar. Esta restricci\'on ser\'a tal que

\begin{equation}\label{restmed}
  f(D_1, D_2, ..., D_n) \geq X_{\text{Propt}}
\end{equation}

donde $X_{\text{Propt}}$ es el n\'umero de c\'elulas cancer\'igenas objetivo y para este caso lo tomaremos como las c\'elulas cancer\'igenas a las que se lleg\'o despu\'es del fraccionamiento est\'andar e hipofraccionamiento de la subsecci\'on anterior (tablas 4.1, 4.2 y 4.3). La desigualdad se tom\'o de esa manera como un primer paso para investigar el tratamiento llevado a cabo por los m\'edicos, es decir, queremos llegar como m\'inimo al n\'umero de c\'elulas cancer\'igenas que se obtienen los tratamientos actuales. La adici\'on de esta restricci\'on no supone problemas al anal\'isis planteado en el cap\'itulo anterior para Nelder-Mead o Evoluci\'on Diferencial ya que en ~\eqref{3.2.1.5} se tienen en cuenta un n\'umero $i$ de restricciones, por lo que solo se añadir\'ia un valor m\'as a la funci\'on de penalizaci\'on.

Sea $X_{\text{ProptA}}$ el n\'umero de c\'elulas cancerigenas alcanzado al final de la minimizaci\'on (el resultado de la minimizaci\'on del n\'umero de c\'elulas cancer\'igenas), los resultados de las dosis optimizadas para los 3 diferentes tamaños de tumor tanto en fraccionamiento est\'andar como hipofraccionamiento son los siguientes (todas las dosis en Gy).

\subsubsection{A1) M\'etodo de Nelder-Mead para Tumor de 1 $\text{cm}^3$}
\normalsize
\vspace{20pt}

\begin{table} [H]
\caption{ $X_0= 1 \ \text{cm}^3$, $X_{\text{ProptCDI}}=781,853$, $X_{\text{ProptCDIS}}=432,199$, Fraccionamiento Est\'andar, Nelder Mead}
\begin{minipage}[t]{0.45 \textwidth}
\begin{tabular}{ccc}
\toprule
\parbox{2cm}{\centering \textbf{N\'umero de Dosis}} & \textbf{CDI} & \textbf{CDIS} \\
\midrule
1  &  1.1361  &  0.8894 \\
2 & 2.2880 & 2.2264 \\ 
3 & 2.1056 & 1.9698 \\ 
4 & 2.1936 & 2.6718 \\ 
5 & 2.1134 & 2.2034 \\ 
6 & 1.8495 & 1.6066 \\ 
7 & 2.7418 & 2.5082 \\ 
8 & 1.2340 & 1.20005 \\ 
9 & 2.1795 & 2.0179 \\ 
10 & 3.2059 & 3.4168 \\ 
11 & 2.2117 & 2.6057 \\ 
12 & 2.0904 & 2.0441 \\ 
\vdots & \vdots & \vdots \\
\multicolumn{3}{c}{} \\
\bottomrule
\end{tabular}
\end{minipage} \hfill
\begin{minipage}[c]{0.45\textwidth}
\begin{tabular}{ccc}
\toprule
\parbox{2cm}{\centering \textbf{N\'umero de Dosis}} & \textbf{CDI} & \textbf{CDIS}  \\
\midrule
13 & 1.4330 & 1.4782 \\
14 &  2.0366  &  2.1977 \\ 
15 & 0.7540 & 0.6307 \\ 
16 & 2.5165 & 2.7665 \\ 
17 & 0.9258 & 0.7906 \\ 
18 & 2.2628 & 2.0212 \\ 
19 & 2.5676 & 2.5062 \\ 
20 & 1.8081 & 2.0110 \\ 
21 & 2.2754 & 2.2669 \\ 
22 & 1.9821 & 2.0682 \\ 
23 & 1.7214 & 1.9442 \\ 
24 & 2.1244 & 2.1159 \\ 
25 & 2.0321 & 1.6114 \\ 
\hline
$\boldsymbol{X_{\textbf{ProptA}}}$ & 781853 & 432495\\
\bottomrule
\end{tabular}

\end{minipage}
\end{table}
 
\newpage
\begin{table} [H]
\caption{$X_0= 1 \ \text{cm}^3$, $X_{\text{ProptCDI}}=3.21 \times 10^6$, $X_{\text{ProptCDIS}}=1.99 \times 10^6$, Hipofraccionamiento, Nelder Mead}

\begin{minipage}[t]{0.45 \textwidth}
\begin{tabular}{ccc}
\toprule
\parbox{2cm}{\centering \textbf{N\'umero de Dosis}} & \textbf{CDI} & \textbf{CDIS} \\
\midrule
1  &  3.8968  &  4.4794 \\
2 & 2.5870 & 3.3221 \\ 
3 & 2.7498 & 2.6316 \\ 
4 & 3.1597 & 2.9009 \\ 
5 & 1.0621 & 1.7860 \\ 
6 & 2.68004 & 2.0046 \\ 
7 & 3.6634 & 3.7172 \\ 
\vdots & \vdots & \vdots \\
\multicolumn{3}{c}{} \\
\bottomrule
\end{tabular}

\end{minipage} \hfill
\begin{minipage}{0.5 \textwidth}
\begin{tabular}{ccc}
\toprule
\parbox{2cm}{\centering \textbf{N\'umero de Dosis}} & \textbf{CDI} & \textbf{CDIS}  \\
\midrule
8 & 1.6686 & 1.0337 \\ 
9 & 2.0070 & 1.4670 \\ 
10 & 3.1805 & 2.5469 \\ 
11 & 4.2046 & 4.4473 \\ 
12 & 1.2585 & 1.5772 \\ 
13 & 1.8767 & 1.8714 \\
14 &  2.4643  &  2.1202 \\ 
15 & 3.2190 & 3.7525 \\ 
\hline
$\boldsymbol{X_{\textbf{ProptA}}}$ & $3.21 \times 10^6$ & $1.9903 \times 10^6$\\
\bottomrule
\end{tabular}

\end{minipage}
\end{table}
\vspace{10pt}
\subsubsection{A2) Evoluci\'on Diferencial para Tumor de 1 $\text{cm}^3$}
\normalsize

\begin{table} [H]
\caption{$X_0= 1 \ \text{cm}^3$, $X_{\text{ProptCDI}}=3.21 \times 10^6$, $X_{\text{ProptCDIS}}=1.99 \times 10^6$, Hipofraccionamiento, Evoluci\'on Diferencial}

\begin{minipage}[t]{0.45 \textwidth}
\begin{tabular}{ccc}
\toprule
\parbox{2cm}{\centering \textbf{N\'umero de Dosis}} & \textbf{CDI} & \textbf{CDIS} \\
\midrule
1  &  1.8056  &  5 \\
2 & 2.4885 & 3.0154 \\ 
3 & 4.576 & 1.9242 \\ 
4 & 0.0181 & 0.5371 \\ 
5 & 3.4602 & 4.2117 \\ 
6 & 1.7035 & 2.5407 \\ 
7 & 2.8284 & 3.5 $\times 10^{-5}$ \\ 
\vdots & \vdots & \vdots \\
\multicolumn{3}{c}{} \\
\bottomrule
\end{tabular}

\end{minipage} \hfill
\begin{minipage}{0.5 \textwidth}
\begin{tabular}{ccc}
\toprule
\parbox{2cm}{\centering \textbf{N\'umero de Dosis}} & \textbf{CDI} & \textbf{CDIS}  \\
\midrule
8 & 1.6505 & 0.6958 \\ 
9 & 3.2658 & 2.6302 \\ 
10 & 1.4420 & 4.2543 \\ 
11 & 4.0523 & 0.5748 \\ 
12 & 3.6799 & 3.5509 \\ 
13 & 0.0094 & 4.9409 \\
14 &  4.9966  &  3.4011\\ 
15 & 3.5273 & 2.2039 \\ 
\hline
$\boldsymbol{X_{\textbf{ProptA}}}$ & $3.2119 \times 10^6$ & $1.99 \times 10^6$\\
\bottomrule
\end{tabular}

\end{minipage}
\end{table}

\begin{table} [H]
\caption{ $X_0= 1 \ \text{cm}^3$, $X_{\text{ProptCDI}}=781853$, $X_{\text{ProptCDIS}}=432199$, Fraccionamiento Est\'andar, Evoluci\'on Diferencial}
\begin{minipage}[t]{0.5 \textwidth}
\begin{tabular}{ccc}
\toprule
\parbox{2cm}{\centering \textbf{N\'umero de Dosis}} & \textbf{CDI} & \textbf{CDIS} \\
\midrule
1  &  0.0012  &  0.0005 \\
2 & 1.9056 & 1.9048 \\ 
3 & 5 & 4.9992 \\ 
4 & 1.9345 & 1.9338 \\ 
5 & 2.8196 & 2.8188 \\ 
6 & 4.5586 & 4.5577 \\ 
7 & 0.8442 & 0.8435 \\ 
8 & 2.9040 & 2.9032 \\ 
9 & 2.0654 & 2.0646 \\ 
10 & 0.3074 & 0.3066 \\ 
11 & 1.8679 & 1.8671 \\ 
12 & 1.7714 & 1.7706 \\ 
\vdots & \vdots & \vdots \\
\multicolumn{3}{c}{} \\
\bottomrule
\end{tabular}
\end{minipage} \hfill
\begin{minipage}[c]{0.45\textwidth}
\begin{tabular}{ccc}
\toprule
\parbox{2cm}{\centering \textbf{N\'umero de Dosis}} & \textbf{CDI} & \textbf{CDIS}  \\
\midrule
13 & 4.3936 & 4.3928 \\
14 &  0.0013  &  0.0006 \\ 
15 & 2.8663 & 2.8655 \\ 
16 & 4.0979 & 4.0970 \\ 
17 & 0.0017 & 0.0009 \\ 
18 & 3.0855 & 3.0847 \\ 
19 & 0.1018 & 0.1010 \\ 
20 & 2.5310 & 2.5302 \\ 
21 & 0.0011 & 0.0004 \\ 
22 & 0.0016 & 0.0008 \\ 
23 & 2.5245 & 2.5237 \\ 
24 & 0.6162 & 0.6154 \\ 
25 & 3.1246 & 3.1238 \\ 
\hline
$\boldsymbol{X_{\textbf{ProptA}}}$ & 781853 & 433901\\
\bottomrule
\end{tabular}

\end{minipage}
\end{table}

\newpage

\subsubsection{B1) M\'etodo de Nelder-Mead para Tumor de 3 $\text{cm}^3$}
\vspace{20pt}

\begin{table} [H]
\caption{ $X_0= 3 \ \text{cm}^3$, $X_{\text{ProptCDI}}=2.34 \times 10^6$, $X_{\text{ProptCDIS}}=1.29 \times 10^6$, Fraccionamiento Est\'andar, Nelder-Mead}
\begin{minipage}[t]{0.45 \textwidth}
\begin{tabular}{ccc}
\toprule
\parbox{2cm}{\centering \textbf{N\'umero de Dosis}} & \textbf{CDI} & \textbf{CDIS} \\
\midrule
1  &  1.3873  &  0.0592 \\
2 & 1.8092 & 0.1565 \\ 
3 & 2.2578 & 4.0109 \\ 
4 & 2.5829 & 3.7847 \\ 
5 & 1.8665 & 0.7203 \\ 
6 & 2.1298 & 3.4121 \\ 
7 & 2.8248 & 4.0976 \\ 
8 & 1.3073 & 0.9125 \\ 
9 & 2.3203 & 0.2389 \\ 
10 & 3.1848 & 2.6805 \\ 
11 & 2.1848 & 2.6483 \\ 
12 & 2.0219 & 3.1390 \\ 
\vdots & \vdots & \vdots \\
\multicolumn{3}{c}{} \\
\bottomrule
\end{tabular}
\end{minipage} \hfill
\begin{minipage}[c]{0.45\textwidth}
\begin{tabular}{ccc}
\toprule
\parbox{2cm}{\centering \textbf{N\'umero de Dosis}} & \textbf{CDI} & \textbf{CDIS}  \\
\midrule
13 & 1.1823 & 0.6169 \\
14 &  2.1162  &  1.2031 \\ 
15 & 1.0672 & 0.4191 \\ 
16 & 2.2502 & 1.9415 \\ 
17 & 1.0358 & 0.0942 \\ 
18 & 1.9511 & 2.1850 \\ 
19 & 2.3454 & 2.7611 \\ 
20 & 1.8136 & 2.0019 \\ 
21 & 2.5759 & 2.5696 \\ 
22 & 1.7259 & 2.7712 \\ 
23 & 1.7699 & 1.9590 \\ 
24 & 1.9748 & 2.5082 \\ 
25 & 2.1193 & 2.6491 \\ 
\hline
$\boldsymbol{X_{\textbf{ProptA}}}$ & $2.34 \times 10^6 $ & $1.29 \times 10^6$ \\
\bottomrule
\end{tabular}

\end{minipage}
\end{table}
\newpage
\begin{table} [H]
\caption{$X_0= 3 \ \text{cm}^3$, $X_{\text{ProptCDI}}=9.66 \times 10^6$, $X_{\text{ProptCDIS}}=6 \times 10^6$, Hipofraccionamiento, Nelder-Mead}

\begin{minipage}[t]{0.45 \textwidth}
\begin{tabular}{ccc}
\toprule
\parbox{2cm}{\centering \textbf{N\'umero de Dosis}} & \textbf{CDI} & \textbf{CDIS} \\
\midrule
1  &  3.8668  &  3.8418 \\
2 & 2.7095 & 2.5412 \\ 
3 & 2.6894 & 2.7539 \\ 
4 & 3.1015 & 3.1090 \\ 
5 & 1.0804 & 0.9982 \\ 
6 & 2.6054 & 2.7977 \\ 
7 & 3.7450 & 3.5465 \\ 
\vdots & \vdots & \vdots \\
\multicolumn{3}{c}{} \\
\bottomrule
\end{tabular}

\end{minipage} \hfill
\begin{minipage}{0.5 \textwidth}
\begin{tabular}{ccc}
\toprule
\parbox{2cm}{\centering \textbf{N\'umero de Dosis}} & \textbf{CDI} & \textbf{CDIS}  \\
\midrule
8 & 1.6729 & 1.6925 \\ 
9 & 1.9612 & 2.0471 \\ 
10 & 3.1545 & 3.2498 \\ 
11 & 4.2320 & 4.1810 \\ 
12 & 1.2888 & 1.2885 \\ 
13 & 1.8193 & 1.9580 \\
14 &  2.5317  &  2.4579 \\ 
15 & 3.1922 & 3.2015 \\ 
\hline
$\boldsymbol{X_{\textbf{ProptA}}}$ & $9.6606 \times 10^6$ & $6 \times 10^6$\\
\bottomrule
\end{tabular}

\end{minipage}
\end{table}
\vspace{10pt}

\subsubsection{B2) Evoluci\'on Diferencial para Tumor de 3 $\text{cm}^3$}

\begin{table} [H]
\caption{$X_0= 3 \ \text{cm}^3$, $X_{\text{ProptCDI}}=9.66 \times 10^6$, $X_{\text{ProptCDIS}}=6 \times 10^6$, Hipofraccionamiento, Evoluci\'on Diferencial}

\begin{minipage}[t]{0.48 \textwidth}
\begin{tabular}{ccc}
\toprule
\parbox{2cm}{\centering \textbf{N\'umero de Dosis}} & \textbf{CDI} & \textbf{CDIS} \\
\midrule
1  &  2.8277  &  2.5651 \\
2 & 0.6789 & 0.7065 \\ 
3 & 4.2979 & 3.5723 \\ 
4 & 0.0814 & 2.8311 \\ 
5 & 2.9576 & $7.14 \times 10^{-5}$ \\ 
6 & 3.9729 & 1.6250 \\ 
7 & 0.0005 & 2.2076 \\ 
\vdots & \vdots & \vdots \\
\multicolumn{3}{c}{} \\
\bottomrule
\end{tabular}

\end{minipage} \hfill
\begin{minipage}{0.5 \textwidth}
\begin{tabular}{ccc}
\toprule
\parbox{2cm}{\centering \textbf{N\'umero de Dosis}} & \textbf{CDI} & \textbf{CDIS}  \\
\midrule
8 & 3.7129 & 1.6025 \\ 
9 & 1.1722 & 2.4491 \\ 
10 & 4.7191 & 5 \\ 
11 & 0.3851 & 1.5179 \\ 
12 & 0.0003 & 4.0408 \\ 
13 & 4.7296 & 1.5052 \\
14 &  4.9122  &  4.8539 \\ 
15 & 4.8216 & 4.9939 \\ 
\hline
$\boldsymbol{X_{\textbf{ProptA}}}$ & $9.66 \times 10^6$ & $6 \times 10^6$\\
\bottomrule
\end{tabular}

\end{minipage}
\end{table}


\begin{table} [H]
\caption{ $X_0= 3 \ \text{cm}^3$, $X_{\text{ProptCDI}}=2.34 \times 10^6$, $X_{\text{ProptCDIS}}=1.29 \times 10^6$, Fraccionamiento Est\'andar, Evoluci\'on Diferencial}
\begin{minipage}[t]{0.48 \textwidth}
\begin{tabular}{ccc}
\toprule
\parbox{2cm}{\centering \textbf{N\'umero de Dosis}} & \textbf{CDI} & \textbf{CDIS} \\
\midrule
1  &  0.7311  &  0.0513 \\
2 & 2.9820 & 2.7402 \\ 
3 & 0.8630 & 3.2385 \\ 
4 & 3.7462 & 3.2807 \\ 
5 & 0.9006 & 4.9747 \\ 
6 & 2.6499 & 0.8398 \\ 
7 & 1.4931 & 5 \\ 
8 & 3.6534 & 0.3617 \\ 
9 & 2.7869 & 2.0457 \\ 
10 & 4.1562 & 0.5651 \\ 
11 & 0.0647 & 1.2935 \\ 
12 & 0.5082 & 5 \\ 
\vdots & \vdots & \vdots \\
\multicolumn{3}{c}{} \\
\bottomrule
\end{tabular}
\end{minipage} \hfill
\begin{minipage}[c]{0.5\textwidth}
\begin{tabular}{ccc}
\toprule
\parbox{2cm}{\centering \textbf{N\'umero de Dosis}} & \textbf{CDI} & \textbf{CDIS}  \\
\midrule
13 & 2.3276 & 0.4244 \\
14 &  3.0926  &  0.8452 \\ 
15 & 3.3146 & 2.929 \\ 
16 & 1.7019 & 0.8864 \\ 
17 & 0.0582 & 0.3958 \\ 
18 & 2.8912 & 4.4232 \\ 
19 & 2.7682 & 2.8125 \\ 
20 & 3.4875 & 0.6921 \\ 
21 & 3.1198 & 0.6576 \\ 
22 & 0.5834 & 0.3861 \\ 
23 & 0.6560 & 2.5259 \\ 
24 & 0.7064 & 0.0399 \\ 
25 & 0.2511 & 2.8975 \\ 
\hline
$\boldsymbol{X_{\textbf{ProptA}}}$ & $2.34 \times 10^6$ & $1.29 \times 10^6$\\
\bottomrule
\end{tabular}

\end{minipage}
\end{table}
\newpage
\subsubsection{C1) M\'etodo de Nelder-Mead para Tumor de 5 $\text{cm}^3$}

\vspace{20pt}

\begin{table} [H]
\caption{ $X_0= 5 \ \text{cm}^3$, $X_{\text{ProptCDI}}=3.91 \times 10^6$, $X_{\text{ProptCDIS}}=2.16 \times 10^6$, Fraccionamiento Est\'andar, Nelder-Mead}
\begin{minipage}[t]{0.48 \textwidth}
\begin{tabular}{ccc}
\toprule
\parbox{2cm}{\centering \textbf{N\'umero de Dosis}} & \textbf{CDI} & \textbf{CDIS} \\
\midrule
1  &  1.2957  &  1.1366 \\
2 & 2.6346 & 2.2877 \\ 
3 & 2.3907 & 2.1054 \\ 
4 & 2.7295 & 2.1933 \\ 
5 & 1.6963 & 2.1132 \\ 
6 & 1.6571 & 1.8495 \\ 
7 & 2.5608 & 2.7412 \\ 
8 & 1.1545 & 1.2345 \\ 
9 & 1.7347 & 2.1793 \\ 
10 & 3.5227 & 3.2049 \\ 
11 & 1.7420 & 2.2115 \\ 
12 & 2.248 & 2.0902 \\ 
\vdots & \vdots & \vdots \\
\multicolumn{3}{c}{} \\
\bottomrule
\end{tabular}
\end{minipage} \hfill
\begin{minipage}[c]{0.5\textwidth}
\begin{tabular}{ccc}
\toprule
\parbox{2cm}{\centering \textbf{N\'umero de Dosis}} & \textbf{CDI} & \textbf{CDIS}  \\
\midrule
13 & 1.8425 & 1.4333 \\
14 &  2.1871  &  2.0364 \\ 
15 & 0.1492 & 0.7594 \\ 
16 & 2.7312 & 2.5161 \\ 
17 & 0.4750 & 0.9265 \\ 
18 & 2.4928 & 2.2625 \\ 
19 & 2.7176 & 2.5671 \\ 
20 & 1.8751 & 1.8081 \\ 
21 & 1.8383 & 2.2751 \\ 
22 & 2.2833 & 1.9820 \\ 
23 & 1.8714 & 1.7215 \\ 
24 & 2.3032 & 2.1242 \\ 
25 & 1.5996 & 2.0320 \\ 
\hline
$\boldsymbol{X_{\textbf{ProptA}}}$ & $3.91 \times 10^6$ & $2.16 \times 10^6$\\
\bottomrule
\end{tabular}

\end{minipage}
\end{table}
\newpage

\begin{table} [H]
\caption{$X_0= 5 \ \text{cm}^3$, $X_{\text{ProptCDI}}=1.61 \times 10^7$, $X_{\text{ProptCDIS}}=1 \times 10^7$, Hipofraccionamiento, Nelder-Mead}

\begin{minipage}[t]{0.48 \textwidth}
\begin{tabular}{ccc}
\toprule
\parbox{2cm}{\centering \textbf{N\'umero de Dosis}} & \textbf{CDI} & \textbf{CDIS} \\
\midrule
1  &  4.0007  &  3.7169 \\
2 & 2.7194 & 2.6373 \\ 
3 & 2.7634 & 2.6960 \\ 
4 & 3.2021 & 2.7927 \\ 
5 & 1.0645 & 1.3018 \\ 
6 & 2.5208 & 2.3902 \\ 
7 & 3.9532 & 3.3391 \\ 
\vdots & \vdots & \vdots \\
\multicolumn{3}{c}{} \\
\bottomrule
\end{tabular}

\end{minipage} \hfill
\begin{minipage}{0.5 \textwidth}
\begin{tabular}{ccc}
\toprule
\parbox{2cm}{\centering \textbf{N\'umero de Dosis}} & \textbf{CDI} & \textbf{CDIS}  \\
\midrule
8 & 1.5425 & 1.7453 \\ 
9 & 1.9549 & 2.0268 \\ 
10 & 3.1379 & 3.4008 \\ 
11 & 4.1360 & 4.2754 \\ 
12 & 1.2504 & 1.5895 \\ 
13 & 1.7115 & 1.8685 \\
14 &  2.4506  &  3.0697 \\ 
15 & 3.2285 & 2.8288 \\ 
\hline
$\boldsymbol{X_{\textbf{ProptA}}}$ & $1.61 \times 10^7$ & $1 \times 10^7$\\
\bottomrule
\end{tabular}

\end{minipage}
\end{table}
\vspace{10pt}

\subsubsection{C2) Evoluci\'on Diferencial para Tumor de 5 $\text{cm}^3$}

\begin{table} [H]
\caption{$X_0= 5 \ \text{cm}^3$, $X_{\text{ProptCDI}}=1.61 \times 10^7$, $X_{\text{ProptCDIS}}=1 \times 10^7$, Hipofraccionamiento, Evoluci\'on Diferencial}

\begin{minipage}[t]{0.48 \textwidth}
\begin{tabular}{ccc}
\toprule
\parbox{2cm}{\centering \textbf{N\'umero de Dosis}} & \textbf{CDI} & \textbf{CDIS} \\
\midrule
1  &  0.0979  &  0.0164 \\
2 & 5 & 4.3258 \\ 
3 & 5 & 0.9926 \\ 
4 & 3.2513 & 4.1397 \\ 
5 & 4.5777 & 0.7214 \\ 
6 & 0.6695 & 4.1147 \\ 
7 & 3.3999 & 2.3025 \\ 
\vdots & \vdots & \vdots \\
\multicolumn{3}{c}{} \\
\bottomrule
\end{tabular}

\end{minipage} \hfill
\begin{minipage}{0.5 \textwidth}
\begin{tabular}{ccc}
\toprule
\parbox{2cm}{\centering \textbf{N\'umero de Dosis}} & \textbf{CDI} & \textbf{CDIS}  \\
\midrule
8 & 0.2671 & 4.9636 \\ 
9 & 2.6259 & 3.3589 \\ 
10 & 0.1929 & 2.2039 \\ 
11 & 2.9213 & 1.1354 \\ 
12 & 1.9172 & 2.8005 \\ 
13 & 2.8370 & 4.7938 \\
14 &  3.1548  &  3.2773 \\ 
15 & 3.5053 & 0.2754 \\ 
\hline
$\boldsymbol{X_{\textbf{ProptA}}}$ & $1.61 \times 10^7$ & $1 \times 10^7$\\
\bottomrule
\end{tabular}

\end{minipage}
\end{table}

\begin{table} [H]
\caption{ $X_0= 5 \ \text{cm}^3$, $X_{\text{ProptCDI}}=3.91 \times 10^6$, $X_{\text{ProptCDIS}}=2.16 \times 10^6$, Fraccionamiento Est\'andar, Evoluci\'on Diferencial}
\begin{minipage}[t]{0.48 \textwidth}
\begin{tabular}{ccc}
\toprule
\parbox{2cm}{\centering \textbf{N\'umero de Dosis}} & \textbf{CDI} & \textbf{CDIS} \\
\midrule
1  &  0.0905  &  1.282 \\
2 & 2.1106 & 0.2555 \\ 
3 & 0.0003 & 4.7816 \\ 
4 & 0.0003 & 0.4404 \\ 
5 & 2.4665 & 3.4796 \\ 
6 & 4.9995 & 3.0439 \\ 
7 & 0.0005 & 3.4388 \\ 
8 & 4.0832 & 2.4072 \\ 
9 & 4.9995 & 3.7489 \\ 
10 & 0.6721 & 0.5378 \\ 
11 & 0.0007 & 1.0724 \\ 
12 & 3.4327 & 1.4535 \\ 
\vdots & \vdots & \vdots \\
\multicolumn{3}{c}{} \\
\bottomrule
\end{tabular}
\end{minipage} \hfill
\begin{minipage}[c]{0.5\textwidth}
\begin{tabular}{ccc}
\toprule
\parbox{2cm}{\centering \textbf{N\'umero de Dosis}} & \textbf{CDI} & \textbf{CDIS}  \\
\midrule
13 & 0.5506 & 0.1773 \\
14 &  0.0009  &  1.7040 \\ 
15 & 5 & 0.9745 \\ 
16 & 4.9689 & 0.9657 \\ 
17 & 1.4157 & 3.1834 \\ 
18 & 4.0386 & 2.5464 \\ 
19 & 3.5479 & 1.1409 \\ 
20 & 0.0003 & 1.1610 \\ 
21 & 1.6387 & 3.8265 \\ 
22 & 1.7595 & 1.8750 \\ 
23 & 1.6828 & 1.1390 \\ 
24 & 1.6347 & 1.1223 \\ 
25 & 0.0013 & 3.7278 \\ 
\hline
$\boldsymbol{X_{\textbf{ProptA}}}$ & $3.91 \times 10^6$ & $2.16 \times 10^6$\\
\bottomrule
\end{tabular}

\end{minipage}
\end{table}

Para los tres tamaños de tumor los resultados son satisfactorios. Mediante Nelder-Mead, para CDI en el tumor de 1 $\text{cm}^3$ y fraccionamiento est\'andar, $X_{\text{ProptA}}$ llega al valor de $X_{\text{ProptCDIS}}$ , el mismo m\'etodo para CDIS solo presenta una diferencia de 296 c\'elulas (tabla 4.4) y en hipofraccionamiento para CDIS existe una diferencia de 300 c\'elulas (tabla 4.5). Para el mismo tamaño de tumor evoluci\'on diferencial presenta una diferencia de 1900 c\'elulas respecto del valor fijado en CDI en Hipofraccionamiento. Para los dem\'as resultados del tumor de  1 $\text{cm}^3$ y evoluci\'on diferencial CDIS presenta una diferencia de 1702 c\'elulas, CDI alcanza el valor fijado. Dentro del an\'alisis del tumor de 3 $\text{cm}^3$, ambos m\'etodos llegan al valor estipulado salvo en Nelder Mead para CDI donde se presenta una diferencia de 600 c\'elulas. Finalmente para el tumor de  5 $\text{cm}^3$ de igual forma ambos m\'etodos alcanzan el valor fijado de n\'umero de c\'elulas. Bajo un esquema general, podemos decir que ambos m\'etodos nos dan resultados bastante buenos ya que las diferencias son del orden de $10^{-3} \ \text{mm}^3$ y de $10^{-4} \ \text{mm}^3$.

Otro aspecto importante es la distribuci\'on general de las dosis optimizadas. Recordemos que minimizamos el n\'umero de c\'elulas cancer\'igenas en la 15 y 25 sesi\'on de tratamiento, sin embargo la ecuaci\'on de nuestro modelo en esas sesiones en espec\'ifico, depende del valor de cada una de las sesiones previas, es decir, el modelo toma en consideraci\'on cada uno de los d\'ias de tratamiento previos, en base a ello la optimizaci\'on se realiza tal que cada una de las dosis est\'a fijada a un d\'ia del tratamiento (d\'ia 1 - dosis d1, etc) por lo que las dosis no son intercambiables. En resumidas cuentas, las dosis encontradas son tales que minimizan el n\'umero de c\'elulas cancerigenas a un valor prefijado de tal manera que se debe de seguir la distribuci\'on ordenada de las dosis. 

Asi mismo (y ya que se toma en cuenta la dependencia de cada uno de los d\'ias de tratamiento previos a la \'ultima sesi\'on) podemos observar que para los resultados de Nelder-Mead, las dosis siguen un patr\'on relativo a una "dosis media" ($\approx$ 2 Gy para fraccionamiento est\'andar y $\approx$ 2.66 Gy para hipofraccionamiento) esto ocasiona que si una dosis llega a ser muy alta, la siguiente dosis tendr\'a un valor mucho menor con el prop\'osito de compensar que el valor final de c\'elulas cancer\'igenas cumpla la restricci\'on impuesta (el criterio de parada que hemos mencionado antes), si la dosis llega a ser pequeña, se observa que la siguiente dosis tratar\'a de compensar el resultado final aumentando su valor. Lo anterior puede deberse a la forma en que se encuentran los m\'inimos para cada m\'etodo ya que si observamos la distribuci\'on de dosis de Evoluci\'on Diferencial, estas no tienen un patr\'on definido (contrario a lo que se aprecia en el caso de Nelder-Mead). 

Si graficamos la dosis vs el n\'umero de d\'ia del tratamiento (para Nelder-Mead, como ya lo habamos mencionado) podemos apreciar mejor el argumento del p\'arrafo anterior y observamos como las dosis fluctuan a lo largo de una dosis media. Se muestran los casos para 1  $\text{cm}^3$ y 5  $\text{cm}^3$ 

\begin{figure}[h!]
    \centering
    \includegraphics[scale=0.43]{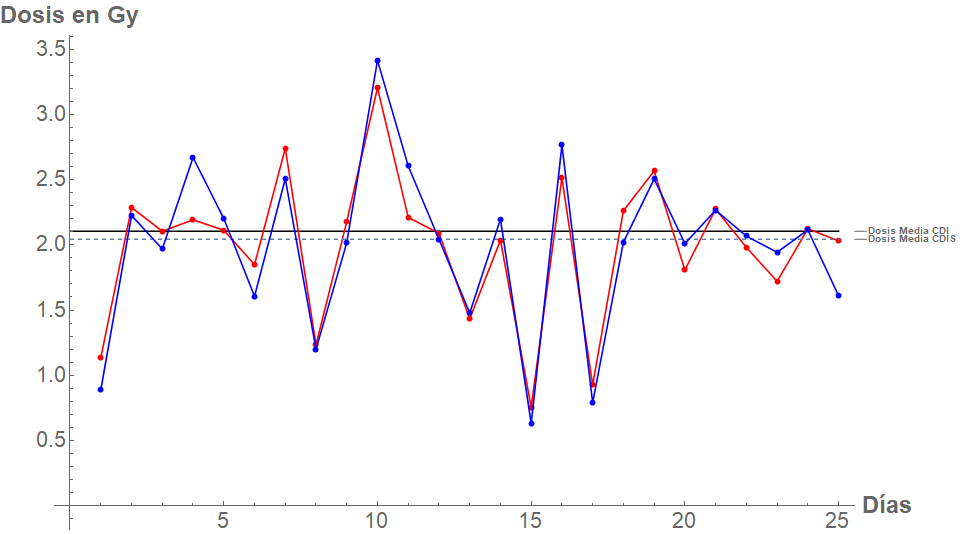}
    \caption{Comportamiento de las dosis a lo largo de los d\'ias del tratamiento con Nelder-Mead para el tumor de 1 $\text{cm}^3$ y fraccionamiento est\'andar, en rojo CDI y en azul CDIS; dosis media: l\'inea solida CDI, l\'inea punteada CDIS }
    
\end{figure}

\begin{figure}[h!]
    \centering
    \includegraphics[scale=0.43]{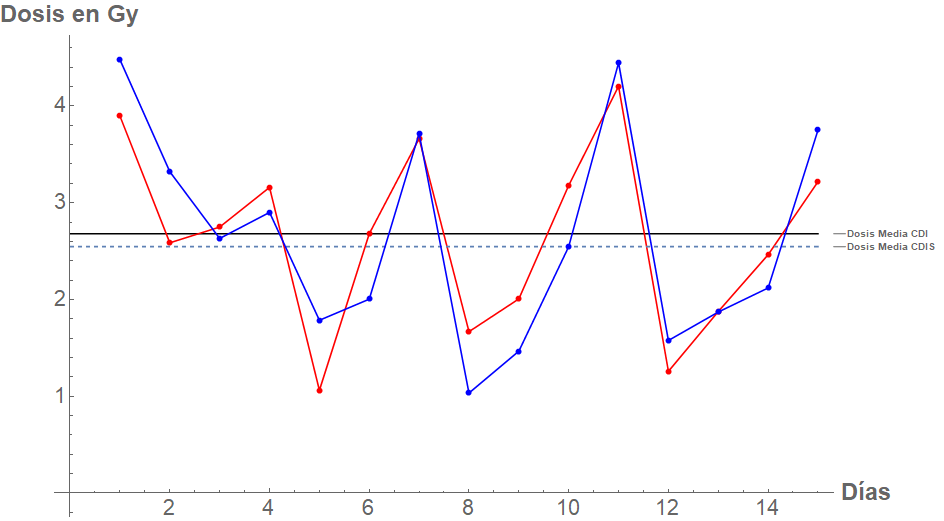}
    \caption{Comportamiento de las dosis a lo largo de los d\'ias del tratamiento con Nelder-Mead para el tumor de 1 $\text{cm}^3$ e hipofraccionamiento, en rojo CDI y en azul CDIS; dosis media: l\'inea solida CDI, l\'inea punteada CDIS }
    
\end{figure}
\newpage
\begin{figure}[h!]
    \centering
    \includegraphics[scale=0.4]{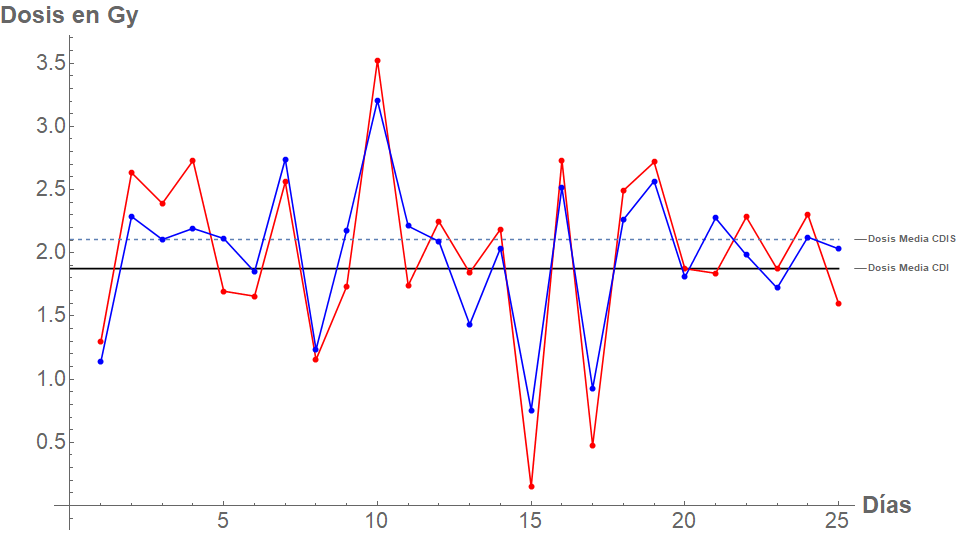}
    \caption{Comportamiento de las dosis a lo largo de los d\'ias del tratamiento con Nelder-Mead para el tumor de 5 $\text{cm}^3$ y fraccionamiento est\'andar, en rojo CDI y en azul CDIS; dosis media: l\'inea solida CDI, l\'inea punteada CDIS }
\end{figure}

\begin{figure}[h!]
    \centering
    \includegraphics[scale=0.4]{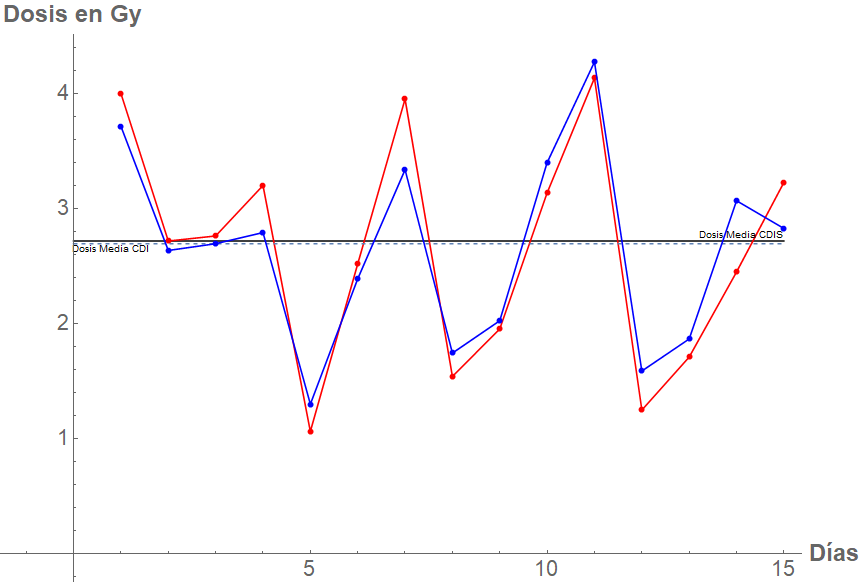}
    \caption{Comportamiento de las dosis a lo largo de los d\'ias del tratamiento con Nelder-Mead para el tumor de 5 $\text{cm}^3$ e hipofraccionamiento, en rojo CDI y en azul CDIS; dosis media: l\'inea solida CDI, l\'inea punteada CDIS }
    \end{figure}

\newpage
Adicionalmente podemos cambiar el n\'umero de c\'elulas cancer\'igenas al que queremos llegar, dependiendo del valor ambos m\'etodos podr\'ian alcanzar $X_{\text{ProptA}}$ o un valor cercano a el. Reiteramos que los valores del n\'umero de c\'elulas cancer\'igenas se tomaron de tal modo que se pudiera comparar el n\'umero que se obtiene al final de un tratamiento con dosis est\'andar (usadas actualmente) y dosis optimizadas.  

Ahora que hemos obtenido las dosis optimizadas para cada tamaño de tumor y tratamiento, veamos como se comporta el aumento y disminuci\'on de c\'elulas cancer\'igenas tal y como vimos en la secci\'on 4.1.1, es decir, simularemos el tratamiento con estas dosis. 

\subsubsection{Simulaci\'on del Tratamiento con Dosis Optimizadas}

\begin{figure}[h!]
    \centering
    \includegraphics[scale=0.6]{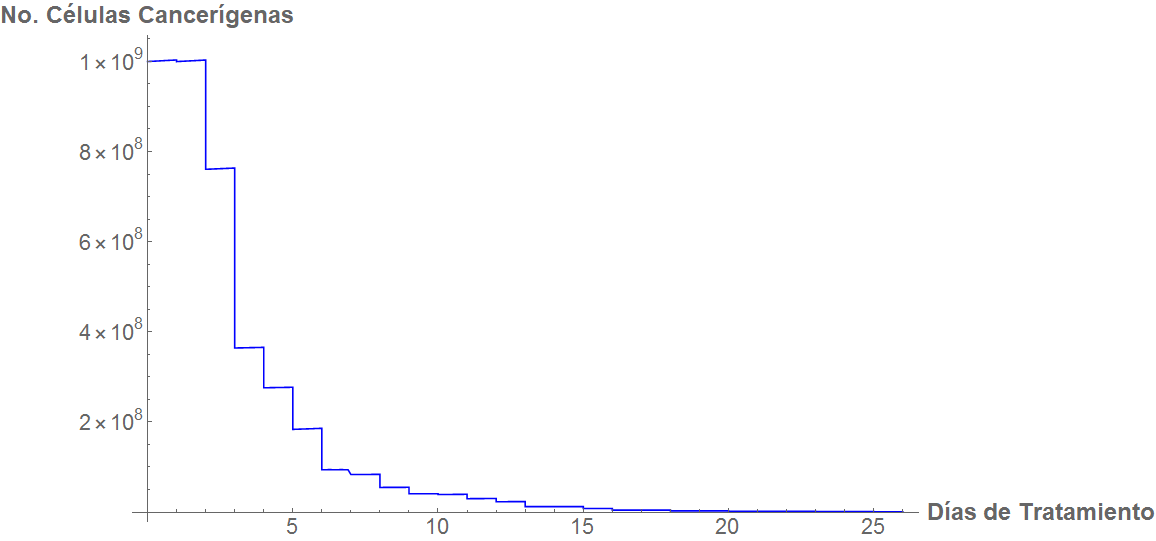}
    \caption{No. de c\'elulas cancer\'igenas durante el fraccionamiento est\'andar (2 Gy por d\'ia) para CDI, tamaño inicial del tumor de 1 $\text{cm}^3$, M\'etodo de Evoluci\'on Diferencial }
    \label{OptBCCL2FraccStand50gyEvo}
\end{figure}

\begin{figure}[H]
    \centering
    \includegraphics[scale=0.6]{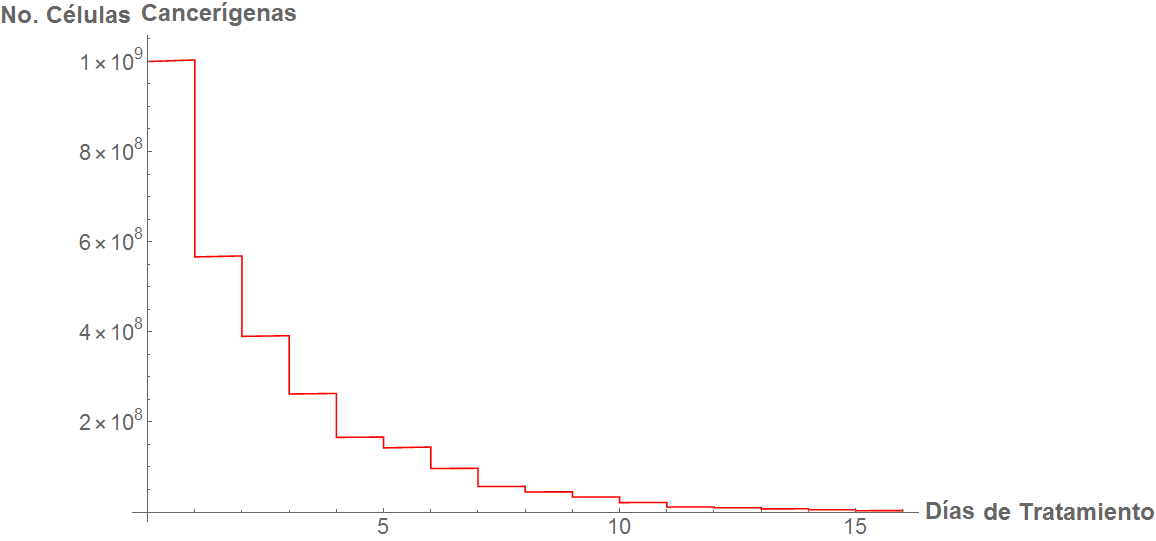}
    \caption{No. de c\'elulas cancer\'igenas durante el hipofraccionamiento (2.66 Gy por d\'ia) para CDI, tamaño inicial del tumor de 1 $\text{cm}^3$, M\'etodo de Nelder-Mead }
    \label{OptBCCL2Hypofracc40gyNeld}
\end{figure}

\begin{figure}[h!]
    \centering
    \includegraphics[scale=0.6]{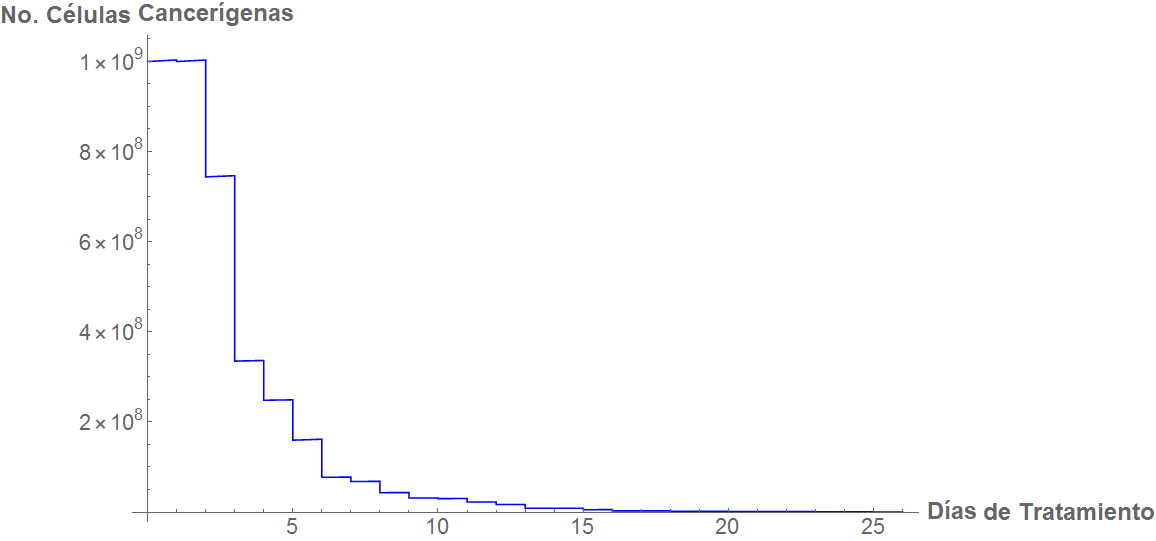}
    \caption{No. de c\'elulas cancer\'igenas durante el fraccionamiento est\'andar (2 Gy por d\'ia) para CDIS, tamaño inicial del tumor de 1 $\text{cm}^3$, M\'etodo de Evoluci\'on Diferencial }
    \label{OptBCCL3FraccStand50gyEvo}
\end{figure}

\begin{figure}[H]
    \centering
    \includegraphics[scale=0.6]{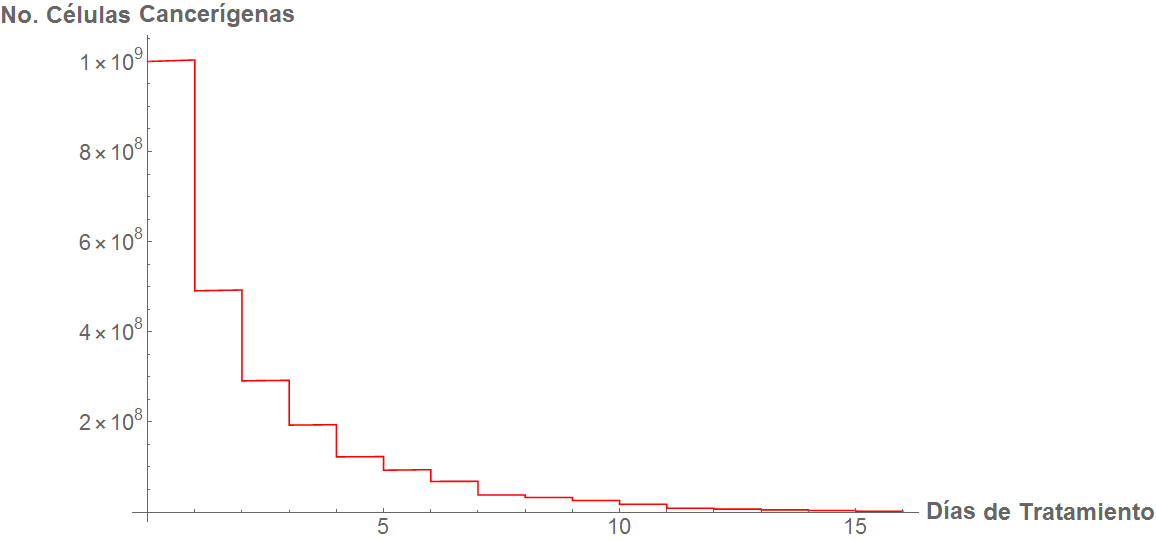}
    \caption{No. de c\'elulas cancer\'igenas durante el hipofraccionamiento (2.66 Gy por d\'ia) para CDIS, tamaño inicial del tumor de 1 $\text{cm}^3$, M\'etodo de Nelder-Mead }
    \label{OptBCCL3Hypofracc40gyNeld}
\end{figure}

\begin{figure}[h!]
    \centering
    \includegraphics[scale=0.6]{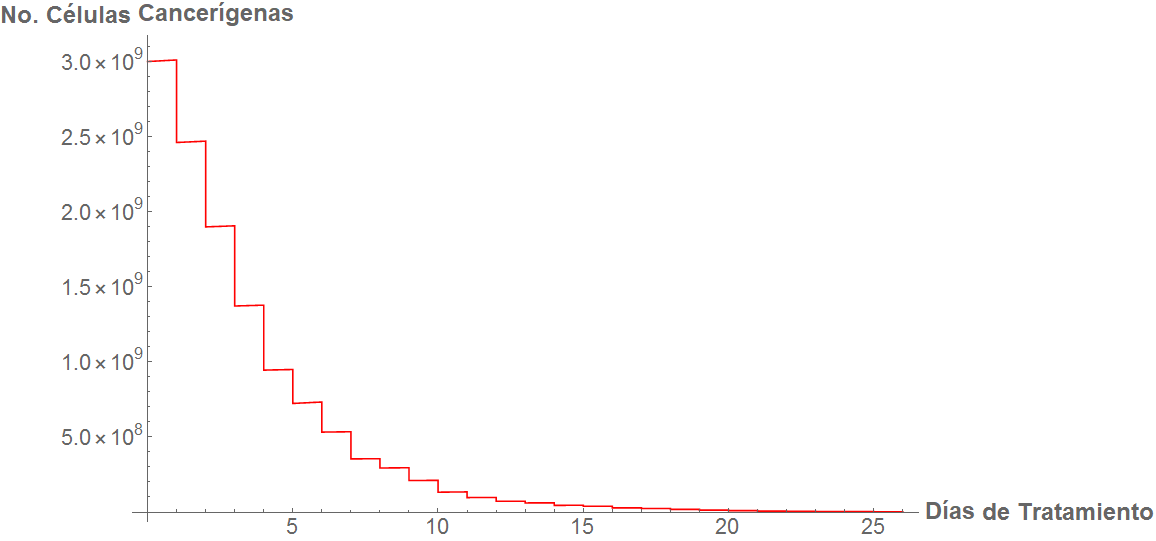}
    \caption{No. de c\'elulas cancer\'igenas durante el fraccionamiento est\'andar (2 Gy por d\'ia) para CDI, tamaño inicial del tumor de 3 $\text{cm}^3$, M\'etodo de Nelder-Mead }
    \label{OptBCCL2FraccStand50gy3cmNeld}
\end{figure}

\begin{figure}[H]
    \centering
    \includegraphics[scale=0.6]{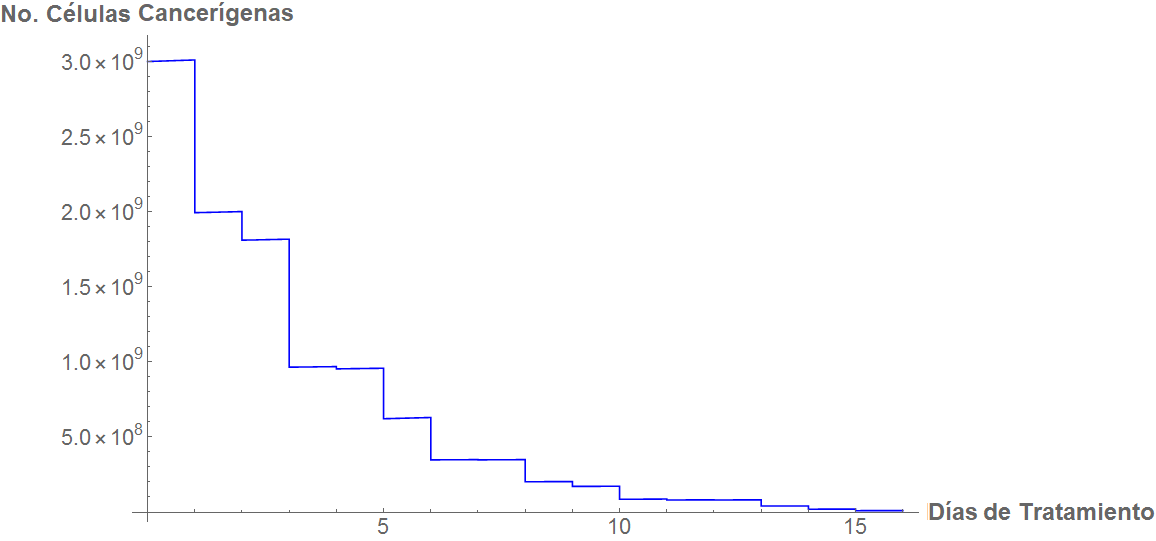}
    \caption{No. de c\'elulas cancer\'igenas durante el hipofraccionamiento (2.66 Gy por d\'ia) para CDI, tamaño inicial del tumor de 3 $\text{cm}^3$, M\'etodo de Evoluci\'on Diferencial }
    \label{OptBCCL2Hypofracc40gy3cmEvo}
\end{figure}

\begin{figure}[h!]
    \centering
    \includegraphics[scale=0.6]{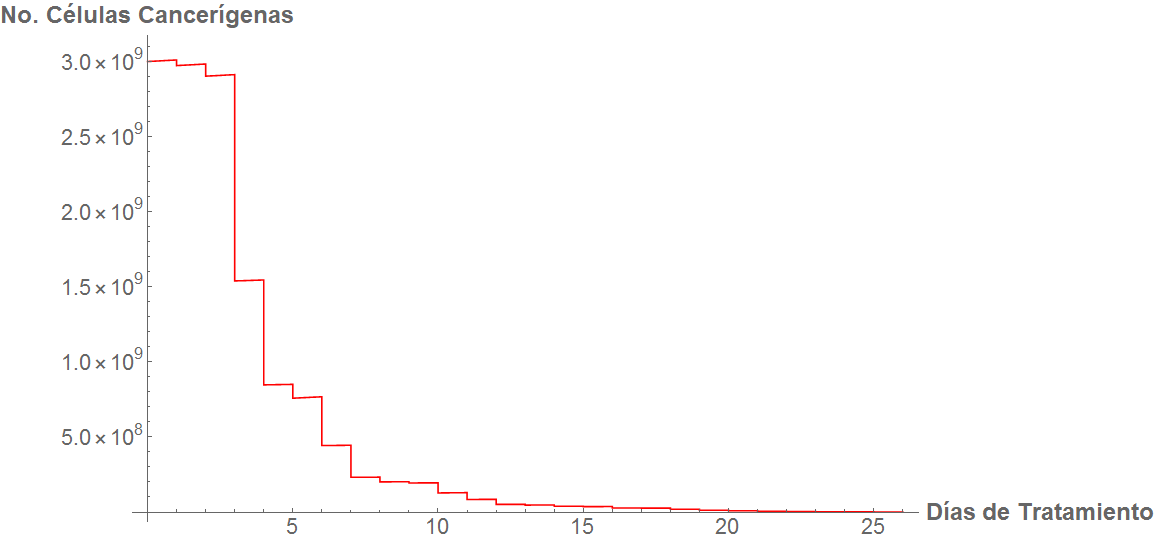}
    \caption{No. de c\'elulas cancer\'igenas durante el fraccionamiento est\'andar (2 Gy por d\'ia) para CDIS, tamaño inicial del tumor de 3 $\text{cm}^3$, M\'etodo de Nelder-Mead }
    \label{OptBCCL3FraccStand50gy3cmNeld}
\end{figure}

\begin{figure}[H]
    \centering
    \includegraphics[scale=0.6]{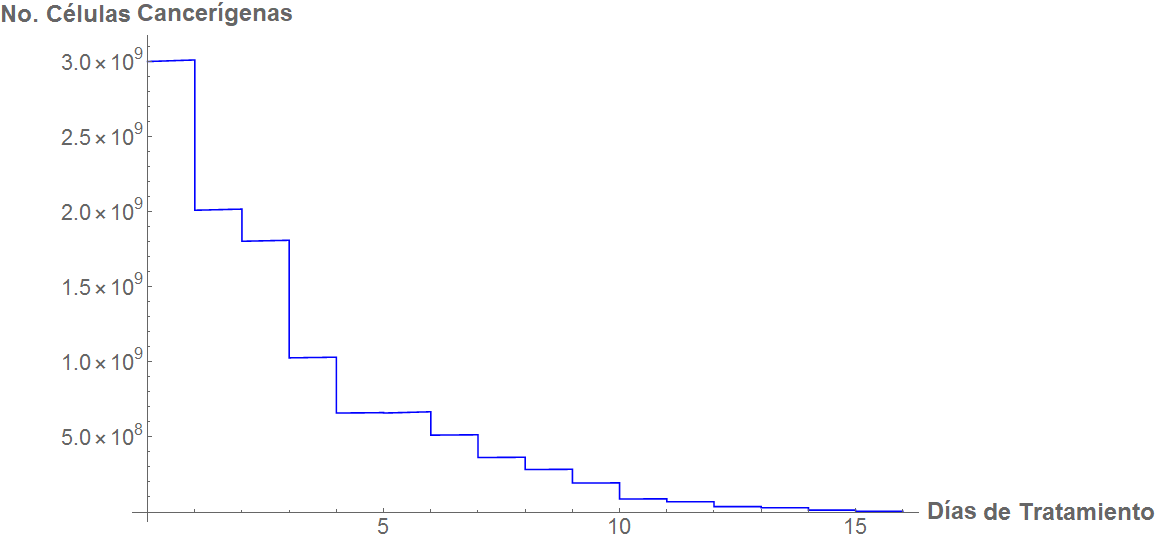}
    \caption{No. de c\'elulas cancer\'igenas durante el hipofraccionamiento (2.66 Gy por d\'ia) para CDIS, tamaño inicial del tumor de 3 $\text{cm}^3$, M\'etodo de Evoluci\'on Diferencial }
    \label{OptBCCL3Hypofracc40gy3cmEvo}
\end{figure}

\begin{figure}[h!]
    \centering
    \includegraphics[scale=0.6]{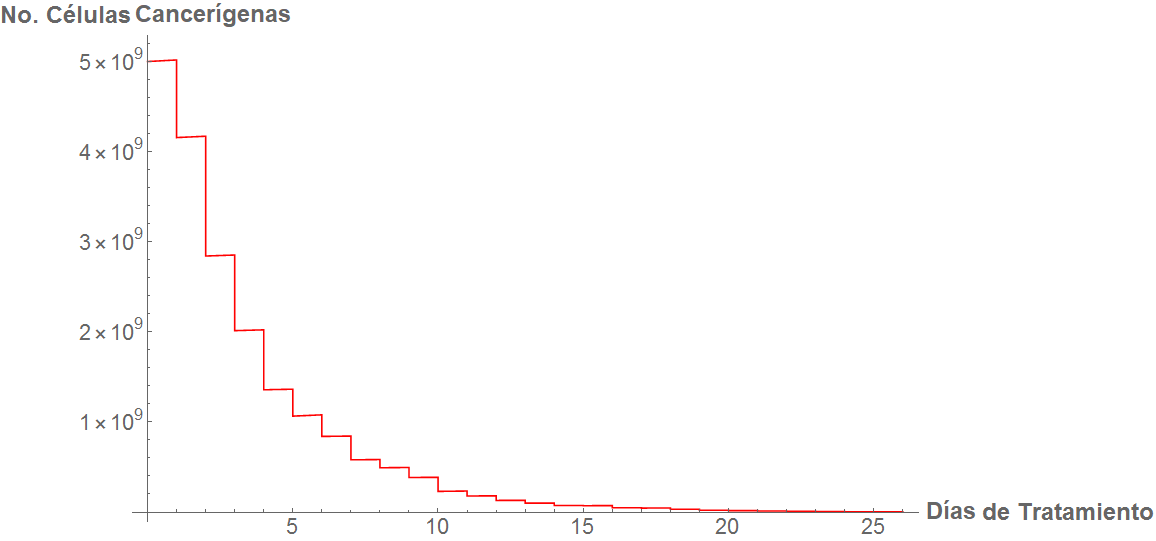}
    \caption{No. de c\'elulas cancer\'igenas durante el fraccionamiento est\'andar (2 Gy por d\'ia) para CDI, tamaño inicial del tumor de 5 $\text{cm}^3$, M\'etodo de Nelder-Mead }
    \label{OptBCCL2FraccStand50gy5cmNeld}
\end{figure}

\begin{figure}[H]
    \centering
    \includegraphics[scale=0.6]{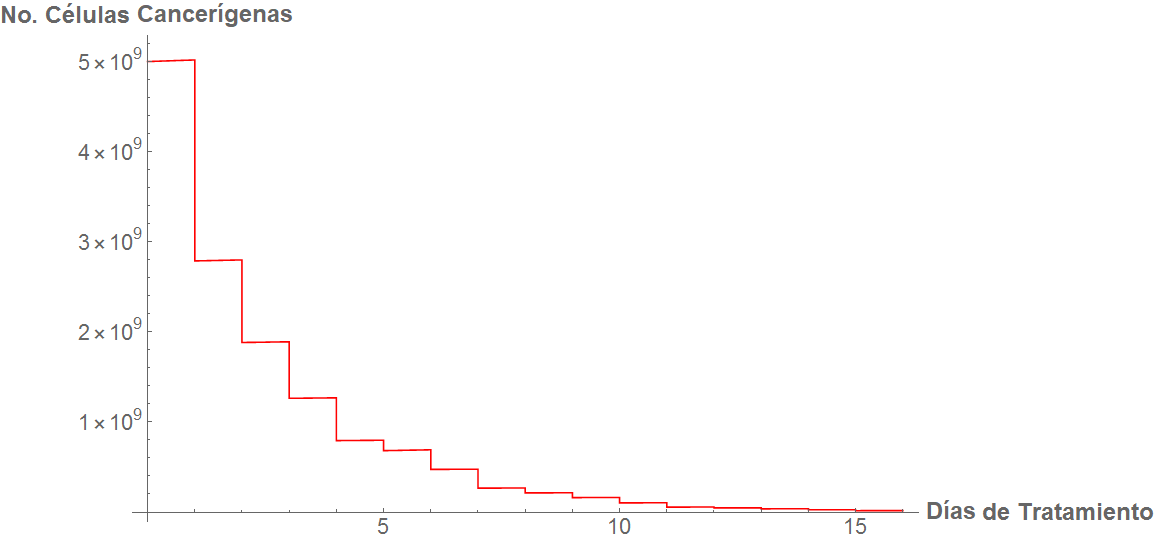}
    \caption{No. de c\'elulas cancer\'igenas durante el hipofraccionamiento (2.66 Gy por d\'ia) para CDI, tamaño inicial del tumor de 5 $\text{cm}^3$, M\'etodo de Nelder-Mead }
    \label{OptBCCL2Hypofracc40gy5cmNeld}
\end{figure}

\begin{figure}[h!]
    \centering
    \includegraphics[scale=0.6]{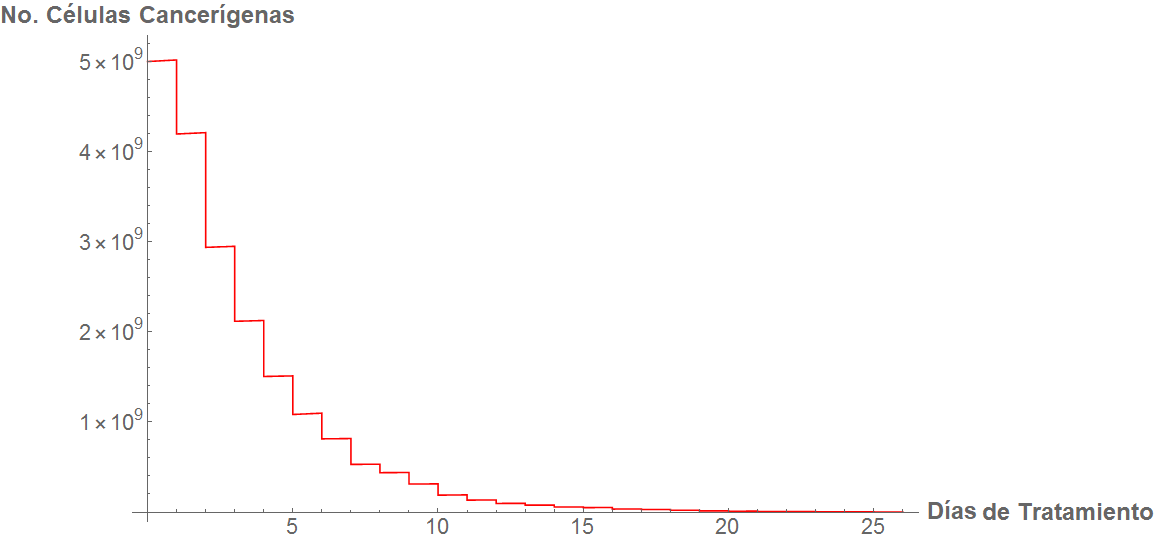}
    \caption{No. de c\'elulas cancer\'igenas durante el fraccionamiento est\'andar (2 Gy por d\'ia) para CDIS, tamaño inicial del tumor de 5 $\text{cm}^3$, M\'etodo de Nelder-Mead }
    \label{OptBCCL3FraccStand50gy5cmNeld}
\end{figure}

\begin{figure}[H]
    \centering
    \includegraphics[scale=0.6]{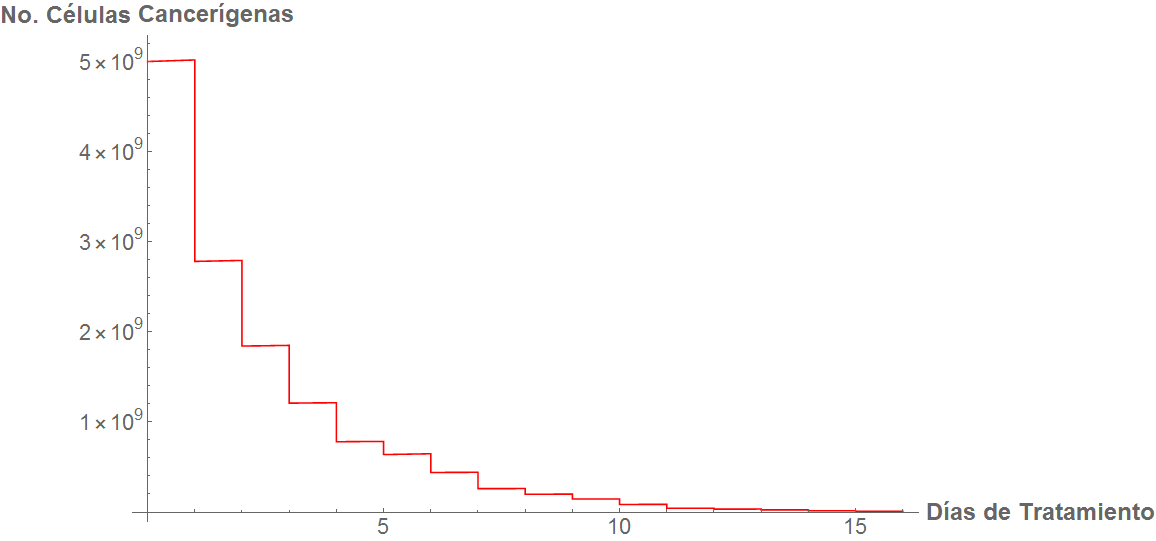}
    \caption{No. de c\'elulas cancer\'igenas durante el hipofraccionamiento (2.66 Gy por d\'ia) para CDIS, tamaño inicial del tumor de 5 $\text{cm}^3$, M\'etodo de Nelder-mead }
    \label{OptBCCL3Hypofracc40gy5cmNeld}
\end{figure}

Se opt\'o por mostrar el m\'etodo para los casos en que tuvo mejores resultados para cada tipo de tumor, en el \'ultimo caso (tumor de 5 $\text{cm}^3$) se mostr\'o el resultado para Nelder-Mead ya que es el que presenta un comportamiento bastante estable. En general se observa que Nelder-Mead, al oscilar en valores de dosis respecto de un "valor promedio", presenta una tendencia estable. Por otro lado Evoluci\'on Diferencial al presentar una distribuci\'on de dosis no tan regular, se observa que el tratamiento no es tan estable, las dosis altas obtenidas mediante este m\'etodo ocasionan que exista una disminuci\'on s\'ubita del n\'umero de c\'elulas cancer\'igenas (por ejemplo la figura 4.16). Si bien las dosis obtenidas por ambos m\'etodos logran reducir el n\'umero de c\'elulas cancer\'igenas hasta el valor prefijado (o cercano a el) no se puede ignorar el efecto biol\'ogico que pudiera ocasionar al paciente (por ejemplo evoluci\'on diferencial en ocasiones arroja dosis altas). Con esto en mente y el hecho de que los valores de la dosificaci\'on est\'andar est\'en muy cercanos a los obtenidos mediante la optimizaci\'on (para el caso de Nelder-Mead), nos da una pista sobre que el tratamiento actual se encuentra muy pr\'oximo a valores \'optimos te\'oricos. Adem\'as los especialistas en el \'area lograron acordar los valores para dosis est\'andares dados conocimientos espec\'ificos sobre fisiolog\'ia del organismo bajo radiaci\'on, efectos secundarios, etc. Por lo que sus resultados se encuentran en alguna frontera de las dosis \'optimas ''te\'oricas reales". Para investigar m\'as sobre este punto tomaremos el algoritmo de Nelder Mead solo que ahora especificaremos los puntos iniciales del simplex (daremos un vector de inicializaci\'on) ya que el objetivo es encontrar las dosis \'optimas estos puntos iniciales corresponder\'an a las dosis est\'andar (usadas actualmente) que son de 2 Gy para fraccionamiento
est\'andar y de 2.66 Gy para hipofraccionamiento.

Con estos datos obtenemos los siguientes resultados:

\subsubsection{* Nelder-Mead con inicializaci\'on para Tumor de 1 $\text{cm}^3$}
\begin{table} [H]
\caption{ $X_0= 1 \ \text{cm}^3$, $X_{\text{ProptCDI}}=781853$, $X_{\text{ProptCDIS}}=432199$, Fraccionamiento Est\'andar, Nelder Mead}
\begin{minipage}[t]{0.45 \textwidth}
\begin{tabular}{ccc}
\toprule
\parbox{2cm}{\centering \textbf{N\'umero de Dosis}} & \textbf{CDI} & \textbf{CDIS} \\
\midrule
1  &  1.6231  & 1.9829 \\
2 & 1.8510 & 2.5527 \\ 
3 & 2.3865 & 1.7291 \\ 
4 & 2.4724 & 1.9146 \\ 
5 & 1.7419 & 1.7403 \\ 
6 & 1.8072 & 1.8591 \\ 
7 & 2.6402 & 2.7301 \\ 
8 & 1.2117 & 1.4499 \\ 
9 & 1.8691 & 2.1713 \\ 
10 & 2.9274 & 3.1899 \\ 
11 & 2.1661 & 2.2703 \\ 
12 & 2.5434 & 2.2143 \\ 
\vdots & \vdots & \vdots \\
\multicolumn{3}{c}{} \\
\bottomrule
\end{tabular}
\end{minipage} \hfill
\begin{minipage}[c]{0.5\textwidth}
\begin{tabular}{ccc}
\toprule
\parbox{2cm}{\centering \textbf{N\'umero de Dosis}} & \textbf{CDI} & \textbf{CDIS}  \\
\midrule
13 & 1.5089 & 1.3929 \\
14 &  2.2683  &  2.4747 \\ 
15 & 0.8596 & 1.1898 \\ 
16 & 2.0432 & 2.4724 \\ 
17 & 1.1526 & 1.3659 \\ 
18 & 2.245 & 2.2377 \\ 
19 & 2.3248 & 1.9371 \\ 
20 & 2.0717 & 1.5824 \\ 
21 & 2.1001 & 1.7490 \\ 
22 & 1.8284 & 1.5910 \\ 
23 & 1.7194 & 2.0828 \\ 
24 & 2.1677 & 1.4492 \\ 
25 & 2.2765 & 2.4804 \\ 
\hline
$\boldsymbol{X_{\textbf{ProptA}}}$ & 781853 & 432199\\
\bottomrule
\end{tabular}

\end{minipage}
\end{table}
\newpage

\begin{table} [H]
\caption{$X_0= 1 \ \text{cm}^3$, $X_{\text{ProptCDI}}=3.21 \times 10^6$, $X_{\text{ProptCDIS}}=1.9903 \times 10^6$, Hipofraccionamiento, Nelder-Mead}

\begin{minipage}[t]{0.48 \textwidth}
\begin{tabular}{ccc}
\toprule
\parbox{2cm}{\centering \textbf{N\'umero de Dosis}} & \textbf{CDI} & \textbf{CDIS} \\
\midrule
1  &  2.9666  &  3.4238 \\
2 & 3.0205 & 3.3966 \\ 
3 & 2.1979 & 2.0865 \\ 
4 & 2.2021 & 1.9331 \\ 
5 & 1.2710 & 1.1561 \\ 
6 & 2.7546 & 2.5552 \\ 
7 & 2.2428 & 2.4024 \\ 
\vdots & \vdots & \vdots \\
\multicolumn{3}{c}{} \\
\bottomrule
\end{tabular}

\end{minipage} \hfill
\begin{minipage}{0.5 \textwidth}
\begin{tabular}{ccc}
\toprule
\parbox{2cm}{\centering \textbf{N\'umero de Dosis}} & \textbf{CDI} & \textbf{CDIS}  \\
\midrule
8 & 2.3486 & 1.9825 \\ 
9 & 2.1518 & 1.9401 \\ 
10 & 4.0034 & 3.8438 \\ 
11 & 4.4891 & 4.2920 \\ 
12 & 1.5453 & 2.0480 \\ 
13 & 2.3446 & 2.6674 \\
14 &  3.5050  &  3.4090 \\ 
15 & 2.6513 & 2.5772 \\ 
\hline
$\boldsymbol{X_{\textbf{ProptA}}}$ & $3.21 \times 10^6$ & $1.9903 \times 10^6$\\
\bottomrule
\end{tabular}

\end{minipage}
\end{table}
\vspace{10pt}

\subsubsection{* Nelder-Mead con inicializaci\'on para Tumor de 5 $\text{cm}^3$}

\begin{table} [H]
\caption{$X_0= 5 \ \text{cm}^3$, $X_{\text{ProptCDI}}=1.61 \times 10^7$, $X_{\text{ProptCDIS}}=1 \times 10^7$, Hipofraccionamiento, Nelder-Mead}

\begin{minipage}[t]{0.48 \textwidth}
\begin{tabular}{ccc}
\toprule
\parbox{2cm}{\centering \textbf{N\'umero de Dosis}} & \textbf{CDI} & \textbf{CDIS} \\
\midrule
1  &  2.7643  &  3.5244 \\
2 & 3.5126 & 3.4377 \\ 
3 & 2.4875 & 2.4155 \\ 
4 & 2.2411 & 2.1169 \\ 
5 & 1.2911 & 1.2266 \\ 
6 & 2.7938 & 2.4679 \\ 
7 & 2.2441 & 1.8992 \\ 
\vdots & \vdots & \vdots \\
\multicolumn{3}{c}{} \\
\bottomrule
\end{tabular}

\end{minipage} \hfill
\begin{minipage}{0.5 \textwidth}
\begin{tabular}{ccc}
\toprule
\parbox{2cm}{\centering \textbf{N\'umero de Dosis}} & \textbf{CDI} & \textbf{CDIS}  \\
\midrule
8 & 2.6012 & 2.4199 \\ 
9 & 2.2527 & 2.1303 \\ 
10 & 3.8243 & 3.6504 \\ 
11 & 4.1167 & 4.5802 \\ 
12 & 1.7112 & 1.4685 \\ 
13 & 2.2698 & 2.2918 \\
14 &  3.2095  &  3.7173 \\ 
15 & 2.3650 & 2.3111 \\ 
\hline
$\boldsymbol{X_{\textbf{ProptA}}}$ & $1.61 \times 10^7$ & $1 \times 10^7$\\
\bottomrule
\end{tabular}

\end{minipage}
\end{table}


\begin{table} [H]
\caption{ $X_0= 5 \ \text{cm}^3$, $X_{\text{ProptCDI}}=3.91 \times 10^6$, $X_{\text{ProptCDIS}}=2.16 \times 10^6$, Fraccionamiento Est\'andar, Nelder-Mead}
\begin{minipage}[t]{0.48 \textwidth}
\begin{tabular}{ccc}
\toprule
\parbox{2cm}{\centering \textbf{N\'umero de Dosis}} & \textbf{CDI} & \textbf{CDIS} \\
\midrule
1  &  1.5684  & 1.4911 \\
2 & 2.0498 & 1.7928 \\ 
3 & 2.1622 & 1.9924 \\ 
4 & 2.3221 & 2.5110 \\ 
5 & 2.0184 & 1.9999 \\ 
6 & 1.9067 & 2.1190 \\ 
7 & 2.6149 & 2.6332 \\ 
8 & 1.2663 & 1.5224 \\ 
9 & 2.1594 & 2.2878 \\ 
10 & 3.1086 &2.7938 \\ 
11 & 2.1685 & 2.0109 \\ 
12 & 2.5942 & 2.2944 \\ 
\vdots & \vdots & \vdots \\
\multicolumn{3}{c}{} \\
\bottomrule
\end{tabular}
\end{minipage} \hfill
\begin{minipage}[c]{0.5\textwidth}
\begin{tabular}{ccc}
\toprule
\parbox{2cm}{\centering \textbf{N\'umero de Dosis}} & \textbf{CDI} & \textbf{CDIS}  \\
\midrule
13 & 1.2632 & 1.3684 \\
14 &  2.1397  &  1.9786 \\ 
15 & 0.8296 & 1.1074 \\ 
16 & 2.1859 & 2.3118 \\ 
17 & 1.1453 & 1.3568 \\ 
18 & 2.1106 & 2.2048 \\ 
19 & 2.4451 & 2.0834 \\ 
20 & 2.0539 & 1.6069 \\ 
21 & 2.0118 & 2.2336 \\ 
22 & 1.8624 & 1.7028 \\ 
23 & 1.5853 & 1.9221 \\ 
24 & 2.0675 & 2.3167 \\ 
25 & 2.1706 & 2.1732 \\ 
\hline
$\boldsymbol{X_{\textbf{ProptA}}}$ & $3.9156 \times 10^6$ & $2.16 \times 10^6$\\
\bottomrule
\end{tabular}

\end{minipage}
\end{table}

Solo se muestran los casos de 1 $\text{cm}^3$ y 5 $\text{cm}^3$ por simplicidad. Es notable el hecho de que las dosis se aproximan al valor de 2 Gy para fraccionamiento est\'andar y de 2.66 Gy para hipofraccionamiento como se muestra en las siguientes gr\'aficas que corresponden a las tablas 4.16 a 4.19.

\newpage

\begin{figure}[h!]
    \centering
    \includegraphics[scale=0.4]{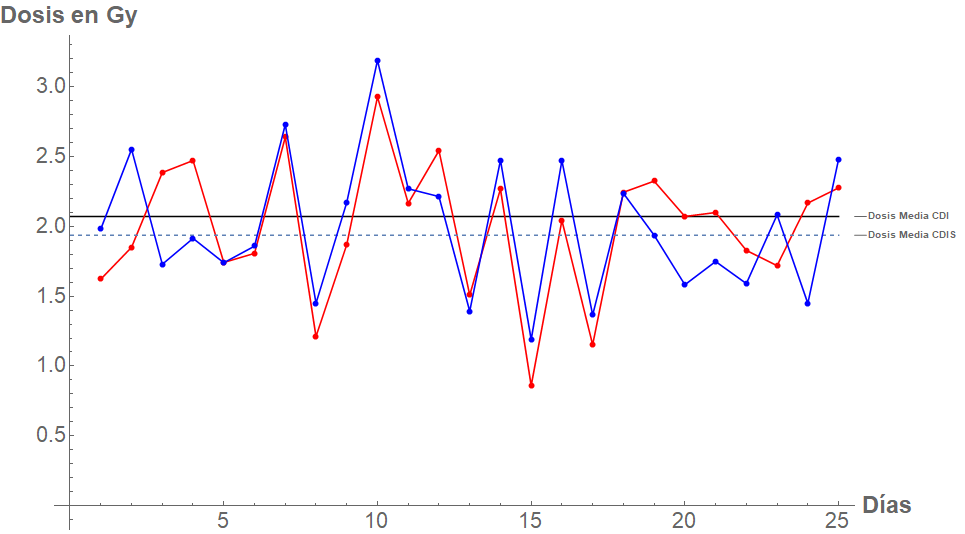}
    \caption{Comportamiento de las dosis a lo largo de los d\'ias del tratamiento con Nelder-Mead para el tumor de 1 $\text{cm}^3$, fraccionamiento est\'andar y vector de inicializaci\'on, en rojo CDI y en azul CDIS; dosis media: l\'inea solida CDI, l\'inea punteada CDIS }
    
\end{figure}

\begin{figure}[h!]
    \centering
    \includegraphics[scale=0.4]{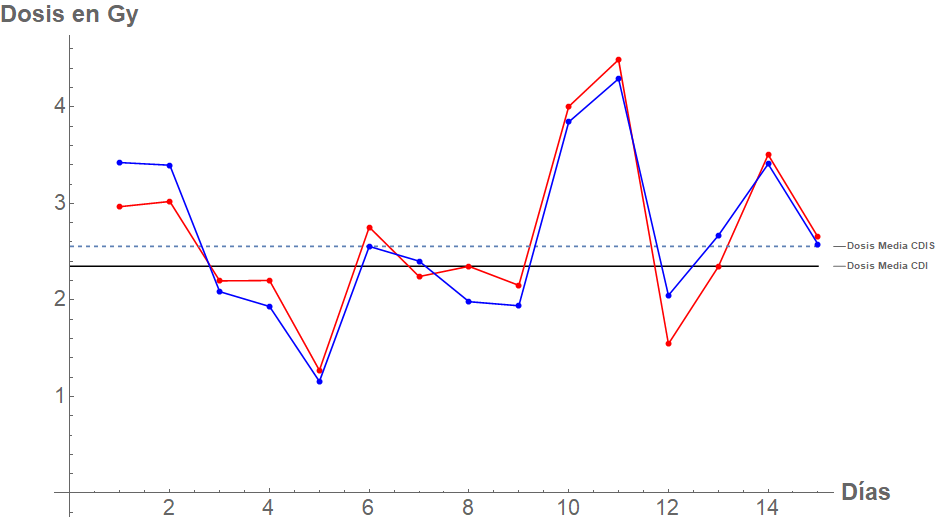}
    \caption{Comportamiento de las dosis a lo largo de los d\'ias del tratamiento con Nelder-Mead para el tumor de 1 $\text{cm}^3$, hipofraccionamiento y vector de inicializaci\'on, en rojo CDI y en azul CDIS; dosis media: l\'inea solida CDI, l\'inea punteada CDIS }
    
\end{figure}

\begin{figure}[h!]
    \centering
    \includegraphics[scale=0.4]{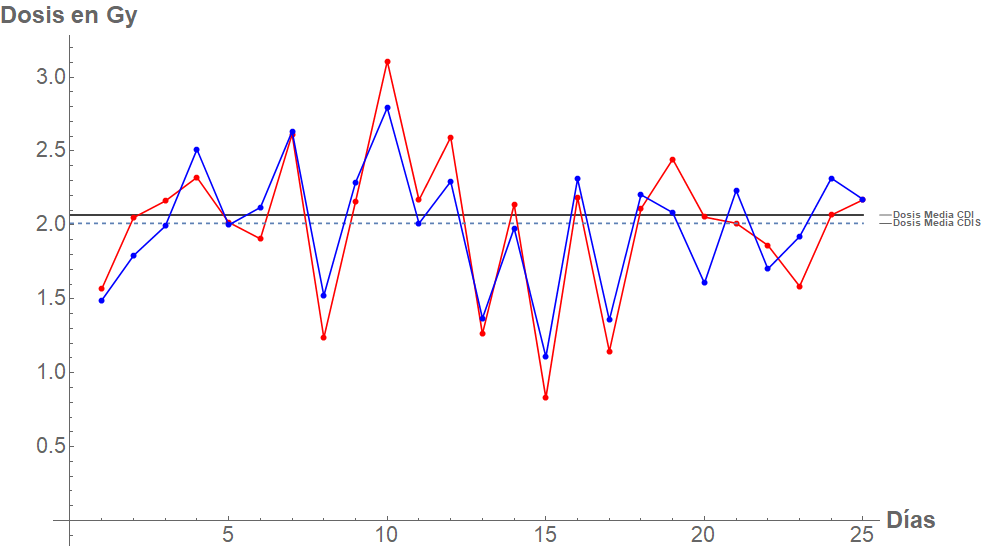}
    \caption{Comportamiento de las dosis a lo largo de los d\'ias del tratamiento con Nelder-Mead para el tumor de 5 $\text{cm}^3$, fraccionamiento est\'andar y vector de inicializaci\'on, en rojo CDI y en azul CDIS; dosis media: l\'inea solida CDI, l\'inea punteada CDIS }
\end{figure}

\begin{figure}[h!]
    \centering
    \includegraphics[scale=0.38]{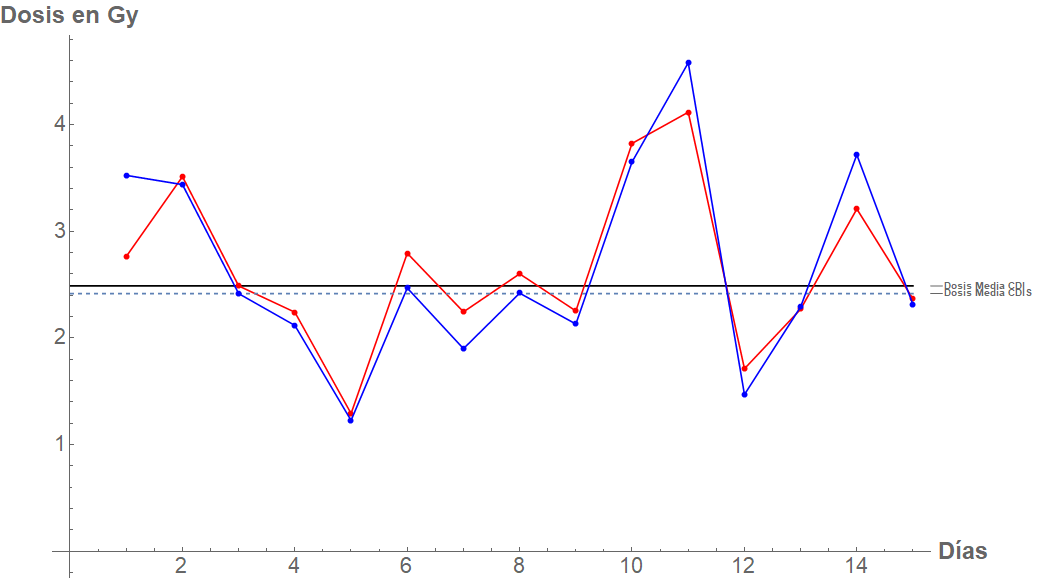}
    \caption{Comportamiento de las dosis a lo largo de los d\'ias del tratamiento con Nelder-Mead para el tumor de 5 $\text{cm}^3$, hipofraccionamiento y vector de inicializaci\'on, en rojo CDI y en azul CDIS; dosis media: l\'inea solida CDI, l\'inea punteada CDIS }
    \end{figure}
\newpage
Para continuar con el an\'alisis (y por comparaci\'on) utilizaremos ahora un m\'etodo de optimizaci\'on basado en gradientes llamado M\'etodo de Punto Interior. Este algoritmo modifica las condiciones de KKT y utiliza un par\'ametro de perturbaci\'on a modo de penalizar las soluciones que no est\'en dentro de una regi\'on factible mediante la utilizaci\'on de una especie de funci\'on de barrera. No es nuestra intenci\'on desarrollar la metodolog\'ia de todos los posibles m\'etodos de optimizaci\'on ya que la forma de implementarlos es muy parecida al pertenecer solo a dos tipos: basados en gradientes y de busqueda directa. Para un an\'alisis detallado sobre el m\'etodo de punto interior se puede consultar \cite{48}. Ya que nos interesa la comparaci\'on dando la inicializaci\'on o vector de inicializaci\'on solo mostraremos resultados para este caso.

\subsubsection{* M\'etodo de Punto Interior con inicializaci\'on para Tumor de 1 $\text{cm}^3$}

\begin{table} [H]
\caption{ $X_0= 1 \ \text{cm}^3$, $X_{\text{ProptCDI}}=781853$, $X_{\text{ProptCDIS}}=432199$, Fraccionamiento Est\'andar, Punto Interior}
\begin{minipage}[t]{0.48 \textwidth}
\begin{tabular}{ccc}
\toprule
\parbox{2cm}{\centering \textbf{N\'umero de Dosis}} & \textbf{CDI} & \textbf{CDIS} \\
\midrule
1  &  1.9913  & 1.9909 \\
2 & 1.9913 & 1.9909 \\ 
3 & 1.9913 & 1.9909 \\ 
4 & 1.9913 & 1.9909 \\ 
5 & 1.9913 & 1.9909 \\ 
6 & 2.0993 & 2.0126 \\ 
7 & 1.9913 & 1.9909 \\ 
8 & 1.9913 & 1.9909 \\ 
9 & 1.9913 & 1.9909 \\ 
10 & 1.9913 &1.9909 \\ 
11 & 2.0093 & 2.0126 \\ 
12 & 1.9913 & 1.9909 \\ 
\vdots & \vdots & \vdots \\
\multicolumn{3}{c}{} \\
\bottomrule
\end{tabular}
\end{minipage} \hfill
\begin{minipage}[c]{0.5\textwidth}
\begin{tabular}{ccc}
\toprule
\parbox{2cm}{\centering \textbf{N\'umero de Dosis}} & \textbf{CDI} & \textbf{CDIS}  \\
\midrule
13 & 1.9913 & 1.9909 \\
14 &  1.9913  &  1.9909 \\ 
15 & 1.9913 & 1.9909 \\ 
16 & 2.0093 & 2.0126 \\ 
17 & 1.9913 & 1.9909 \\ 
18 & 1.9913 & 1.9909 \\ 
19 & 1.9913 & 1.9909 \\ 
20 & 1.9913 & 1.9909 \\ 
21 & 2.0093 & 2.0126 \\ 
22 & 1.9913 & 1.9909 \\ 
23 & 1.9913 & 1.9909 \\ 
24 & 1.9913 & 1.9909 \\ 
25 & 1.9913 & 1.9909 \\ 
\hline
$\boldsymbol{X_{\textbf{ProptA}}}$ & $781853$ & $432199$\\
\bottomrule
\end{tabular}

\end{minipage}
\end{table}

\newpage


\begin{table} [H]
\caption{$X_0= 1 \ \text{cm}^3$, $X_{\text{ProptCDI}}=3.21 \times 10^6$, $X_{\text{ProptCDIS}}=1.9903 \times 10^6$, Hipofraccionamiento, Punto Interior}

\begin{minipage}[t]{0.48 \textwidth}
\begin{tabular}{ccc}
\toprule
\parbox{2cm}{\centering \textbf{N\'umero de Dosis}} & \textbf{CDI} & \textbf{CDIS} \\
\midrule
1  &  2.6510  &  2.6523 \\
2 & 2.6510 & 2.6523 \\ 
3 & 2.6510 & 2.6523 \\ 
4 & 2.6510 & 2.6523 \\ 
5 & 2.6510 & 2.6523 \\ 
6 & 2.6622 & 2.6627 \\ 
7 & 2.6510 & 2.6523 \\ 
\vdots & \vdots & \vdots \\
\multicolumn{3}{c}{} \\
\bottomrule
\end{tabular}

\end{minipage} \hfill
\begin{minipage}{0.5 \textwidth}
\begin{tabular}{ccc}
\toprule
\parbox{2cm}{\centering \textbf{N\'umero de Dosis}} & \textbf{CDI} & \textbf{CDIS}  \\
\midrule
8 & 2.6510 & 2.6523 \\ 
9 & 2.6510 & 2.6523 \\ 
10 & 2.6510 & 2.6523 \\ 
11 & 2.6622 & 2.6627 \\ 
12 & 2.6510 & 2.6523 \\ 
13 & 2.6510 & 2.6523 \\
14 &  2.6510  &  2.6523 \\ 
15 & 2.6510 & 2.6523 \\ 
\hline
$\boldsymbol{X_{\textbf{ProptA}}}$ & $3.21 \times 10^6$ & $1.9903 \times 10^6$\\
\bottomrule
\end{tabular}

\end{minipage}
\end{table}
\vspace{10pt}

\subsubsection{* M\'etodo de Punto Interior con inicializaci\'on para Tumor de 5 $\text{cm}^3$}

\begin{table} [H]
\caption{$X_0= 5 \ \text{cm}^3$, $X_{\text{ProptCDI}}=1.61 \times 10^7$, $X_{\text{ProptCDIS}}=1 \times 10^7$, Hipofraccionamiento, Punto Interior}

\begin{minipage}[t]{0.48 \textwidth}
\begin{tabular}{ccc}
\toprule
\parbox{2cm}{\centering \textbf{N\'umero de Dosis}} & \textbf{CDI} & \textbf{CDIS} \\
\midrule
1  &  2.6488  &  2.6496 \\
2 & 2.6488 & 2.6496 \\ 
3 & 2.6488 & 2.6496 \\ 
4 & 2.6488 & 2.6496 \\ 
5 & 2.6488 & 2.6496 \\ 
6 & 2.6620 & 2.6613 \\ 
7 & 2.6488 & 2.6496 \\ 
\vdots & \vdots & \vdots \\
\multicolumn{3}{c}{} \\
\bottomrule
\end{tabular}

\end{minipage} \hfill
\begin{minipage}{0.5 \textwidth}
\begin{tabular}{ccc}
\toprule
\parbox{2cm}{\centering \textbf{N\'umero de Dosis}} & \textbf{CDI} & \textbf{CDIS}  \\
\midrule
8 & 2.6488 & 2.6496 \\ 
9 & 2.6488 & 2.6496 \\ 
10 & 2.6488 & 2.6496 \\ 
11 & 2.6620 & 2.6613 \\ 
12 & 2.6488 & 2.6496 \\ 
13 & 2.6488 & 2.6496 \\
14 &  2.6488  &  2.6496 \\ 
15 & 2.6488 & 2.6496 \\ 
\hline
$\boldsymbol{X_{\textbf{ProptA}}}$ & $1.61 \times 10^7$ & $1 \times 10^7$\\
\bottomrule
\end{tabular}

\end{minipage}
\end{table}


\begin{table} [H]
\caption{ $X_0= 5 \ \text{cm}^3$, $X_{\text{ProptCDI}}=3.91 \times 10^6$, $X_{\text{ProptCDIS}}=2.16 \times 10^6$, Fraccionamiento Est\'andar, Punto Interior}
\begin{minipage}[t]{0.48 \textwidth}
\begin{tabular}{ccc}
\toprule
\parbox{2cm}{\centering \textbf{N\'umero de Dosis}} & \textbf{CDI} & \textbf{CDIS} \\
\midrule
1  &  1.9907  & 1.9907 \\
2 & 1.9907 & 1.9907 \\ 
3 & 1.9907 & 1.9907 \\ 
4 & 1.9907 & 1.9907 \\ 
5 & 1.9907 & 1.9907 \\ 
6 & 2.0095 & 2.0120 \\ 
7 & 1.9907 & 1.9907 \\ 
8 & 1.9907 & 1.9907 \\ 
9 & 1.9907 & 1.9907 \\ 
10 & 1.9907 &1.9907 \\ 
11 & 2.0095 & 2.0120 \\ 
12 & 1.9907 & 1.9907 \\ 
\vdots & \vdots & \vdots \\
\multicolumn{3}{c}{} \\
\bottomrule
\end{tabular}
\end{minipage} \hfill
\begin{minipage}[c]{0.5\textwidth}
\begin{tabular}{ccc}
\toprule
\parbox{2cm}{\centering \textbf{N\'umero de Dosis}} & \textbf{CDI} & \textbf{CDIS}  \\
\midrule
13 & 1.9907 & 1.9907 \\
14 &  1.9907  &  1.9907 \\ 
15 & 1.9907 & 1.9907 \\ 
16 & 2.0095 & 2.0120 \\ 
17 & 1.9907 & 1.9907 \\ 
18 & 1.9907 & 1.9907 \\ 
19 & 1.9907 & 1.9907 \\ 
20 & 1.9907 & 1.9907 \\ 
21 & 2.0095 & 2.0120 \\ 
22 & 1.9907 & 1.9907 \\ 
23 & 1.9907 & 1.9907 \\ 
24 & 1.9907 & 1.9907 \\ 
25 & 1.9907 & 1.9907 \\ 
\hline
$\boldsymbol{X_{\textbf{ProptA}}}$ & $3.91 \times 10^12$ & $2.16 \times 10^6$\\
\bottomrule
\end{tabular}

\end{minipage}
\end{table}

Una r\'apida observaci\'on a las dosis arrojadas por el m\'etodo del punto interior con el vector de inicializaci\'on (dosis est\'andar) basta para darnos cuenta que, al igual que el m\'etodo del Nelder-Mead, las dosis optimizadas est\'an muy cercanas a las dosis usadas actualmente, al menos para c\'ancer de mama. Podemos decir que las dosis actuales se encontraron sin tomar en cuenta efectos sobre c\'elulas sanas y que estas dosis se encuentran muy pr\'oximas a una distribuci\'on de dosis \'optimas para los casos de carcinoma ductal invasivo (CDI) y carcinoma ductal in situ (CDIS). Los resultados obtenidos en este cap\'itulo dan pie a continuar la investigaci\'on experimental aplicando el modelo de tratamiento planteado en el presente trabajo. An\'alisis y comparaci\'on de diferentes tejidos (comerciales y obtenidos de pacientes, tanto in vivo como in vitro) se precisan para completar el area de investigaci\'on abierta en la presente tesis.

\subsection{Tratamiento con Optimizaci\'on de las Dosis}

Ya que hemos visto el comportamiento de los tratamientos con dosis cercanas a las que son utilizadas por los m\'edicos, sustituiremos la restricci\'on \eqref{restmed} por

\begin{equation}\label{restopt}
f(D_1, D_2, ..., D_n)\leq X_{\text{Propt}}.
\end{equation}

De este modo obtendremos resultados de dosis que logran reducir el n\'umero de c\'elulas cancer\'igenas a un valor m\'as bajo que al que llegan los m\'edicos (anteriormente pediamos que el n\'umero de c\'elulas alcanzado fuera mayor o igual a los valores alcanzados con el tratamiento de los m\'edicos). Por simplicidad y dado que ya observamos el comportamiento general de la optimizaci\'on con varios tamaños de tumor, solo lo relizaremos para el tumor de 1 $\text{cm}^3$.

\subsubsection{* M\'etodo de Nelder Mead para Tumor de 1 $\text{cm}^3$}

\begin{table} [H]
\caption{$X_0= 1 \ \text{cm}^3$, $X_{\text{ProptCDI}}=3.21 \times 10^6$, $X_{\text{ProptCDIS}}=1.99 \times 10^6$, Hipofraccionamiento, Nelder Mead}

\begin{minipage}[t]{0.48 \textwidth}
\begin{tabular}{ccc}
\toprule
\parbox{2cm}{\centering \textbf{N\'umero de Dosis}} & \textbf{CDI} & \textbf{CDIS} \\
\midrule
1  &  0  &  0 \\
2 & 3.9999 & 3.9999 \\ 
3 & 3.9999 & 3.9999 \\ 
4 & 3.9999 & 3.9999 \\ 
5 & 0 & 0 \\ 
6 & 0 & 3.9999 \\ 
7 & 3.9999 & 3.9999 \\ 
\vdots & \vdots & \vdots \\
\multicolumn{3}{c}{} \\
\bottomrule
\end{tabular}

\end{minipage} \hfill
\begin{minipage}{0.5 \textwidth}
\begin{tabular}{ccc}
\toprule
\parbox{2cm}{\centering \textbf{N\'umero de Dosis}} & \textbf{CDI} & \textbf{CDIS}  \\
\midrule
8 & 3.9999 & 0 \\ 
9 & 3.9999 & 3.9999 \\ 
10 & 3.9999 & 3.9999 \\ 
11 & 3.9999 & 0 \\ 
12 & 3.9999 & 3.9999 \\ 
13 & 3.9999 & 3.9999 \\
14 &  0  &  0 \\ 
15 & 0 & 3.9999 \\ 
\hline
$\boldsymbol{X_{\textbf{ProptA}}}$ & $2.9079 \times 10^6$ & $1.7936 \times 10^6$\\
\bottomrule
\end{tabular}

\end{minipage}
\end{table}

\begin{table} [H]
\caption{ $X_0= 1 \ \text{cm}^3$, $X_{\text{ProptCDI}}=781853$, $X_{\text{ProptCDIS}}=432199$, Fraccionamiento Est\'andar, Nelder Mead}
\begin{minipage}[t]{0.48 \textwidth}
\begin{tabular}{ccc}
\toprule
\parbox{2cm}{\centering \textbf{N\'umero de Dosis}} & \textbf{CDI} & \textbf{CDIS} \\
\midrule
1  &  3.9999  & 3.9999 \\
2 & 0 & 0 \\ 
3 & 0 & 3.9999 \\ 
4 & 3.9999 & 0 \\ 
5 & 0 & 3.9999 \\ 
6 & 3.9999 & 0 \\ 
7 & 1.9999 & 0 \\ 
8 & 3.9999 & 1.9999 \\ 
9 & 0 & 0 \\ 
10 & 3.9999 & 0 \\ 
11 & 0 & 3.9999 \\ 
12 & 0 & 3.9999 \\ 
\vdots & \vdots & \vdots \\
\multicolumn{3}{c}{} \\
\bottomrule
\end{tabular}
\end{minipage} \hfill
\begin{minipage}[c]{0.5\textwidth}
\begin{tabular}{ccc}
\toprule
\parbox{2cm}{\centering \textbf{N\'umero de Dosis}} & \textbf{CDI} & \textbf{CDIS}  \\
\midrule
13 & 3.9999 & 3.9999 \\
14 &  0  &  3.9999 \\ 
15 & 0 & 0 \\ 
16 & 3.9999 & 0 \\ 
17 & 3.9999 & 3.9999 \\ 
18 & 3.9999 & 0 \\ 
19 & 0 & 3.9999 \\ 
20 & 0 & 3.9999 \\ 
21 & 3.9999 & 3.9999 \\ 
22 & 3.9999 & 3.9999 \\ 
23 & 0 & 0 \\ 
24 & 0 & 0 \\ 
25 & 3.9999 & 0 \\ 
\hline
$\boldsymbol{X_{\textbf{ProptA}}}$ & $678009$ & $370953 $\\
\bottomrule
\end{tabular}

\end{minipage}
\end{table}

En este caso las dosis marcadas como 0 corresponden a valores muy pequeños del orden de $\approx 10^{-10}$. Estos valores pueden verse como d\'ias de descanso entre los dias de aplicaci\'on de dosis de radiaci\'on. Adicionalmente, añadimos una restricci\'on de modo que las dosis se encuentren en un intervalo de 0 a 4 Gy. Por lo que obtuvimos una distribuci\'on de dosis que es completamente diferente a la utilizada por los m\'edicos y que logra reducir el n\'umero de c\'elulas cancer\'igenas a valores m\'as pequeños que los obtenidos con las dosis utilizadas actualmente. Estas dosis son \'optimas en el sentido de que presentan mejores resultados respecto del n\'umero final de c\'elulas adem\'as de que nos muestran que pueden existir m\'as d\'ias de descanso que los fines de semana. Ya que Nelder Mead, al parecer, nos report\'o resultados m\'as acordes con lo visto actualmente, solo mostraremos resultados de este m\'etodo de optimizaci\'on.
A continuaci\'on mostramos las gr\'aficas correspondientes.

\begin{figure}[h!]
    \centering
    \includegraphics[scale=0.45]{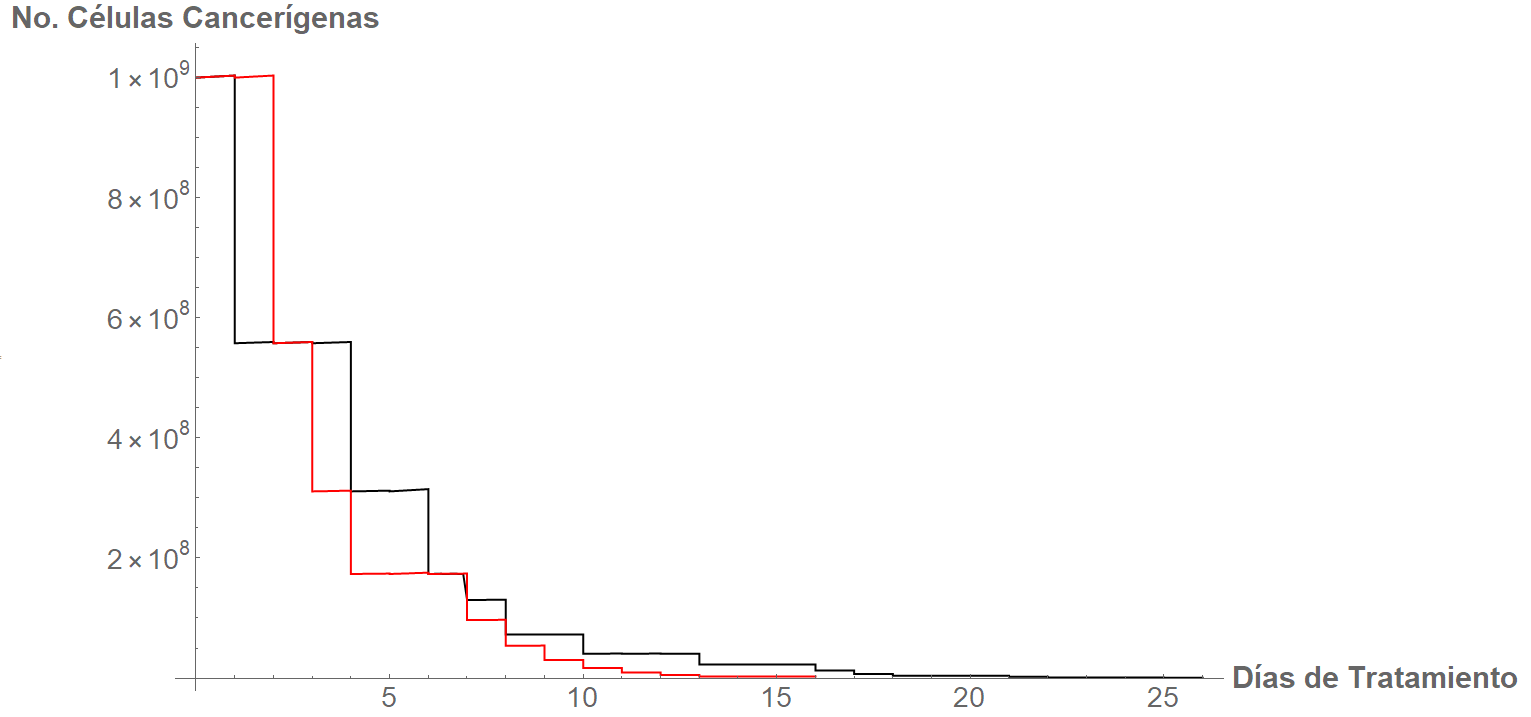}
    \caption{N\'umero de c\'elulas cancer\'igenas a lo largo de los d\'ias del tratamiento con Nelder-Mead para el tumor de 1 $\text{cm}^3$ y CDI, fraccionamiento est\'andar en negro y en rojo hipofraccionamiento }
\end{figure}

\begin{figure}[h!]
    \centering
    \includegraphics[scale=0.45]{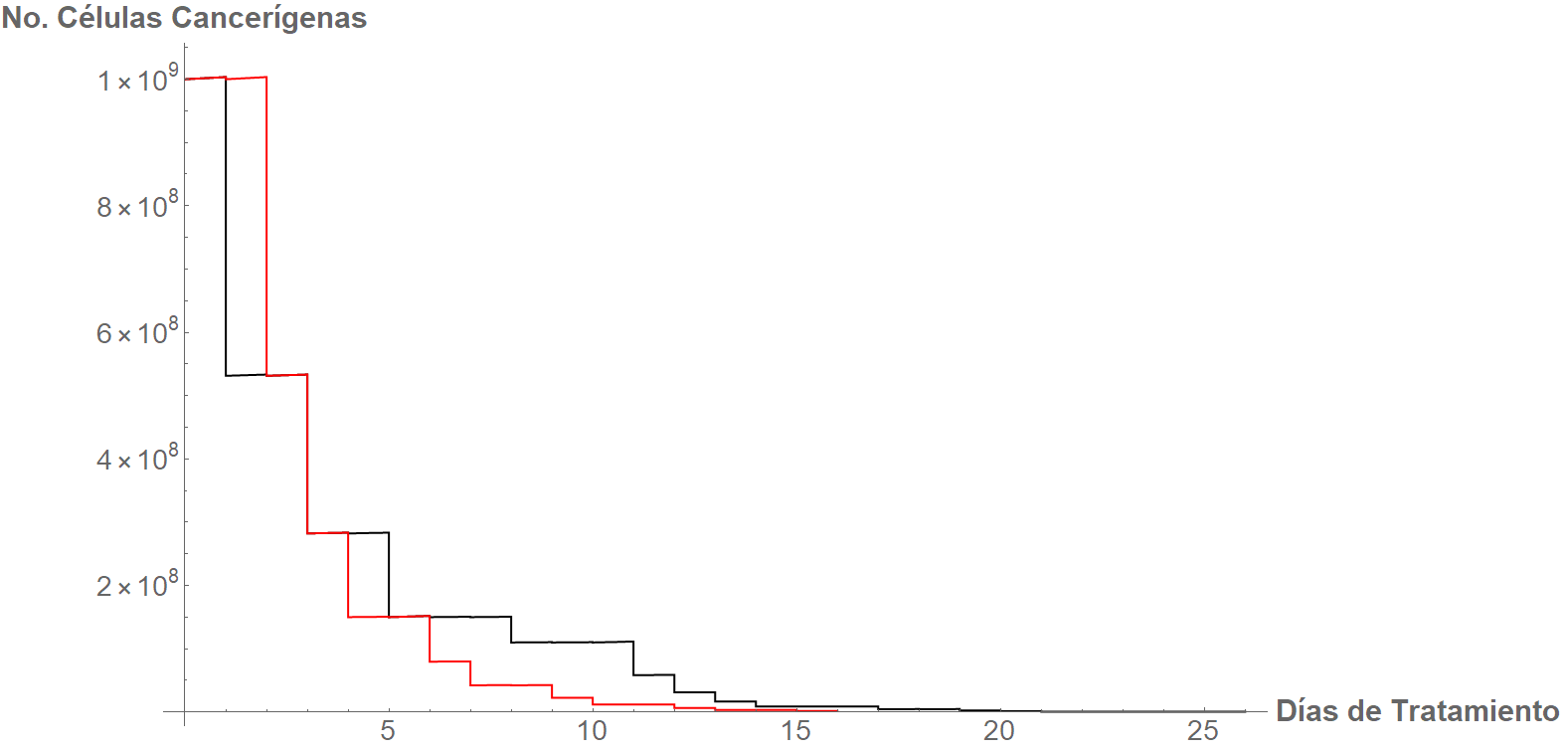}
    \caption{N\'umero de c\'elulas cancer\'igenas a lo largo de los d\'ias del tratamiento con Nelder-Mead para el tumor de 1 $\text{cm}^3$ y CDIS, fraccionamiento est\'andar en negro y en rojo hipofraccionamiento }
\end{figure}
\chapter{Conclusi\'on}

El objetivo de este trabajo consisti\'o en la propuesta de una metodolog\'ia de forma que se pudieran alcanzar dos objetivos: un an\'alisis del modelo de tratamiento actualmente usado para radioterapia tomando en cuenta las dosis usadas en un tratamiento est\'andar y la propuesta de una nueva distribuci\'on de dosis de radiaci\'on de tal forma que se lograra minimizar el n\'umero de c\'elulas cancer\'igenas. Para la propuesta de esta metodolog\'ia se opt\'o por usar un modelo de crecimiento de Gompertz simple combinado con una nueva forma de la fracci\'on de supervivencia celular (para describir el efecto de la radiaci\'on sobre el tejido) propuesta en \cite{20}. La soluci\'on (2.0.16) nos muestra la evoluci\'on del n\'umero de c\'elulas cancer\'igenas a lo largo de todo un tratamiento de radioterapia y puede ayudarnos a visualizar c\'omo influye la terapia sobre un paciente dado un determinado tipo de tumor. Debido a la forma en que se construy\'o este resultado, en teor\'ia  podemos caracterizar cualquier tipo de tumor dando informaci\'on sobre su crecimiento y su respuesta a la radiaci\'on (mediante la constante $\gamma$ de la fracci\'on de supervivencia celular (\ref{2.5.2.8})) .

Si lo que queremos es obtener el menor n\'umero de c\'elulas cancer\'igenas que sean posibles, procedemos a aplicar un algoritmo de optimizaci\'on a fin de encontrar las dosis \'optimas tales que minimizan el tamaño del tumor. Para tal efecto se decidi\'o utilizar m\'etodos de optimizaci\'on de b\'usqueda directa, es decir, suponiendo que no sabemos nada sobre el vector de dosis \'optimas. Con estos m\'etodos se logr\'o reducir el tiempo de computaci\'on para hallar las dosis. Como una primera parte, utilizamos la restricci\'on \eqref{restmed} a fin de investigar lo que sucede con los tratamientos usados actualmente por los m\'edicos ya que dicha desigualdad supone alcanzar como m\'inimo el n\'umero de c\'elulas cancerigenas de los tratamientos actuales. Con esto en mente, obtuvimos soluciones a las dosis que fluctuaban bastante (para el caso de Evoluci\'on Diferencial) y que se pueden apreciar en el cap\'itulo anterior. Si bien es cierto que las dosis son \'optimas en el sentido de que minimizan el tamaño del tumor, tendr\'ia que tenerse en cuenta los efectos fisiol\'ogicos producidos en el paciente. Para el otro m\'etodo de busqueda directa, Nelder-Mead, obtuvimos resultados un poco m\'as razonables ya que estas dosis se acercan m\'as a lo que comunmente se utiliza en un tratamiento actual. Los tratamientos actuales utilizan una distribuci\'on de dosis la cual ha sido fijada debido a todo un siglo de prueba y error tratando diferentes tumores, tan es as\'i que la investigaci\'on de nuevos tipos de fraccionamiento siguen en pie. A pesar de ello y del hecho de que no existe una herramienta anal\'itica para determinar la optimalidad de las dosis, el resultado emp\'irico obtenido por los m\'edicos (y que se basa en considerar los efectos que produce la radiaci\'on en el organismo \cite{49}) ha tenido buenos resultados para los tumores que hoy en d\'ia se tratan con radioterapia.

Los resultados (de las dosis) obtenidos en el presente trabajo dan un indicio de que las dosis utilizadas en el tratamiento actual para c\'ancer de mama se encuentran en alguna vecindad de la soluci\'on con dosis optimizadas, de hecho podr\'iamos decir que las dosis que encontramos tambi\'en se encuentran en alguna vecidad de las dosis \'optimas totales. Aunque decir lo anterior es muy arriesgado ya que las caracter\'isticas del tumor como del paciente influyen en c\'omo la enfermedad reacciona al tratamiento. Como apoyo a este argumento, en las tablas 4.4 a 4.15 (para Nelder-Mead), as\'i como las figuras 4.10 a 4.13, podemos observar  que las dosis se encuentran muy cercanas a los valores de 2 Gy y 2.66 Gy (fraccionamiento est\'andar e hipofraccionamiento respectivamente). Algo interesante a recalcar es que estos resultados se obtuvieron sin tomar en cuenta un conjunto de puntos de inicializaci\'on o vector de inicializaci\'on (puntos a partir de los cuales nos gustar\'ia encontrar el resultado de la optimizaci\'on). Si en cambio, damos un conjunto de puntos iniciales para el m\'etodo de Nelder-Mead (dichos puntos iniciales corresponder\'ian a los valores 2 y 2.66) volvemos a obtener valores cercanos a las dosis actuales (tablas 4.16 a 4.19 y figuras 4.26 a 4.29). Si adem\'as de esto probamos realizar la optimizaci\'on con otro m\'etodo (m\'etodo de punto interior el cual es basado en gradientes) y con los mismos puntos de inicializaci\'on, obtenemos unos resultados bastante interesantes (tablas 4.20 a 4.23) ya que el resultado de las dosis ahora arroja que son casi constantes y difieren en menor medida de los valores 2 y 2.66. Podemos concluir entonces que efectivamente las dosis actuales para c\'ancer de mama no son del todo erroneas, tampoco son \'optimas pero se encuentran en una vecindad del conjunto de soluciones \'optimas.

Como ya comentamos, las dosis utilizadas hoy en d\'ia para tratar diversos tumores han sido estandarizadas debido a su amplia trayectoria combinada con resultados cl\'inicos. Dichas dosis fueron fijadas principalmente debido a dos causas: su potencial efecto para tratar de no causar daño al tejido sano y el efecto percibido en tejidos con respuesta tard\'ia a radiaci\'on \cite{51}. El dividir la dosis total en pequeñas fracciones a lo largo de varios d\'ias tambien se apoya del hecho de que logra explotar algun(as) de las 4 Rs en radioterapia: reparaci\'on, redistribuci\'on, repoblaci\'on y reoxigenaci\'on; de modo que se logra obtener alg\'un beneficio terap\'eutico. A pesar de esto, la administraci\'on de dosis est\'andares de radiaci\'on sigue siendo clasificada como categor\'ia 2A de recomendaci\'on segun la National Comprehensive Cancer Network. Esta categor\'ia se refiere a que existe un bajo nivel de evidencia y se ha acordado que puede existir alguna intervenci\'on/modificaci\'on en las dosificaciones de radiaci\'on. Es por ello que los ensayos cl\'inicos siguen en pie y fue debido a lo anterior que surgieron dosificaciones como el hipo e hiper fraccionamiento, en resumidas cuentas no existe un protocolo fijo a seguir.

Debido a lo anterior, la restricci\'on \eqref{restopt} fue impuesta con el fin de llegar a dosis que nos redujeran m\'as el n\'umero de c\'elulas en comparaci\'on con la dosificaci\'on actual. Lo que obtenemos es una dosificaci\'on totalmente diferente y que nos arroja mayores d\'ias de descanso que el fraccionamiento est\'andar e hipofraccionamiento. Todo esto da pie a una l\'inea de investigaci\'on sobre los resultados del modelo desarrollado en este trabajo y la restriccion \eqref{restopt}.

La relevancia de este trabajo radica en que hemos propuesto una herramienta que puede servir parar sustentar el porque de la utilizaci\'on de ciertas dosis. Hay que recordar que se utiliz\'o el ejemplo de c\'ancer de mama para este trabajo pero el resultado de las ecuaciones es extensible para aplicarse en cualquier tipo de tumor (previamente caracterizado mediante las constantes correspondientes). Es por ello que el futuro de esta investigaci\'on requiere necesariamente la validaci\'on correspondiente y por etapas: tejido in vitro, in vivo, animales, ensayos cl\'inicos. Claramente esto supone un extenso trabajo debido a la cantidad de datos que se tiene que usar y m\'as a\'un la cantidad de datos que se obtendr\'an. Debido a esto, reiteramos la importancia de los resultados obtenidos en esta tesis. Paralelamente y de acuerdo a lo que se observa en las figuras 4.4 a 4.9 y 4.30-4.31, probamos anal\'iticamente el resultado emp\'irico (cl\'inico) de que el Hipofraccionamiento es una mejor alternativa al fraccionamiento est\'andar \cite{51} (para varios tipos de c\'ancer se est\'a adoptando el hipofraccionamiento pero nuestro resultado claramente se enfoca a c\'ancer de mama).

Finalmente, la l\'inea de investigaci\'on abierta en este trabajo da pie a varias formas de continuar con su desarrollo, una forma ser\'ia el complicar un poco el problema y tomar en cuenta lo que sucede en el caso de considerar c\'elulas sanas, es decir obtener unas dosis que tomen en cuenta el efecto para ambos tipos de tejido. Otra forma consistir\'ia en considerar c\'omo se distribuyen espacialmente las c\'elulas cancer\'igenas y a partir de ello obtener de igual forma las dosis pertinentes.


\appendix
\chapter{Obtenci\'on de la Fracci\'on de Supervivencia Celular con Entrop\'ia de Tsallis}

Para aplicar el principio de m\'axima entrop\'ia a Tsallis, debemos tener algunas consideraciones. Postulamos que exista una cantidad de radiaci\'on absorbida $D_0< \infty$ (o su canticad equivalente, el efecto de m\'inima aniquilaci\'on $E_0=\alpha_0 D_0$) despues de la cual ninguna c\'elula pueda sobrevivir. Recordemos la forma de la entrop\'ia de Tsallis

\begin{equation}
    S_q = \frac{1}{q-1} \left[ 1-\int_{0}^{E_0} p^q (E) dE \right].
\end{equation}

Imponemos dos condiciones adicionales referentes a la normalizaci\'on

\begin{equation}
    \int_{0}^{E_0} p(E)dE=1
\end{equation}

y el valor q-medio

\begin{equation}
    \int_{0}^{E_0}p^{q} (E)EdE= \left < E\right>_q < \infty.
\end{equation}

Con estas restricciones podemos hacer uso del m\'etodo de los multiplicadores de Lagrange. De esta forma construimos el funcional que queda

\begin{equation}
    \frac{1-\int_{0}^{E_0} p^q (E) dE}{1-q}+a_q \int_{0}^{E_0} p(E)dE + b_q \int_{0}^{E_0}p^{q} (E)EdE
\end{equation}

Maximizando bajo las condiciones requeridas, encontramos los valores de $E_0$, $a_q$ y $b_q$

\large
\begin{equation}
    \begin{array}{cc}
         E_0 =& \frac{2-q}{1-q} \left( \frac{\left< E \right>_q }{2-q} \right)^{\frac{1}{2-q}}  \\
         & \\
        a_q =& -\frac{q}{1-q}   \left( \frac{\left< E \right>_q }{2-q} \right)^{\frac{1-q}{2-q}} \\
        & \\
        b_q = & -\frac{1}{2-q}   \left( \frac{\left< E \right>_q }{2-q} \right)^{-\frac{1}{2-q}} 
    \end{array}
\end{equation}

\normalsize
De las ecuaciones anteriores podemos obtener la funci\'on de densidad de probabilidad $p(E)$

\begin{equation}
    p(E)=\left( \frac{2-q}{\left< E \right>_q } \right)^{\frac{1}{2-q}} \left( 1-\frac{1-q}{2-q} \left( \frac{2-q}{\left< E \right>_q} \right)^{1/2-q} E \right)^{\frac{1}{1-q}}
\end{equation}

Se conoce que la fracci\'on de supervivencia celular es \cite{20}

\begin{equation}
    F_s (E)= \int_{E}^{\infty} p(x)dx. 
\end{equation}

Sustituyendo lo encontrado anteriormente para la densidad de probabilidad e integrando en sobre el intervalo $[E,E_0]$, obtenemos

\begin{equation}
    F_s (E) = \int_{E}^{E_0} p(x)dx= \left( 1-\frac{E}{E_0} \right)^{\frac{2-q}{1-q}}
\end{equation}

para $E< E_0$. Si tomamos en cuenta la ecuaci\'on $D_0 =E_0 / \alpha$ y $E=\alpha D$ e identificando $\gamma=\frac{2-q}{1-q}$, finalmente podemos escribir (2.24) como

\begin{equation}
    F_s (D)= \left\{ \begin{array}{cc}
        (1-\frac{D}{D_0})^{\gamma} & \forall \ D < D_0,  \\
        0 & \forall \ D \geq D_0 
    \end{array} \right.
\end{equation}

\chapter{Obtenci\'on de la Fracci\'on de Supervivencia Celular con Entrop\'ia de Boltzman-Gibbs}

Podemos obtener la fracci\'on de supervivencia celular del modelo lineal de radiobiolog\'ia considerando la entrop\'ia de Boltzman-Gibbs, sea 

\begin{equation}
    S= \int_{\Omega}ln \left[ p(E) \right]p(E)dE
\end{equation}

donde que $E$ es una forma adimensional de la dosis de radiaci\'on, $p(E)$ la densidad de probabilidad de muerte celular y $\Omega$ todos los estados accesibles de $E$. Entonces para aplicar el principio de m\'axima entrop\'ia necesitamos satisfacer un par de condiciones relacionadas con la normalizaci\'on de $p(E)$

\begin{equation}
    \int_{\Omega} p(E)dE=1
\end{equation}

y la existencia del valor medio finito de $E$

\begin{equation}
    \int_{\Omega}p(E)dE= \left<E \right>.
\end{equation}

Lo que hacemos es encontrar la forma de la distribuci\'on $p(E)$ tal que la entrop\'ia se haga m\'axima sujeta a las condiciones anteriormente dadas. Esto es un problema de multiplicadores de Lagrange. A pesar de ello se conoce que la distribuci\'on exponencial es la hace m\'axima la entrop\'ia en un intervalo $[0,\infty]$ (de modo que $\Omega \in [0,\infty]$) dado que existe un valor medio finito \cite{50}. De ese modo se obtiene que 

\begin{equation}
    p(E)= \dfrac{1}{\left< E \right>} e^{-\frac{E}{\left< E \right>}},
\end{equation}

por lo que la probabilidad de supervivencia celular es

\begin{equation}
    F_s=e^{-\frac{E}{\left< E \right>}}
\end{equation}

Sabiendo que la relaci\'on entre $E$ y $D$ es $E=\alpha_0 D$, podemos proponer que 

\begin{equation}
    \alpha= \dfrac{\alpha_0}{\left< E\right>}= \dfrac{1}{\left< D \right>}
\end{equation}

por lo que finalmente la fracci\'on de supervivencia celular queda

\begin{equation}
    F_s = e^{-\alpha D}
\end{equation}

que corresponde al modelo lineal de la fracci\'on de supervivencia celular usado en radiobiolog\'ia.
\chapter{M\'etodos de Optimizaci\'on}

\section{Teorema - Condiciones de Kuhn-Tucker}

Sea $f_0(\boldsymbol{x})$ la funci\'on objetivo, diferenciable. Sea $f_i(\boldsymbol{x})$ ($i=1,...,k$) el conjunto de restricciones asociadas al problema. Si  $f_0(\boldsymbol{x})$ alcanza un m\'inimo local en algun punto $\boldsymbol{x_0}$ tal que dicho punto pertenece al conjunto de variables de las que depende el problema, entonces existen multiplicadores de Lagrange $\boldsymbol{\lambda}$ de modo que las siguientes condiciones se satisfacen

\begin{equation} \label{eqn:3.2.1.3}
    \begin{array}{ccrcl}
        \dfrac{\partial f_0(\boldsymbol{x}_0)}{\partial x_j} & + & \sum \limits_{i=1}^{k} \lambda_i \dfrac{\partial f_i(\boldsymbol{x}_0)}{\partial x_j} & = & 0 \ \ \ \ (j=1,...,l)   \\
        & & f_i (\boldsymbol{x_0})&\leq &0 \ \ \ \ (i=1,...,k) \\
        & & \lambda_i f_i (\boldsymbol{x_0})&= & 0 \ \ \ \ (i=1,...,k) \\
        & & \lambda_i & \geq & 0 \ \ \ \ (i=1,...,k) \\
    \end{array}
\end{equation}

Donde $\boldsymbol{x}$ denota el conjunto de variables de las que depende el problema, $\boldsymbol{x_0}$ el conjunto de puntos donde la funci\'on alcanza el m\'inimo local y $\boldsymbol{\lambda}$ el conjunto de los multiplicadores asociados al problema. 
Una vez obtenidas todas las condiciones de KT se procede a resolver el sistema de ecuaciones e inecuaciones a fin de obtener los valores $\boldsymbol{x_0}$ que hagan m\'inima la funci\'on. \\
Las condiciones de Kuhn-Tucker (KT) son condiciones necesarias para encontrar el m\'inimo local de un problema de optimizaci\'on con restricciones (del tipo antes mencionado), para una prueba del teorema anterior se puede consultar \cite{36}.
\\
Con las condiciones arriba mencionadas podemos obtener un sistema de ecuaciones e inecuaciones y obtener los valores \'optimos para las dosis de radiaci\'on. Sin embargo aun queda el restringir m\'as el problema referente a las dosis, adem\'as de que la suma de todas sea menor o igual a una dosis m\'axima, estas deben ser mayores o iguales a cero para que tengan sentido f\'isico. \\
Si al problema de optimizaci\'on le imponemos la restricci\'on de que las variables sean no negativas tendr\'iamos

\begin{equation}\label{3.2.1.5}
\begin{array}{ccccc}
    \text{min} &  & f_0 (\boldsymbol{x}) & &\\
    
    \text{s.a.}&  & f_i (\boldsymbol{x}) & \leq & 0 \  \ (i=1,2,...,k), \\
     & &  -x_j & \leq & 0 \  \ (j=1,2,...,l).
\end{array}
\end{equation}

Podemos formar la funci\'on Lagrangiana como

\begin{equation}\label{eqn:3.2.1.4}
    \psi (\boldsymbol{x},\boldsymbol{\lambda},\boldsymbol{\mu})= f_0 (\boldsymbol{x}) + \sum \limits_{i=1}^{k} \lambda_i f_i (\boldsymbol{x}) + \sum \limits_{j=1}^{l} \mu_j (-x_j).
\end{equation}

Aplicando la derivada de las condiciones ~\eqref{eqn:3.2.1.3} obtenemos

\begin{equation} \label{3.2.1.6}
    \dfrac{\partial \psi}{\partial x_j} = \dfrac{\partial f_0 (\boldsymbol{x_0})}{\partial x_j} + \sum \limits_{i=1}^{k} \lambda_i \dfrac{\partial f_i (\boldsymbol{x_0})}{\partial x_j} - \mu_j =0 \ \ (j=1,...,l)
\end{equation}

de la misma forma 

\begin{equation} \label{3.2.1.7}
    \begin{array}{rc}
         \dfrac{\partial \psi}{\partial \lambda_i}=f_i (\boldsymbol{x_0})& \leq 0 \ \ (i=1,2,...,k) \\
         \lambda_i \dfrac{\partial \psi}{\partial \lambda_i}=\lambda_i f_i (\boldsymbol{x_0}) & =0 \ \ (i=1,2,...,k) \\
         \dfrac{\partial \psi}{\partial \mu_j}=-x_j & \leq 0 \ \ (j=1,2,...,l) \\
         \mu_j \dfrac{\partial \psi}{\partial \mu_j} =\mu_j (-x_j) & =0 \ \ (j=1,2,...,l) \\
         \lambda_i & \geq 0 \ \ (i=1,2,...,k) \\
         \mu_j & \geq 0 \ \ (j=1,2,...,l)
    \end{array}
\end{equation}
\section{M\'etodo de Nelder Mead}

Sea $f(x_1, x_2, ...,x_n)$ la funci\'on a minimizar, lidiamos con un espacio $n$ dimensional, como \textbf{primer paso} generamos un simplex aleatorio de $n+1$ puntos: $x_1,x_2,...,x_{n+1}$. Los vertices del simplex se acomodar\'an de tal forma que 

\begin{equation}
    f(x_1)\leq f(x_2)\leq ... \leq f(x_{n+1})
\end{equation}

y donde $f(x_1)$ representa el mejor punto o punto mas \'optimo y $f(x_{n+1})$ el peor punto. Por simplicidad trabajemos con solo 3 puntos y denotemos: $x_p$ como el mejor punto, $x_s$ como el segundo mejor punto, $x_t$ como el peor punto. El simplex quedar\'ia

\begin{figure}[h!]
    \centering
    \includegraphics[scale=0.5]{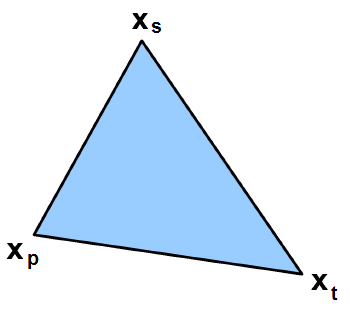}
    \label{nd1}
\end{figure}

El \textbf{segundo paso} del algoritmo consiste en encontrar el centroide considerando todos los puntos excepto el peor. Matem\'aticamente tenemos

\begin{equation}
    x=\dfrac{1}{n} \sum \limits_{i=1}^{n} x_i
\end{equation}

\begin{figure}[h!]
    \centering
    \includegraphics[scale=0.5]{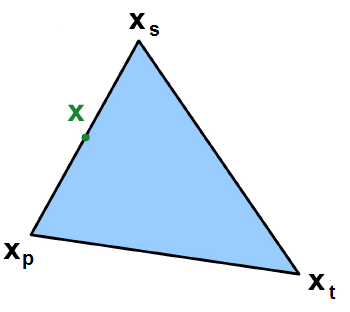}
    \label{nd2}
\end{figure}

El \textbf{tercer paso} se denomina \textbf{transformaci\'on} donde en primer lugar calcularemos un punto de reflexi\'on $x_r$, es decir, el que se refleja del centroide

\begin{equation}
    x_r = x + \rho (x-x_{n-1})
\end{equation}

donde $\rho$ es el par\'ametro de reflexi\'on y habitualmente se toma como 1

\begin{figure}[h!]
    \centering
    \includegraphics[scale=0.5]{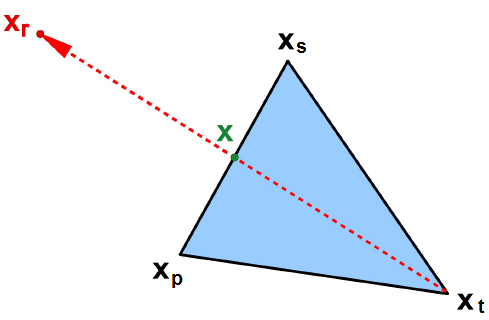}
    \label{nd3}
\end{figure}

Posteriormente calculamos el valor de $f(x_r)$ y tendremos 4 posibilidades: 1) $x_r$ es mejor que $x_n$ pero peor que $x_1$; 2) $x_r$ es mejor que todos los demas puntos; 3) $x_r$ es mejor que $x_{n+1}$ pero peor que todos los dem\'as puntos; 4) $x_r$ es peor que todos los demas puntos.
\\

\textbf{Caso 1:} $x_r$ mejor que $x_n$ pero peor que $x_1$, $f(x_1)\leq f(x_r) \leq f(x_n)$.\\
En este caso se crear\'a un nuevo simplex con v\'ertices $(x_1,x_2,...,x_r)$ o en otras palabras, la \textbf{reflexi\'on} suceder\'a.

\begin{figure}[h!]
    \centering
    \includegraphics[scale=0.5]{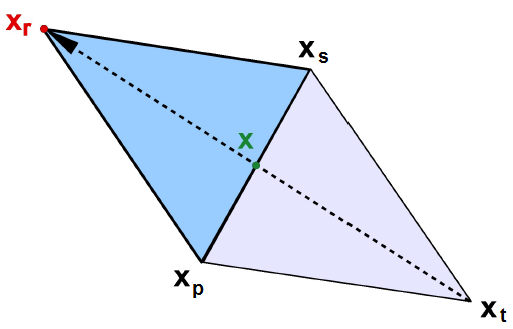}
    \label{nd4}
\end{figure}

\textbf{Caso 2:} $x_r$  mejor que todos los dem\'as puntos, $f(x_r) \leq f(x_1) \leq f(x_2) \leq ... \leq f(x_{n+1})$.

Si esto se cumple, calculamos la \textbf{expansi\'on} hacia el punto $x_e$

\begin{equation}
    x_e = x+\chi (x_r - x)
\end{equation}

donde $\chi$ es el par\'ametro de expansi\'on y debe cumplirse que $\chi > 1$, un valor estandar que se usa es $\chi=2$.

\begin{figure}[h!]
    \centering
    \includegraphics[scale=0.6]{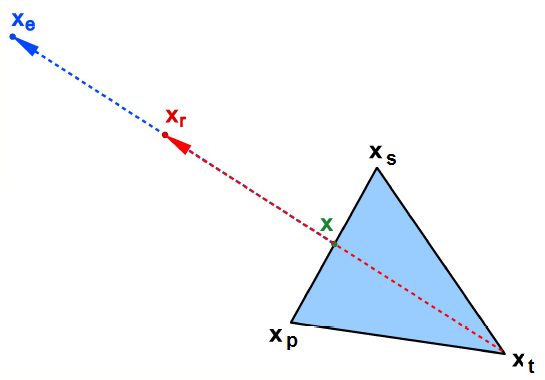}
    \label{nd5}
\end{figure}

Dentro de este caso existen dos posibilidades adicionales: a) Si $f(x_e) \geq f(x_r)$ ($f(x_e)$ es peor que $f(x_r)$) entonces hay que tomar $x_r$ y aceptar la \textbf{reflexi\'on}; b) Si $f(x_e) \leq f(x_r)$ ($f(x_e)$ es mejor que $f(x_r)$) entonces tomamos $x_e$ como nuevo mejor punto, es decir, aceptamos la expansi\'on y formamos un nuevo simplex con v\'ertices $(x_1,x_2,...,x_n,x_e)$

\begin{figure}[h!]
    \centering
    \includegraphics[scale=0.6]{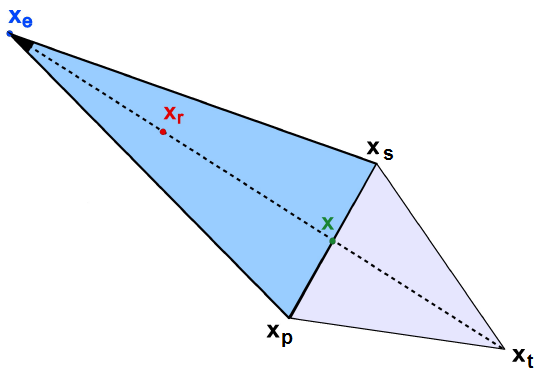}
    \label{nd6}
\end{figure}
\vspace{15pt}
\textbf{Caso 3:} $x_r$ mejor que $x_{n+1}$ pero peor que todos los dem\'as puntos, $f(x_n) \leq f(x_r) \leq f(x_{n+1})$.\\
Si esto se cumple realizamos una \textbf{contracci\'on exterior} hacia el centroide 

\begin{equation}
    x_{oc}=x+\gamma(x_r - x)
\end{equation}

donde $\gamma$ es el par\'ametro de contracci\'on tal que $0< \gamma < 1$, un valor comunmente tomado es $\gamma=1/2$. 

\begin{figure}[h!]
    \centering
    \includegraphics[scale=0.6]{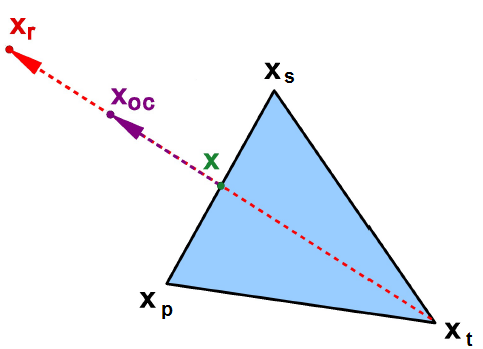}
    \label{nd7}
\end{figure}

Si $f(x_{oc}) \leq f(x_r)$ ($f(x_{oc})$ mejor que $f(x_r)$) entonces aceptamos la contracci\'on exterior como nuevo mejor punto y procedemos a formar un nuevo simplex con v\'ertices ($x_1,x_2,...,x_n,x_{oc}$)

\begin{figure}[h!]
    \centering
    \includegraphics[scale=0.6]{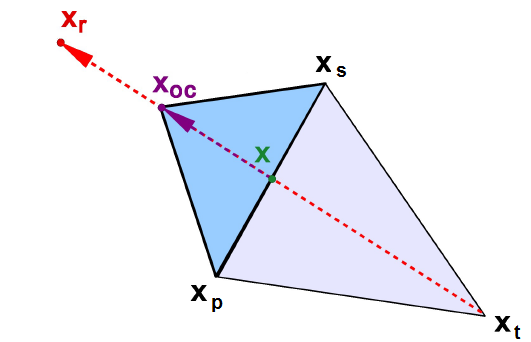}
    \label{nd8}
\end{figure}

si lo anterior no se cumple, se procede a realizar una contracci\'on total del simplex, como veremos mas adelante.

\textbf{Caso 4:} $x_r$  peor que todos los demas puntos, $f(x_r)\geq f_{n+1}$. \\
Si este es el caso, realizaremos una \textbf{contracci\'on interior} (hacia dentro del simplex)

\begin{equation}
    x{ic}=x+\gamma(x-x_{n+1})
\end{equation}

el $\gamma$ sigue siendo el mismo par\'ametro de contracci\'on. 

\begin{figure}[h!]
    \centering
    \includegraphics[scale=0.6]{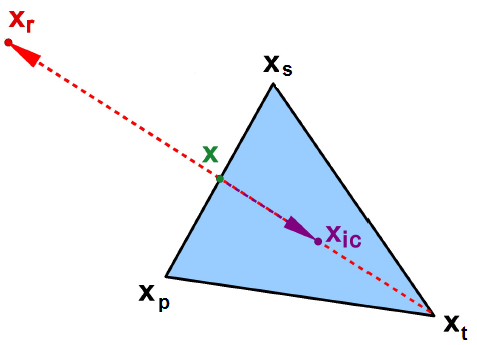}
    \label{nd9}
\end{figure}

Si $f(x_i) \leq f(x_{n+1})$ ($f(x_i)$ es mejor que $f(x_{n+1})$) entonces la \textbf{contracci\'on interior} es aceptada y se forma un nuevo simplex con vertices ($x_1,x_2,...,x_n,x_{ic}$). 

\begin{figure}[h!]
    \centering
    \includegraphics[scale=0.6]{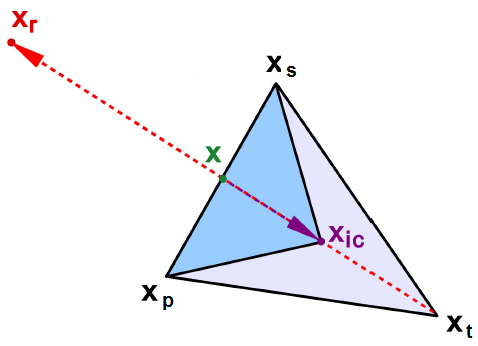}
    \label{nd10}
\end{figure}

Si no se cumple lo anterior entonces se realiza una contracci\'on total del simplex.\\
Si los casos anteriores fallan y se procede a una contracci\'on total del simplex, el simplex es contraido en direcci\'on a el mejor punto, en este caso $x_{p}$. De este modo un $j$-\'esimo nuevo punto ser\'a 

\begin{equation}
    x_j = x_p + \sigma (x_p-x_j)
\end{equation}

donde $\sigma$ es el par\'ametro de contracci\'on total, un valor habitual es $\sigma=1/2$. Una contracci\'on total tendr\'ia la forma siguiente 

\begin{figure}[h!]
    \centering
    \includegraphics[scale=0.6]{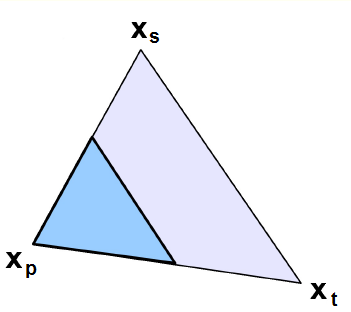}
    \label{nd11}
\end{figure}

El algoritmo continuara hasta que se cumpla algun criterio de parada como que los valores de la funci\'on en los v\'ertices estan muy cerca, cuando los v\'ertices estan muy cercanos unos de otros, cuando se alcanza algun n\'umero l\'imite de iteraciones, etc \cite{361}. 

\section{M\'etodo de Evoluci\'on Diferencial}

Supongamos una poblaci\'on de $NP$ individuos y queremos optimizar una funci\'on de $P$ par\'ametros reales tal que $j=1,2,...,P$. Los vectores de par\'ametros ser\'an

\begin{equation}
    x_{i,G}=[x_{1,i,G},x_{2,i,G},...,x_{P,i,G}] \ \ \ i=1,2,...,NP,
\end{equation}

donde $G$ es el n\'umero de la generaci\'on tal que $G=0,1,2,...,G_{max}$. 

El m\'etodo comienza con la \textbf{inicializaci\'on}, empezando con la generacion $G=0$, definimos l\'imites superiores e inferiores para cada par\'ametro/variable. 

\begin{equation}
\begin{array}{c}
    x_{min}=\{ x_{1,min},x_{2,min},...,x_{P,min} \} \\
    x_{max} = \{ x_{1,max},x_{2,max},...,x_{P,max} \}
\end{array}
\end{equation}

de tal forma que $x_{j,min} \leq x_{j,i,0} \leq x_{j,max}$.

La poblaci\'on para $G=0$ se genera de manera aleatoria y debe cubrir todo el espacio de busqueda. La elecci\'on aleatoria generalmente se lleva a cabo de acuerdo a la siguiente ecuaci\'on \cite{40}

\begin{equation}
x_{j,i}=x_{j,min} + \text{rand}_{i,j}[0,1] (x_{j,max}-x_{j,min})
\end{equation}

y $\text{rand}[0,1]$ es un n\'umero generado aleatoriamente. 

Despues sigue la \textbf{mutaci\'on}, en este paso ocurre un cambio o perturbaci\'on entre los par\'ametros y el espacio de busqueda se amplia. Para un cierto vector de par\'ametros $x_{i,G}$, seleccionamos de manera aleatoria tres vectores distintos $x_{r1,G}$, $x_{r2,G}$ y $x_{r3,G}$ ($r1,r2,r3 \in \{ 1,2,...,NP \}$) y agregamos la diferencia ponderada de dos de los vectores al tercero con el fin de generar un "vector mutado"

\begin{equation}
v_{i,G+1}=x_{r1,G}+F(x_{r2,G}-x_{r3,G})
\end{equation}

donde $F$ es el factor de mutaci\'on tal que $F\in[0,2]$, esta constante controla cuan mayor o menor es el peso de la diferencia de los vectores. 

El siguiente paso se denomina \textbf{recombinaci\'on} o \textbf{cruzamiento}, que se lleva a cabo para aumentar la diversidad entre los individuos de la poblaci\'on (como su nombre lo sugiere). Hacemos uso de un vector de prueba que incorpora elementos de $x_{i,G}$ y $v_{i,G+1}$, se define

\begin{equation}
u_{i,G+1}=(u_{1i,G+1},u_{2i,G+1},...,u_{Pi,G+1})
\end{equation}

donde

\begin{equation}
u_{ji,G+1}=\left \{ \begin{array}{cc}
    v_{ji,G+1} & \text{si} \ \ \text{randn}(j) \leq CR \ \ \text{\'o} \ \ j=\text{randp} \\
    x_{ji,G} & \text{si} \ \ \text{randn(j)}> CR \ \  \text{y} \ \ j \neq \text{randp}
\end{array} \right. \ \ \ \begin{array}{c}j=1,2,...,P \\ i=1,2,...,NP \end{array}
\end{equation}

donde randn($j$) es la $j$-\'esima evaluaci\'on de un generador de n\'umeros aleatorios cuyo resultado esta entre 0 y 1, CR es la constante de cruzamiento tal que $CR \in [0,1]$ y randp es el valor de un \'indice aleatorio tomado de $j \in 1,2,...,P$ y asegura que el vector de prueba tenga al menos un elemento de el vector mutado. 

Finalmente, se concluye con la \textbf{selecci\'on}. En este paso el vector de par\'ametros $x_{i,G}$ se compara con el vector de prueba $u_{i,G+1}$, aquel que evaluado en la funci\'on tenga el menor valor, ser\'a admitido en la siguiente generaci\'on

\begin{equation}
x_{i,G+1}= \left \{ \begin{array}{cc}
    u_{i,G+1} & \text{si} \ \ f(u_{i,G+1}) \leq f(x_{i,G})  \\
     x_{i,G} & \text{c.o.c.} \ \ i=1,2,...,NP 
\end{array} \right.
\end{equation}

La mutaci\'on, recombinaci\'on y cruzamiento continuar\'an hasta que se cumpla o alcance algun criterio de parada, por ejemplo añadiendo un contador al final de cada generaci\'on y parar hasta que el contador sea menor o igual al tamaño de la poblaci\'on.

\end{document}